\documentclass[aps,prd,reprint,superscriptaddress,preprintnumbers,amsmath,nofootinbib]{revtex4-2}
\usepackage{graphicx}% Include figure files
\usepackage{dcolumn}% Align table columns on decimal point
\usepackage{bm}% bold math
\usepackage{braket}
\usepackage{epstopdf}
\usepackage{hyperref}% add hypertext capabilities
\usepackage{color}
\usepackage{cancel}
\usepackage{amsfonts}
\begin{document}
\preprint{RIKEN-iTHEMS-Report-25, FQSP-25-3, YITP-25-110}

%Title of paper
\title{Wavefunction-based operator optimization for two-hadron systems in lattice QCD }
%=======================================================================================

\author{Yan Lyu}
\email{yan.lyu@riken.jp}
\affiliation{RIKEN Center for Interdisciplinary Theoretical and
Mathematical Sciences (iTHEMS), RIKEN, Wako 351-0198, Japan}

\author{Sinya Aoki}
\email{saoki@yukawa.kyoto-u.ac.jp}
\affiliation{Fundamental Quantum Science Program (FQSP), TRIP Headquarters, RIKEN, Wako 351-0198, Japan}
\affiliation{Center for Gravitational Physics and Quantum Information, Yukawa Institute for Theoretical Physics, Kyoto University, Kyoto 606-8502, Japan}

\author{Takumi Doi}
\email{doi@ribf.riken.jp}
\affiliation{RIKEN Center for Interdisciplinary Theoretical and
Mathematical Sciences (iTHEMS), RIKEN, Wako 351-0198, Japan}

\author{Tetsuo Hatsuda}
\email{thatsuda@riken.jp}
\affiliation{RIKEN Center for Interdisciplinary Theoretical and
Mathematical Sciences (iTHEMS), RIKEN, Wako 351-0198, Japan}
\affiliation{Kavli Institute for the Physics and Mathematics of the Universe (Kavli IPMU), WPI, The University of Tokyo, Kashiwa, Chiba 277-8568, Japan}

\author{Kotaro Murakami}
\email{kotaro.murakami@yukawa.kyoto-u.ac.jp}
\affiliation{Department of Physics, Institute of Science Tokyo, 2-12-1 Ookayama, Megro, Tokyo 152-8551, Japan}
\affiliation{RIKEN Center for Interdisciplinary Theoretical and
Mathematical Sciences (iTHEMS), RIKEN, Wako 351-0198, Japan}

\author{Takuya Sugiura}
\email{sugiura@rcnp.osaka-u.ac.jp}
\affiliation{Faculty of Data Science, Rissho University, Kumagaya 360-0194, Japan}
\affiliation{RIKEN Center for Interdisciplinary Theoretical and
Mathematical Sciences (iTHEMS), RIKEN, Wako 351-0198, Japan}

\date{\today}
%=======================================================================================
\begin{abstract}
A systematic way to constructing optimized interpolating operators for two-hadron systems is developed by incorporating inter-hadron spatial wavefunctions.
The wavefunctions can 
be obtained from an iterative process with an appropriate initial guess.
To implement these operators,
a novel quark smearing technique utilizing $Z_3$ noise vectors is proposed, which allows for
effectively incorporating inter-hadron spatial wavefunctions at the source
without using all-to-all quark propagators.
Proof-of-principle application to the $\Omega_{ccc}\Omega_{ccc}$ system
using physical-point lattice configurations with a large size $La\simeq8.1$~fm
demonstrates that optimized operators outperform combinations of limited plane-wave operators in the variational analysis, enabling clear identification of states
around $2m_{\Omega_{ccc}}\simeq 9700$ MeV with the energy gap as narrow as $\sim 5$ MeV. 
A comparison on
correlation functions, effective energies, and HAL QCD potentials 
between unoptimized operators and optimized operators is given, with a special emphasis on the effects from nearby elastic scattering states.
Potential applicability of the optimized operator to various two-hadron systems and its relation to the variational method are also discussed.
\end{abstract}
%=======================================================================================

%\keywords{}

%\maketitle must follow title, authors, abstract, \pacs, and \keywords
\maketitle

%=================================================

\section{Introduction}
Lattice Quantum Chromodynamics (QCD) provides a nonperturbative approach through numerical simulations to  {\it ab-initio} calculations of QCD, the theory of the strong interaction of quarks and gluons in Standard Model.
In recent years, lattice QCD becomes
more and more important in understanding  
low-energy aspects of QCD,  relevant to hadron structures, spectra and interactions,
thanks to rapid developments in both theoretical and computational sides~\cite{FLAG2024}.

A lattice QCD calculation typically involves evaluations of hadronic correlation functions, such as the two-point function $\braket{B(t)\bar B (0)}$ relevant to hadron spectroscopy,
which receive contributions from a tower of states that the creation operator $\bar B$ can excite out of the QCD vacuum.
The ground state hadron can be identified when contributions from all higher states are strongly suppressed $O(e^{-\Delta E^* t})$
at large Euclidean time $t\gg 1/\Delta E^*$, with the typical energy gap to the nearest higher state $\Delta E^*\sim m_\pi$ or $\Delta E^* \sim\Lambda_\text{QCD}$.
Although any operators with given quantum numbers in principle can be used to calculate hadronic correlation functions,
some strongly couple to the ground state hadron while largely suppress  inelastic excited states, leading to better signal.
To obtain such operators for single hadrons,
the smearing technique is employed, in which the quark field is smeared in a way that the hadron operator $B$ constructed from it mimics the spatial profile (``shape'') of the target hadron~\cite{Gusken:1989ad}.
(See Refs~\cite{HadronSpectrum:2009krc,Morningstar:2011ka, Bali:2016lva, Morningstar:2013bda, Larsen:2020rjk} for advanced smearing techniques and
construction of extended hadronic operators).
Using these operators, lattice QCD has determined single hadron spectra that agree remarkably with experiments~\cite{FLAG2021,PACS-CS:2008bkb,Fodor2012,Aoyama:2024cko}.

In two-hadron systems, which are crucially important for understanding hadron interactions, finite nuclei and nuclear matter,
the presence of elastic scattering states poses significant challenges
to lattice QCD calculations.
This is because the energy gap between neighboring elastic scattering states is approximately given by $\Delta E^{\text{el}} \sim \dfrac{4\pi^2}{m_{B} L^2}$, 
shrinking with larger lattice sizes $L$ and heavier hadron mass $m_{B}$.
Such dense states make state identification particularly difficult, as contributions from nearby states decrease slowly as $e^{-\Delta E^{\text{el}} t}$
and only become negligible at sufficiently large Euclidean time $t\gg 1/\Delta E^{\text{el}}$.
Usually, by this time the signal-to-noise ratio has deteriorated significantly due to growing statistical fluctuations, 
in particular for nucleon systems~\cite{Parisi:1983ae,Lepage1989}.
This difficulty indeed led to serious problems on two-nucleon systems
in practice:
bound $NN$ states (in both isospin channels) were claimed by early lattice calculations~\cite{Yamazaki:2012hi,Yamazaki:2015asa, NPLQCD:2011naw, NPLQCD:2012mex,NPLQCD:2013bqy, Orginos:2015aya, Berkowitz:2015eaa} at heavy pion masses, where 
two-nucleon correlation functions constructed using compact operators (two nucleon operators are placed in the same/close spatial positions) were used to extract finite volume energies at Euclidean time region $t\sim1-2~$fm,
which is too short to sufficiently suppress contributions from nearby states and thus leads to a high chance of misidentification of states, as pointed out in Refs.~\cite{Iritani2016,Iritani:2017rlk,Iritani2019Jhep}.
Subsequent calculations~\cite{Francis2019,Horz2021,Amarasinghe:2021lqa, BaSc:2025yhy} 
using operators from the variational method 
found large modification of finite volume energies compared to those obtained from compact operators, 
leading to the absence of $NN$ bound states at heavy pion masses claimed before.
The variational method~\cite{Luscher:1990ck,Blossier:2009kd} 
improves the coupling of two-hadron operators to specific states
by linearly combining a set of 
operators with low-momentum mode such as $\{ B(\vec p) B(-\vec p)\}$ with $\frac{L}{2\pi}\vec p=(0,0,0), (0,0,\pm 1),\cdots$,
making state identification possible at relatively short Euclidean time.

In analogy with operator optimization in single-hadron case,
one may also consider optimizing two-hadron operators by making them more accurately represent the ``shape'' of specific states.
Intuitively, two-hadron states exhibit distinct spatial profiles in coordinate space. In quantum mechanics, the stationary scattering wavefunctions of two particles in the center-of-mass (CM) frame with different energies are indeed orthogonal to each other. 
Early efforts in this direction include two-hadron operators with fixed displacement~\cite{Kurth:2013tua,Berkowitz:2015eaa,Wu:2021xvz} and with exponential dependence~\cite{Amarasinghe:2021lqa}. 
In this paper, 
we make a serious attempt to
optimize two-hadron operators by incorporating inter-hadron spatial information.
This is achieved by addressing two fundamental questions: how to systematically construct such information, and how to effectively incorporate it.
Since conventional quark smearing fails to encode the relative spatial structure between composite hadrons, incorporating this information typically requires the use of computationally expensive all-to-all propagators.

This paper is organized as follows.
In Sec.~\ref{Sec-opt-o}, we explain 
how to optimize the two-hadron operators by incorporating inter-hadron spatial wavefunctions.
Sec.~\ref{Sec-dual-f} discusses the construction of these spatial wavefunctions.
To implement these optimized operators at the source, we propose a novel quark smearing method in Sec.~~\ref{Sec-imp}.
After showing details of our calculation setup in Sec.~\ref{Sec-lat-setup}, 
we demonstrate our method through a proof-of-principle application in Sec.~\ref{Sec-results}.
Sec.~\ref{Sec-summary} is devoted to the summary and discussions.
Details relevant to the content of the main text are given in Appendices~\ref{Sec-err}, ~\ref{Sec-wick-cont},~\ref{Sec-sink-opr} and~\ref{Sec-phaseshift-multi}.
A companion Letter summarizing these results has been published alongside this work~\cite{Lyu:2025lnd}.
%=================================================
\section{Optimized two-hadron operators}\label{Sec-opt-o}
To study two-hadron systems on the lattice, typically, one can measure a correlation function of generating two hadrons at $t_0$ and annihilating them at $t$, i.e.,
\begin{align}
    &\mathcal{F}(\vec r_\text{snk}, t;\vec r_\text{src}, t_0), \nonumber \\
    &=\frac{1}{V^2}\sum_{\vec{x},\vec{y}\in 
 \Lambda}\langle  B(\vec x+\vec r_\text{snk},t) B(\vec x,t)  
  \bar { B}(\vec y+\vec r_\text{src},t_0) \bar { B}(\vec y,t_0)\rangle. \label{Eq-F}
\end{align}
Here, we consider a system of two identical baryons for simplicity, and generalization to distinct hadron pair cases is straightforward.
The summation for spatial indices $\vec x$ (sink) and $\vec y$ (source) over all lattice sites $\Lambda$ with volume $V$ places the system into the CM frame.
Transferring to the moving frame with total momentum $\vec P$
is given by multiplying with appropriate factors $e^{i\vec P \cdot \vec y}$ and $e^{-i\vec P \cdot \vec x}$ at the source and the sink, respectively.

In practice, it is useful to consider the $R$ correlation function 
defined below, 
which isolates the effects of the inter-hadron interactions by subtracting the two-hadron threshold energy.
It also exhibits clear signals compared to those from $\mathcal{F}$ in Eq.~(\ref{Eq-F}), and can be decomposed as contributions from  elastic scattering states as well as possible bound state(s) when the time separation between sink and source is large $t-t_0\gg\frac{1}{\Delta E^*}$ with $\Delta E^*$ being the energy gap to the nearest inelastic states.
\begin{align}\label{Eq-R}
    &\mathcal{R}(\vec r_\text{snk}, t;\vec r_\text{src})  =  \frac{\mathcal{F}(\vec r_\text{snk}, t;\vec r_\text{src}, t_0=0)}{C_B(t)\cdot C_B(t)}\nonumber \\ 
  &=\sum^N_{n=0}\psi_n(\vec r_\text{snk}) \psi^*_n(\vec r_\text{src}) e^{-\Delta E_nt} + O(e^{-\Delta E^*t}), 
\end{align}
where the two point function $C_B(t)$ for the baryon $B$ with mass $m_B$ is defined as,
\begin{align}
    C_B(t) &= \frac{1}{V^2}\sum_{\vec{x},\vec{y}\in 
 \Lambda}\langle   B(\vec x,t) \bar { B}(\vec y,t_0=0)\rangle, \nonumber\\
 &=Z_Be^{-m_B t} + O(e^{-m_{B^*}t}), \label{Eq-C2}
\end{align}
with the wavefunction normalization factor$\sqrt{Z_B}=\frac{1}{V}\sum\limits_{\vec x\in \Lambda}\langle 0|B(\vec x,0)|B, \vec p=0\rangle$, and $m_{B^*}$ being the mass of the possible 1st excited state of the baryon $B$.

The equal-time Nambu-Bethe-Salpeter (NBS) amplitude $\psi_n(\vec r)$ is defined as,
\begin{align} \label{Eq-NBS}
    \psi_n(\vec r)=\frac{1}{Z_BV}\sum_{\vec x\in\Lambda} \langle 0| B(\vec x+\vec r,0) B(\vec x,0)|2B, E_n\rangle,
\end{align}
where $|2B, E_n\rangle$ is the two-baryon state below the inelastic threshold $E_{n} <E^*$ with $E^*$ lying approximately $2m_\pi\simeq 280$ MeV higher.
The NBS amplitude represents nothing but the amplitude to find one baryon at $\vec x$ and another at $\vec x + \vec r$, 
acting as quantum field theory analogy 
of the stationary scattering wavefunction in quantum mechanics~\cite{Ishii2007}.

A key difference between two-hadron systems in Eq.~(\ref{Eq-R}) and single hadron systems in Eq.~(\ref{Eq-C2}) is the presence of elastic scattering states (labeled by $n$) in the former.
These states complicate spectral identification, as the energy gaps scale as $\Delta E^{\text{el}} \sim \frac{4\pi^2}{m_{B} L^2}$, shrinking with larger lattice sizes $L$ or heavier mass $m_{B}$.
To address this challenge, we exploit $\psi_n$ as a function of $\vec r_\text{snk}/ \vec r_\text{src}$ to enhance contributions from states of interests while suppressing others.
This leads us to introducing a set of dual functions $\{\Psi_n(\vec r)\}$ for each state, satisfying,
\begin{align}
    \langle \Psi_n | \psi_m\rangle=\frac{1}{V}\sum_{\vec r\in\Lambda} \Psi^*_n(\vec r) \psi_m(\vec r) = \delta_{nm}.
\end{align}
A set of highly optimized two-hadron operators $\{ O_n(t)\}$
can be defined accordingly as follows,
\begin{align}\label{Eq-O}
     O_n(t) = \frac{1}{V^2}\sum_{\vec x,\vec r\in\Lambda} B(\vec x+\vec r,t)  B(\vec x,t) \Psi^*_n(\vec r),
\end{align}
which strongly couples to the state $|2B, E_n\rangle$.
This results in the correlation functions $R_n(\vec r_\text{snk},t)$ dominated by the $\psi_n(\vec r_\text{snk})e^{-\Delta E_nt}$ term when 
such an operator is employed at the source,
\begin{align}
    R_n(\vec r_\text{snk},t) &=\frac{1}{V}\sum_{\vec x\in\Lambda}\langle  B(\vec x+\vec r_\text{snk},t) B(\vec x,t) \bar O_n(0) \rangle/{C^2_B(t)} ,\nonumber \\ 
    &=\frac{1}{V}\sum_{\vec r_\text{src}\in\Lambda}\mathcal{R}(\vec r_\text{snk}, t;\vec r_\text{src}, t_0) \Psi_n(\vec r_\text{src}) \nonumber \\
    & = \psi_n(\vec r_\text{snk}) e^{-\Delta E_nt}, \label{Eq-Rr}
\end{align}
and the correlation function $R_n(t)$ dominated by the $e^{-\Delta E_n t}$ term when 
the operator is employed at both the source and the sink,
\begin{align}
    R_n(t) &=\langle O_n(t) \bar O_n(0) \rangle/C^2_B(t), \nonumber\\
    &=\frac{1}{V}\sum_{\vec r_\text{snk},\in\Lambda}\Psi^*_n(\vec r_\text{snk}) R_n(\vec r_\text{snk},t)\nonumber \\
    & = e^{-\Delta E_n t},  \label{Eq-Rt}
\end{align}
where the contribution from inelastic states is neglected.
The effective energy of $n$-th state can be derived from $R_n(t)$,
\begin{align}
    \Delta E^{\text{eff}}_n(t) = \ln\left[\frac{R_n(t)}{R_n(t+1)}\right].\label{Eq-Eeff}
\end{align}

The dual function $\Psi_n(\vec r)$ 
can be obtained by using the kernel $\mathcal{K}$ of the NBS amplitudes as follows,
\begin{align} \label{Eq-APsi}
    & \Psi_n(\vec r) =\sum^N_{m=0}\mathcal\psi_m(\vec r){K}^{-1}_{mn},\\
    & \mathcal{K}_{mn} = \langle \psi_m|\psi_n \rangle,
\end{align}
which satisfies the orthogonal relation $\langle \Psi_n|\psi_m \rangle  = \delta_{nm}$.
The exact determination of $\Psi_{n}$
requires a complete knowledge of all NBS amplitudes $\{\psi_n(\vec r)\}$ below the inelastic threshold, which are difficult to be obtained in general. 
In practice, being lack of accurate $\Psi_n (\vec r)$,
conventional operators can only provide approximate reconstructions of the genuine wavefunction:
\begin{enumerate}
    \item[(i).] $\Psi_n(\vec r)\sim \delta(\vec r)$.\\
    This setup was widely used in early studies~\cite{Yamazaki:2012hi,Yamazaki:2015asa, NPLQCD:2011naw, NPLQCD:2012mex,NPLQCD:2013bqy, Orginos:2015aya, Berkowitz:2015eaa}, in which a local smearing function was first tuned according to single baryon, and then applied to two-baryon systems. 
    Intuitively, the delta-like function places two baryon directly on top of each other, which generally does not match with profiles of the low-energy states. 
    The observation that the two-baryon operator constructed with $\delta(\vec r)$ receives nearly equal contributions from all elastic states in the weak interaction limit suggests substantial contaminations from high-energy elastic states, which make isolating low-energy states particularly challenging~\cite{Iritani2016,Iritani:2017rlk,Iritani2019Jhep}.
    \item[(ii).] $\Psi_n(\vec r)\sim 1$. \\
    This setup, which assigns equal weight to all spatial separations $\vec r$ between two baryons, is expected to be a reasonable approximation to the two-baryon system with zero relative momentum $B(\vec p = 0)B(\vec p = 0)$. 
    %To our knowledge, such operators have not been directly applied in practice.
    The wall source operator shares a similar equal-weighting characteristic in the sense that all spatial positions are weighted equally, although wall source is too extended to have well defined local baryons.
    \item[(iii).] $\Psi_n(\vec r)$ approximated by a few leading terms in its Fourier series. \\
    The optimized operator in Eq.~(\ref{Eq-O}) can be defined in momentum space as well,
    \begin{align}
    {O}_n(t) =& \frac{1}{V^4}\sum_{\vec{x},\vec{r}\in \Lambda} \sum_{\vec{p},\vec{q}\in \tilde{\Lambda} }\tilde{B}(\vec p,t) \tilde B(\vec q,t)e^{i\vec{q}\cdot\vec{x} +i\vec{p}\cdot(\vec{x}+\vec{r})} \Psi^*_n(\vec{r}), \nonumber\\
    =& \frac{1}{V^3}\sum_{\vec{p}\in \tilde{\Lambda} }\tilde{B}(\vec p,t)\tilde{B}(-\vec p,t)\sum_{\vec{r}\in V} \Psi^*_n(\vec{r})e^{i\vec{p}\cdot\vec{r}}, \nonumber \\
    =& \frac{1}{V^3}\sum_{\vec{p}\in \tilde{\Lambda} }\tilde{B}(\vec p,t)\tilde{B}(-\vec p,t)\tilde\Psi^*_n(-\vec{p}), \label{Eq-O-p}
    \end{align}
    where $\tilde\Psi(\vec p)$ and $\tilde B(\vec p)$ are Fourier transformation of $\psi(\vec r)$ and $ B(\vec r)$, respectively.
    An approximation is given by a truncation of the summation over $\vec p$ into a few low-momentum modes $\frac{L}{2\pi}\vec p=(0,0,0), (0,0,\pm 1),\cdots$.
    This approach has been commonly adopted in the variational framework~\cite{Luscher:1990ck,Blossier:2009kd}, in which a set of two-hadron operators, typically constructed from those with low-momentum modes $\{B(\vec p)B(-\vec p)\}$, is linearly combined through diagonalization of the hadronic correlation function matrix to optimize operators.
\end{enumerate}

%=================================================
\section{Dual functions from the HAL QCD method}\label{Sec-dual-f}

As discussed in the above section, a crucial component of constructing optimized two-hadron operators is the dual functions $\{\Psi_n(\vec r)\}$.
In this section, we develop a systematic way to constructing $\{\Psi_n(\vec r)\}$ through iterative application of the HAL QCD method.

\subsection{HAL QCD method} \label{Sec-Dual-Func-A}
The HAL QCD method~\cite{Ishii2007,Ishii2012,Aoki2010,Aoki2012,Aoki2020} exploits the fact that the NBS amplitude in Eq.~(\ref{Eq-NBS}) encodes the information of the scattering matrix, or equivalently, the partial wave scattering phaseshift $\delta_l$ once projected to specific channels,
\begin{align}
    \psi_{n,l}(r)\xrightarrow{r\rightarrow\infty} \frac{e^{i\delta_l(p_n)}}{p_nr}\sin\left(p_nr-\frac{l\pi}{2}+\delta_l(p_n)\right),
\end{align}
with $p_n=\sqrt{E_n^2/4-m^2_B}$.
This allows for introducing a local energy-dependent potential, or a non-local but energy-independent potential as follows~\cite{Ishii2007}.
\begin{align}
    \left(\frac{\nabla^2}{m_B} +\frac{p_n^2}{m_B}\right)\psi_{n}(\vec r) = \int d^3 r_0 U(\vec r, \vec r_0) \psi_{n}(\vec r_0).~\label{Eq-HAL}
\end{align}
The potential $U(\vec r,\vec r_0)$ by construction gives the correct scattering phaseshift $\delta_l$ at the momentum $p_n$.

The energy-independent nature of $U(\vec x,\vec y)$ suggests that it 
governs all elastic scattering states, and consequently state identification is not required anymore. This nice property leads to the time-dependent HAL QCD equation~\cite{Ishii2012},
\begin{align}
    \left( \frac{1}{4m_B}\frac{\partial^2}{\partial t^2} -\frac{\partial}{\partial t} + \frac{\nabla^2}{m_B} \right)R_J(\vec r, t) = \int d^3 r_0 U(\vec r, \vec r_0) R_{J}(\vec r_0,t),~\label{Eq-td-HAL}
\end{align}
where $R_J(\vec r, t)$ is the $R$ correlation function calculated with a given (unoptimized) source operator $\bar J(0)$ (such as the wall source operator, the conventional smearing source operator, etc.),
\begin{align}
 R_{J}(\vec r,t)=\frac{1}{V}\sum_{\vec x\in\Lambda}\langle B(\vec x+\vec r,t) B(\vec x,t) \bar J(0) \rangle/{C^2_B(t)}, \label{Eq-RJ}
\end{align}
and receives contributions from multiple elastic scattering states.

To study low-energy scattering properties, 
the derivative expansion to the nonlocal potential $U(\vec r,\vec r_0)$ is employed in practice 
\footnote{
The derivative expansion also fixes a scheme of the nonlocal potential $U(\vec r,\vec r_0)$, since it is not uniquely determined in Eq.~(\ref{Eq-td-HAL}), as shown in Ref.~\cite{Aoki2012}.
Theoretically, potentials from different schemes result in exact same physical observables below the inelastic threshold $E^*$.
}.
In the spin-singlet case, the leading order (LO) expansion is given by,
\begin{align}
    U(\vec r-\vec r_0)=V^\text{LO}(\vec r)\delta(\vec r-\vec r_0),\label{Eq-HAL-LO}
\end{align}
with a local potential,
\begin{align}
    V^\text{LO}(\vec r) = R^{-1}_J(\vec r, t) \left( \frac{1}{4m_B}\frac{\partial^2}{\partial t^2} -\frac{\partial}{\partial t} + \frac{\nabla^2}{m_B} \right)R_J(\vec r, t). \label{Eq-HAL-LOV}
\end{align}
Here, $V^\text{LO}(\vec r)$ should be understood as $V^\text{LO}_J(\vec r)$ with the subscript $J$ stressing the source-operator dependence of the potential.
We note that this potential provides an approximated description to all states contained in $R_J(\vec r, t)$.
The accuracy of the description for a particular state improves with that state's relative contribution to $R_J(\vec r, t)$.
In particular, if $R_J(\vec r, t)$ is dominated by a single state, the potential becomes exact to that state.

The effect from the high-order terms in the expansion can be estimated by the variation of $V^\text{LO}(\vec r)$ extract at different $t$, as the relative contribution of $\psi_n(\vec r)$ to $R_J(\vec r, t)$ evolve with Euclidean time $t$.
Alternatively, one can estimate the effect by extract higher order terms explicitly as follows~\cite{Iritani2019PRD}.
\begin{align}
    U(\vec r-\vec r_0)=\left[V^{\text{N}^2\text{LO}}_0 + V^{\text{N}^2\text{LO}}_2(\vec r)\nabla^2 \right] \delta(\vec r-\vec r_0), \label{Eq-HAL-N2LO}
\end{align}
where each potential is given by,
\begin{align}
    V^{\text{N}^2\text{LO}}_0 (\vec r) &= V^\text{LO}_J(\vec r) + \frac{V^\text{LO}_J(\vec r) - V^\text{LO}_{J'}(\vec r)}{D_1-1},\label{Eq-N2L0-V0}\\
    V^{\text{N}^2\text{LO}}_2(\vec r) & = \frac{V^\text{LO}_J(\vec r) - V^\text{LO}_{J'}(\vec r)}{D_2} \label{Eq-N2L0-V2}.
\end{align}
Here, $V^\text{LO}_J(\vec r)$ and $V^\text{LO}_{J'}(\vec r)$ are the leading order potentials calculated from $R_{J}(\vec r,t)$ with the source operator $\bar J(0)$ and $R_{J'}(\vec r,t)$ with the source operator $\bar J'(0)$, respectively. $D_{1}$ and $D_2$ are defined as follows.
\begin{align}
    D_1 &= \frac{R_{J}(\vec r,t) \nabla^2R_{J'}(\vec r,t)}{R_{J'}(\vec r,t) \nabla^2R_{J}(\vec r,t) }, \\
     D_2 &= \frac{\nabla^2 R_J(\vec r,t)}{R_{J}(\vec r,t)} - \frac{\nabla^2 R_{J'}(\vec r,t)}{R_{J'}(\vec r,t)}.
\end{align}

\subsection{Eigen equations on a finite box}\label{Sec-Dual-Func-B}
Once the potential is obtained by using the time-dependent HAL QCD method,
one can calculate eigen functions and eigen values on a finite box with the obtained potential.
As proposed in our previous studies in Refs.~\cite{Iritani2019Jhep, Lyu2022}, we shall utilize these eigen functions to extract information about the underlying NBS amplitudes $\{\psi_{n}(\vec r)\}$.

Specifically, let us consider the following Schr\"odinger equation defined on a three-dimensional finite box with the periodic boundary condition (PBC) as an eigen value problem,
\begin{align}
    & \left[-\frac{\nabla^2}{m_B} +V^{(0)}(\vec r)\right]\psi^{(0)}_n (\vec r) = \frac{k^2_n}{m_B}\psi^{(0)}_n (\vec r),  \label{Eq-FV}
\end{align}
where $\dfrac{k^2_n}{m_B}$ is the eigen value, which is related to the energy as  
    \begin{align}
    & \varepsilon_n=2\left(\sqrt{m_B^2+k^2_n}-m_B\right), \label{Eq-FV-E}
\end{align}
and $\psi^{(0)}_n (\vec r)$ is the eigen function.
Here, $V^{(0)}(\vec r)$ may be chosen as the LO potential $V^\text{LO}(\vec r)$ in Eq.~(\ref{Eq-HAL-LOV}) calculated from the hadronic function $R_J(\vec r, t)$.
These eigen functions provide good approximation to underlying NBS amplitudes $\{\psi_{n}(\vec r)\}$, and can be used to define optimized two-hadron operators in Eq.~(\ref{Eq-O}).
More specifically, the optimized two-hadron operators can be obtained through an iterative process as follows.
\begin{enumerate}
    \item[(i).] Calculate the hadronic correlation function $R_J(\vec r, t)$ using a given source operator $\bar J(0)$ such as a wall source operator. This initial result is denoted as $R^{(0)}(\vec r, t)$.
    
    \item[(ii).] Solve the eigen functions in Eq.~(\ref{Eq-FV}) with the potential $V^{(0)}(\vec r)$ derived from $R^{(0)}(\vec r, t)$. 
    The resulting eigen functions $\{\psi^{(0)}_n(\vec r)\}$ serve as good approximation to $\{\psi_n(\vec r)\}$ and can be used in Eq.~(\ref{Eq-APsi}) to define dual functions $\{\Psi^{(0)}_n(\vec r)\}$.

    \item[(iii).] Using dual functions $\{\Psi^{(0)}_n(\vec r)\}$, construct the optimized two-hadron operators as in Eq.~(\ref{Eq-O}), and compute the corresponding hadronic correlation functions $\{R^{(1)}_n(\vec r, t)\}$. 

    \item[(iv).] Verify whether obtained $\{R^{(1)}_n(\vec r, t)\}$ are dominated by their respect states, as indicated by a stable spatial profile of $\{R^{(1)}_n(\vec r, t)\}$ against Euclidean time $t$. \\
    If not, define a new set of local potentials $\{V^{(1)}_n(\vec r)\}$ from $\{R^{(1)}_n(\vec r, t)\}$ using Eq.~(\ref{Eq-HAL-LO}), and return to step (ii) to calculate $\{\psi^{(1)}_n(\vec r)\}$, which should be better approximation to the NBS amplitudes $\{\psi_n(\vec r)\}$ than $\{\psi^{(0)}_n(\vec r)\}$, and subsequently to steps (ii), (iii), and (iv). \\
    This process can be repeated iteratively until convergence is achieved. 
\end{enumerate}
Once the convergence is achieved, 
the resulting optimized two-hadron operators $O_n(t)$ can be used to compute correlation functions in Eqs.~(\ref{Eq-Rr}) and (\ref{Eq-Rt}), and to derive the effective energies in Eq.~(\ref{Eq-Eeff}), which should exhibit stable plateaux against $t$.
In particular, these energies should also agree with the energies
$\varepsilon_n$ in Eq.~(\ref{Eq-FV-E}) derived from the potentials.

%=================================================
\section{Implementation}\label{Sec-imp}
The optimized operators in Eq.~(\ref{Eq-O}) can be used as both source and sink operators.
In this section, we discuss how to calculate hadronic correlation functions using the optimized operators.

%==========================
\begin{figure}[htbp]
  \centering
  \includegraphics[width=8.7cm]{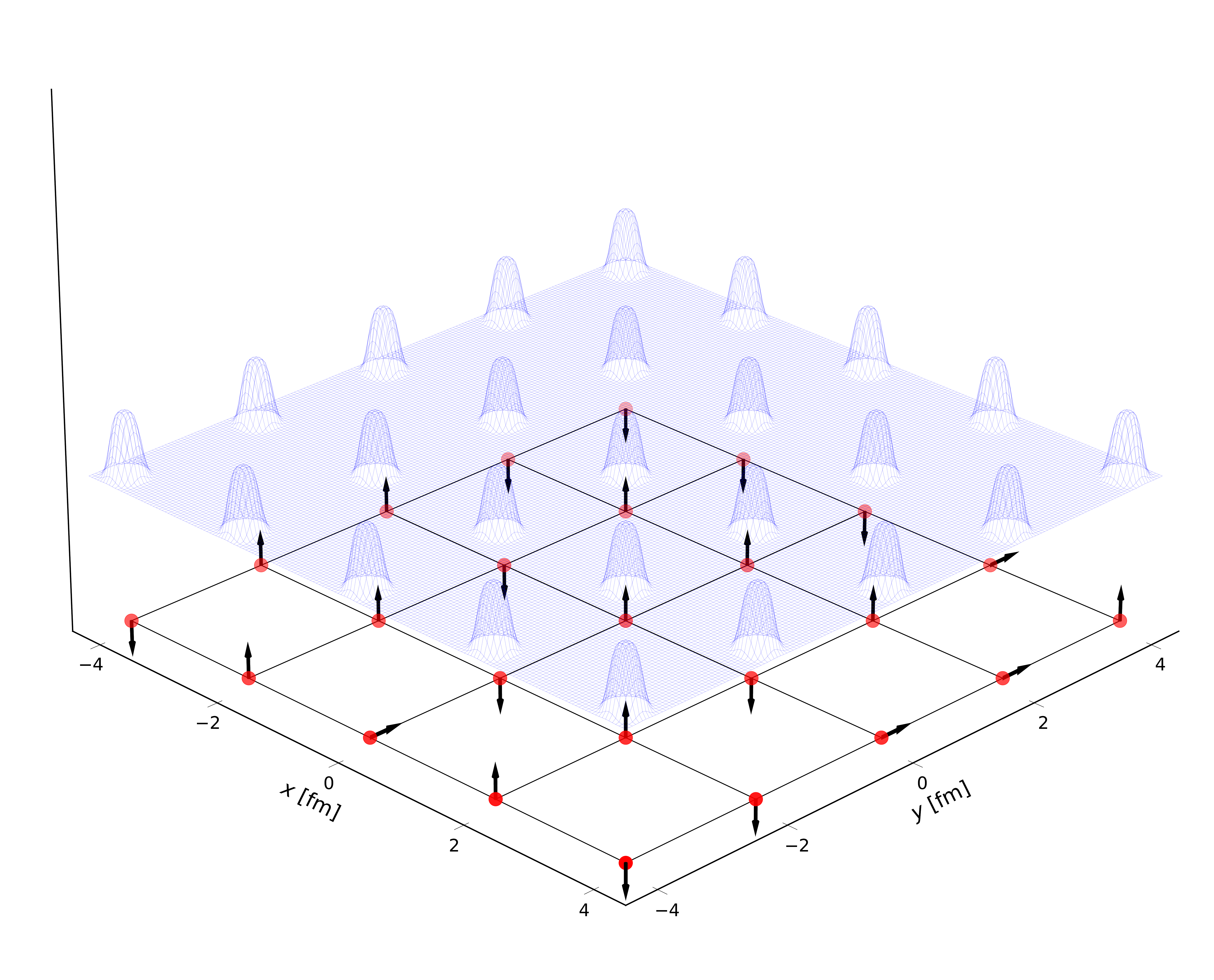}
  \caption{Setup for the novel quark smearing in Eq.~(\ref{Eq-Smear-2}). 
  Each of the multiple supports distributed on a sparse lattice grid (red points) contains a local smearing function $f(\vec r)$ (blue) fine-tuned according to single hadron together with an independent $Z_3$ noise vector (arrows).
  Different supports are weighted differently
  according to the cubic root of inter-hadron spatial wavefunctions $\Psi^{1/3}_n(\vec r)$ in Fig.~\ref{Fig-Psi_01_3d}.
  } \label{Fig-sps-grid}
\end{figure}
%==========================

\subsection{Novel quark smearing for the source operator}\label{Sec-Smear-A}
%==========================
A key challenge to implement the optimized operators comes from the fact that they contain a spatial wavefunction factor $\Psi^*_n(\vec r)$ between two local baryon operators separated by a distance $\vec r$.
In typical lattice QCD calculations,
the spatial index of quark propagators is implicitly contracted with source smearing functions,
while the spatial index at the sink remains explicitly open.
As a result, application of optimized operators is straightforward at the sink but challenging at the source.
A naive way to implement the optimized operators at the source requires 
computationally expensive all-to-all propagators~\cite{Bali:2005fu}.

On the other hand,
the summation of $\vec {r}_\text{src}$ in Eq.~(\ref{Eq-O}) implies that explicit spatial disentanglement at the source may not be necessary, if we can design suitable smearing functions at the source such that the summation over $\vec {r}_\text{src}$ is automatically handled through spatial index contraction when computing the quark propagators.
To this end, we introduce the following two smearing functions
to be used for each hadron respectively,
\begin{align}
   & G(\vec r) = f(\vec r), \label{Eq-Smear-1}\\
   & F_n(\vec r) = \frac{1}{V^{1/3}_\text{sub}}\sum_{\vec r_0\in\Lambda_\text{sub}} Z_3(\vec r_0) \Psi^{1/3}_n(\vec r_0) f(\vec r - \vec r_0)\label{Eq-Smear-2},
\end{align}
where $f(\vec r)$ is a local smearing function (such as the exponential form, the Gaussian form, etc.) fine-tuned for single baryon.
This construction gives $G(\vec r)$ a single support at the origin, while $F_n(\vec r)$ has multiple supports distributed on a sub lattice $\Lambda_\text{sub}$ with volume $V_\text{sub}$, see Fig~\ref{Fig-sps-grid}.
(A reason for a use of the sub lattice will be explained later.) 
Each support of $F_n(\vec r)$ is  weighted by the cubic root of the wavefunction $\Psi^{1/3}_n(\vec r_0)$, and is multiplied by an independent $Z_3$ noise vector~\cite{xQCD:2010pnl}, which takes values from the following set, 
\begin{align}
     Z_3\in \{1, e^{i2/3\pi}, e^{i4/3\pi}\},
\end{align}
and satisfies
\begin{align}
     \langle\langle Z_3(\vec r_0) Z_3(\vec r_1) Z_3(\vec r_2) \rangle\rangle_{Z_3} =\delta(\vec r_0 -\vec r_1)\delta(\vec r_0 -\vec r_2),
\end{align}
with $\langle\langle\cdots\rangle\rangle_{Z_3}$ representing the average for the noise vector. 
This property of $Z_3$ noise vector ensures that when combining three propagators calculated with $F_n(\vec r)$, only terms from identical supports survive while cross terms between different supports vanish statistically. 
Therefore, the optimized source operator $\bar O_n$ is achieved by combining three quark operators smeared with $G(\vec r)$ and another three smeared with $F_n(\vec r)$, 
\begin{align}
    (\bar q_G)^3(\bar q_{F_n})^3 = &\frac{1}{V_\text{sub}}\bar{ B}(\vec 0)\sum_{\vec r_0\in\Lambda_\text{sub}}\Psi_n(\vec r_0) \bar{ B}(\vec r_0),\nonumber\\
    \simeq& \frac{1}{V}\bar{ B}(\vec 0)\sum_{\vec r_0\in\Lambda}\Psi_n(\vec r_0) \bar{B}(\vec r_0)~\label{Eq-sp-src},
\end{align}
where the smeared quark operators are defined as $\bar{q}_G=\sum\limits_{\vec r\in\Lambda} \bar{q}(\vec r) G(\vec r)$, similar for $\bar{q}_{F_n}$, and are assumed to be suitably combined in color and spinor spaces to form two baryon operators.
The sub lattice $\Lambda_\text{sub}$ may be defined through sparsening the original lattice grid $\Lambda$ of size $L$, e.g.,
\begin{align}
     \Lambda_\text{sub}=\{l\vec r~|~\vec r\in \mathbb{Z}^3, 0\leq r_{x,y,z}<L/l \} \label{Eq-Sub-Lattice},
\end{align}
where an integer $l$ must be a divisor of $L$.
As discussed in Refs.~\cite{Li:2020hbj,Detmold:2019fbk}, the summation over 
the full lattice $\Lambda$ can be well approximated by a summation over a sub-lattice $\Lambda_\text{sub}$ due to the strong correlation among the nearby lattice points.

The above proposal provides an efficient way of implementing optimized two-hadron operators at the source. Here, we list some comments for the purpose of practical use:
\begin{enumerate}
    \item[(i).] The difference between a summation over sub lattice $\Lambda_\text{sub}$ and that over full lattice $\Lambda$ in Eq.~(\ref{Eq-sp-src}) comes from contaminations of high momentum modes, which can be seen clearly from the following identity,
    \begin{align}\label{Eq-Summation}
    & \frac{1}{V_\text{sub}}\sum_{\vec{r}\in\Lambda_\text{sub}}\cdots = \frac{1}{V} \sum_{\vec{m}\in\Gamma}\sum_{\vec{r}\in\Lambda}e^{i\frac{2\pi}{l}\vec{m}\cdot\vec{r}} \cdots,
    \end{align} 
    with $\Gamma=\{\vec m\in \mathbb{Z}^3 |~ 0\leq m_{x,y,z}<l\}$,
    where contributions from $\vec m\neq0$ still remain. 
    Therefore, one should make sure that the Euclidean time $t$ is sufficiently large (in particular $t\gg \dfrac{m_Bl^2}{4\pi^2}$) to suppress these contamination. 
    Alternatively, one may consider constructing sub lattice $\Lambda_\text{sub}$ by randomly selecting sites in  $\Lambda$.
    
    \item[(ii).] Combining three propagators calculated with the smearing function $F_n(\vec r)$ in Eq.~(\ref{Eq-Smear-2}) results in $V_\text{sub}$ terms (those with identical supports) acting as signal and remaining $V^3_\text{sub}-V_\text{sub}$ terms (those with different supports) acting as noise.
    The latter together with fluctuations of the lattice gauge configurations contribute to the final statistical uncertainties $\epsilon$,
    \begin{align}
    \epsilon=\sqrt{\frac{\sigma^2_g}{N_g} + \frac{\sigma^2_{Z_3}}{N_{Z_3}N_g}} \label{Eq-Tot-Stat-Err},
    \end{align}
    where $\sigma^2_g$ ($\sigma^2_{Z_3}$) is the variance associated with gauge configurations ($Z_3$ noise), and $N_g$ ($N_{Z_3}$) is the number of gauge configurations ($Z_3$ noises).
    In practice, $\Lambda_\text{sub}$ should be fine enough
    to minimize contaminations from high momentum mode shown in Eq.~(\ref{Eq-Summation}), while at the same time sparse enough to reduce the variance $\sigma^2_{Z_3}$. 
    A concrete variance analysis is given in Appendix~\ref{Sec-err}.

    \item[(iii).] The coordinate $\vec 0$ in Eq.~(\ref{Eq-sp-src}) actually represents the reference coordinate $\vec y$ in Eq.~(\ref{Eq-F}).
     While the summation over $\vec x$ at the sink combined with a conservation of the total momentum is enough to place the two-hadron system into the CM frame, the additional summation over $\vec y$ enhances total statistics. 
    To achieve this summation,
    one can randomly shift the reference coordinate for each configuration, and the  sub lattice is constructed subsequently according to the reference coordinate.
    
\end{enumerate}

\subsection{The calculation of correlation functions} 
The hadronic correlation functions are calculated with the ``block algorithm'' combined with the unified contraction algorithm (UCA)~\cite{Doi2013}.
The block algorithm constructs a baryon sink by combining three quark propagators into a single block, which enables an efficient calculation of $\sum\limits_{\vec x\in\Lambda} B(\vec x+\vec r,t) B(\vec x,t)$ at the sink via the convolution technique.
The UCA performs the evaluation of hadronic correlation functions by simultaneously handling quark permutations and color/spin index contractions through a unified index list, significantly reducing computational costs.
Since the UCA assumes identical quark smearing for all quarks of the same flavor at the source, which does not hold in our case due to the distinct smearing functions in Eqs.~(\ref{Eq-Smear-1}) and (\ref{Eq-Smear-2}).
To address this, we classify the quark permutations into two categories: those involving quark exchanges between the two baryons and those without. 
Similar classifications for $N\Lambda$ are discussed in Refs.~\cite{Nemura:2015yha, Nemura:2022wcs}.
By constructing appropriate index lists for each case, we adapt the UCA framework to accommodate the different smearing functions while maintaining computational efficiency.
See Appendix~\ref{Sec-wick-cont} for more details.

%=================================================
\section{Lattice setup}\label{Sec-lat-setup}

We adopt the HAL-conf-2023 lattice QCD gauge configurations in this study, newly generated by the HAL QCD Collaboration using the supercomputer Fugaku at
RIKEN~\cite{Aoyama:2024cko}.
The nonperturbatively improved Wilson action with stout smearing
is employed for ($2+1$) light quarks at the physical point ($m_\pi\simeq137$ MeV),
together with the Iwasaki gauge action.
The lattice size is $L^4=96^4$ with the lattice spacing $a^{-1}\simeq 2338.8$~MeV, leading to $La\simeq8.1$ fm, which is large enough to
accommodate two-hadron systems on the lattice. The key properties of HAL-conf-2023 are summarized in Table~\ref{tab-conf}.

As will become clear in the following section, we take $\Omega_{ccc}\Omega_{ccc}$ system in $^1S_0$ channel as an typical example throughout this paper to demonstrate how our method works.
To simulate this  system,  we employ the relativistic heavy quark action for the charm quark~\cite{Aoki2003} in order to remove the cutoff errors associated with the charm quark mass up to next-to-next-to-leading order.
The charm quark mass is set to be very close to its physical value
~\cite{Namekawa2017},
which results in a spin-averaged $1S$ charmonium mass $M_\text{av}=(m_{\eta_c} + 3m_{J/\psi})/4\simeq3102$ MeV ($1\%$ large than the physical value~\cite{PDG2020}), and the $\Omega_{ccc}$ mass $m_{\Omega_{ccc}}=4847.3(1)$ MeV.

%==========================
\begin{table}[htbp]
\caption{Properties of the HAL-conf-2023 lattice QCD gauge configurations. Data taken from Ref.~\cite{Aoyama:2024cko}.}
\begin{tabular}{cccccc}
  \hline\hline
$a$ [fm] ~~~~~&$L$~~~&$m_\pi$ [MeV]~~~&$m_K$ [MeV]~~~&$La$ [fm]~~~ &$m_\pi L$\\
    \hline
0.0844(1)~~~&96~~~&137.1(3)~~~&501.8(7)~~~&8.10(1)~~~&5.63(1) \\
 \hline\hline
\end{tabular} \label{tab-conf}
\end{table}
%==========================  

\begin{figure}[htbp]
  \centering
  \includegraphics[width=8.7cm]{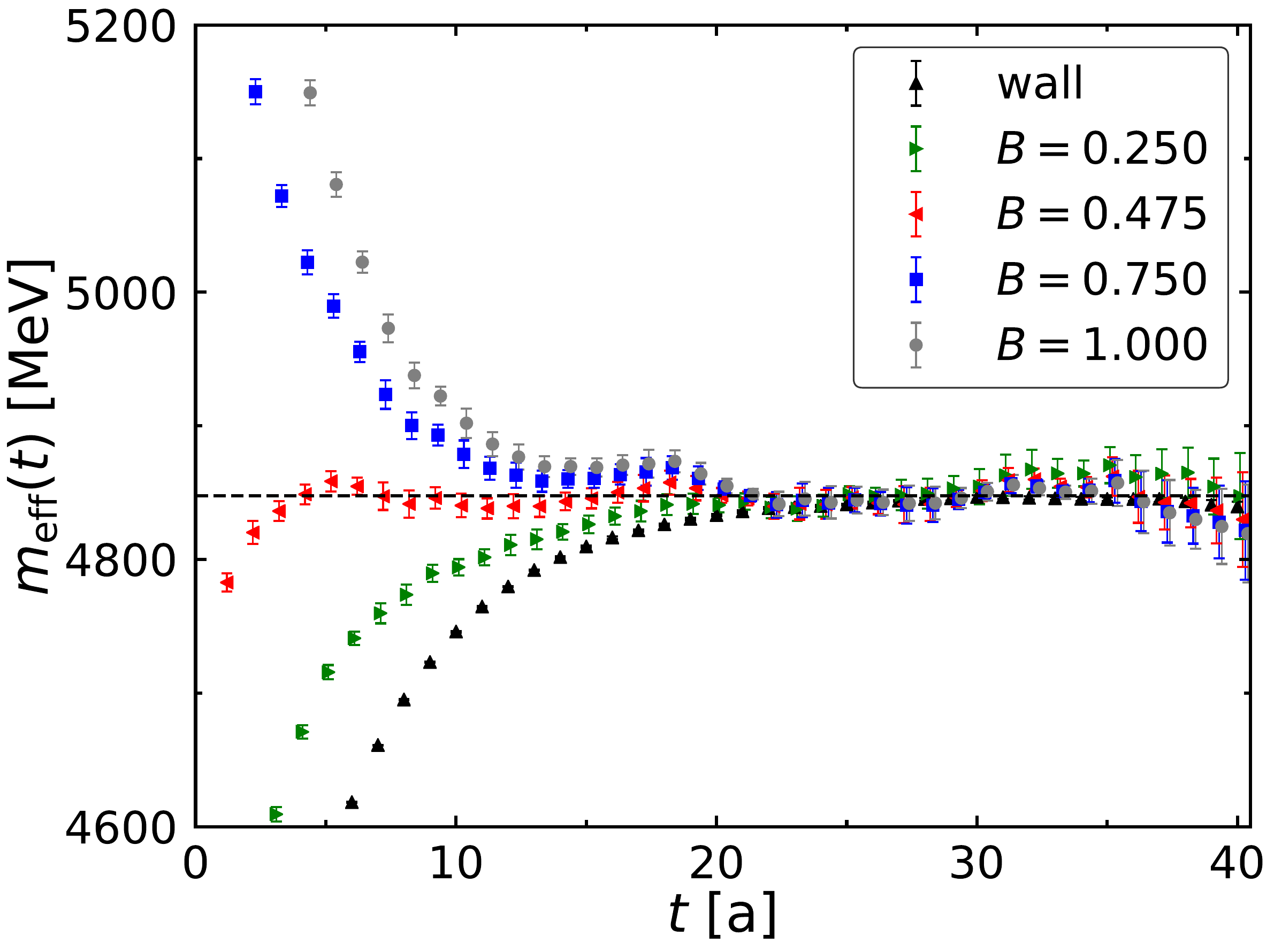}
  \caption{Determination of the smearing parameter in Eq.~(\ref{Eq-Smear-Exp}). The effective mass of single $\Omega_{ccc}$ obtained with the smearing parameter $B=0.475~[a^{-1}]$ exhibits early plateau. Results from the wall source are also shown by black triangles for reference.
  The black dashed line represents the $m_{\Omega_{ccc}}=4847.3$ MeV obtained from Fig.~\ref{Fig-M-Omg}.
  } \label{Fig-M-Omg-tune}
\end{figure}

\begin{figure}[htbp]
  \centering
  \includegraphics[width=8.7cm]{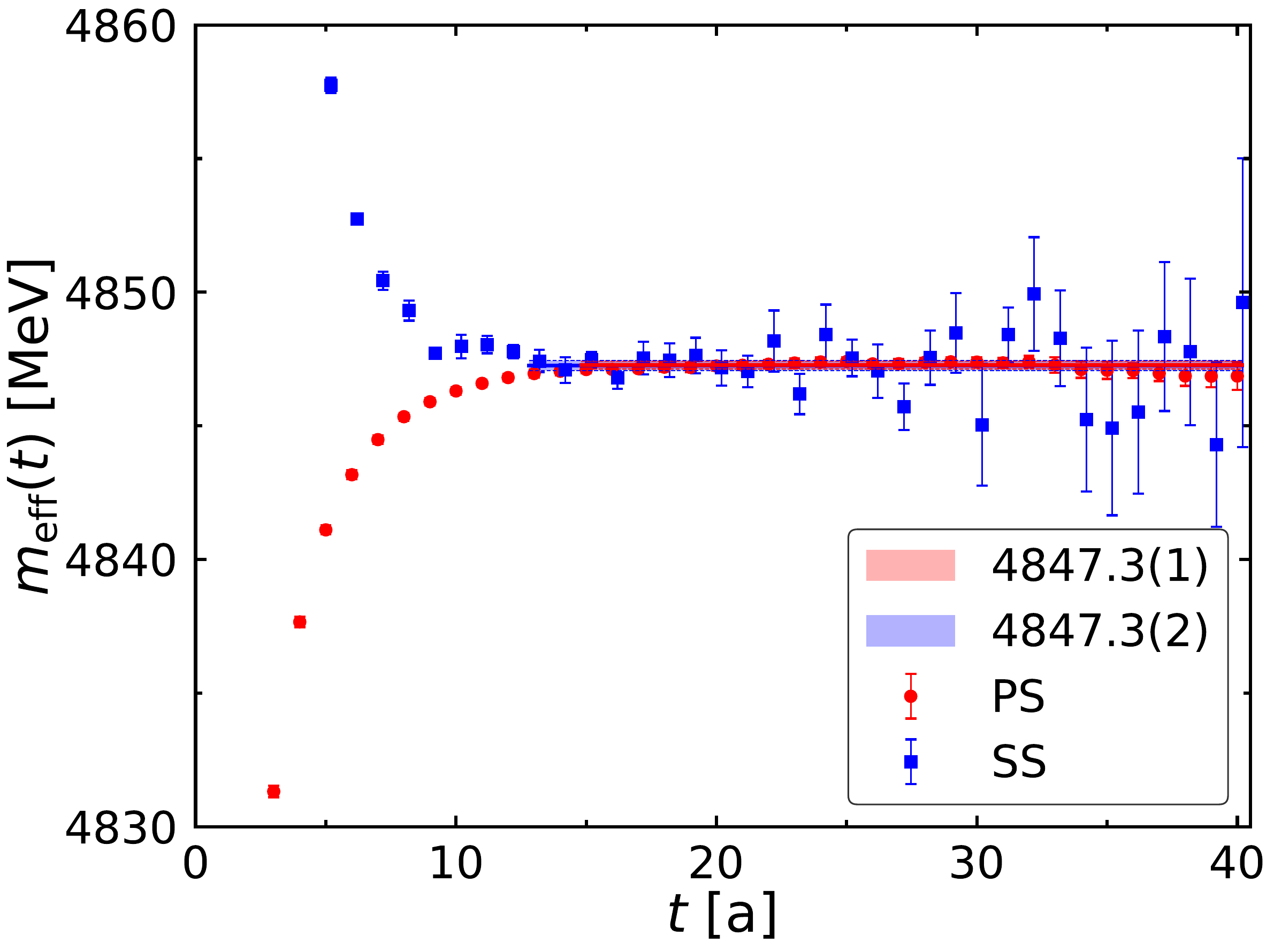}
  \caption{The effective mass of $\Omega_{ccc}$, where
the ``PS" represents results from point sink and smeared source, while
the ``SS" represents those from smeared sink and smeared source. 
 The exponential-type smearing function in Eq.~(\ref{Eq-Smear-Exp}) is used with the smearing parameter determined in Fig.~\ref{Fig-M-Omg-tune}. 
 Although both provide consistent results, the ``PS'' provides better signals than the ``SS''.
 The red (blue) band is from the single-state fit to the ``PS'' (`SS'') correlator.
  } \label{Fig-M-Omg}
\end{figure}

A local baryon operator is adopted at the sink as,
\begin{align}
      \Omega_{ccc}(x)=\varepsilon_{lmn}c_{l}^T(x)\mathcal{C}\gamma_k c_m(x) c_{n,\alpha}(x), \label{Eq-pt-O}
\end{align}
where $l$, $m$, and $n$ stand for color indices, $\gamma_k$ being the Dirac matrix, $\alpha$ being the spinor index, and $\mathcal C=\gamma_4\gamma_2$ being the charge conjugation matrix.
On the other hand, the baryon operator at the source is constructed
by the smeared quark with the Coulomb gauge fixing.
We replace $c(x)$ in Eq.~\eqref{Eq-pt-O} by $c(\vec x,t)\rightarrow \sum\limits_{\vec y\in\Lambda} f(\vec x-\vec y)c(\vec y,t)$, where we employ two types of smearing functions $f(\vec r)$, 
the wall smearing and the exponential smearing, defined as
\begin{align}
    & f^\text{wall}(\vec r) = 1, \label{Eq-Wall}\\
    & f^\text{expo}(\vec r) = e^{-B|\vec r|} \label{Eq-Smear-Exp}.
\end{align}
The wall smearing $f^\text{wall}(\vec r)$ is used as the initial input to start the operator optimization procedure outlined in Sec.~\ref{Sec-opt-o}, while the exponential smearing $f^\text{expo}(\vec r)$ is adopted as the smearing function in Eqs.~\eqref{Eq-Smear-1} and \eqref{Eq-Smear-2}.%
\footnote{Strictly speaking, the NBS amplitude $\psi_n(\vec r)$ defined with the point-like baryon operator in Eq.~(\ref{Eq-pt-O}) is different from that defined with the baryon operator constructed from smeared quarks.
However, the difference is expected to be non-negligible only at short range $|\vec r| \lesssim \sqrt{\langle r^2\rangle}$ with $\sqrt{\langle r^2\rangle}$ being the root-mean-square radius of single baryon.
For $\Omega_{ccc}$, the previous lattice calculation in Ref.~\cite{Can2015} shows $\sqrt{\langle r^2\rangle}\simeq 0.28$ fm, implying the difference is likely very small in practice. }
The parameter $B$ is set to be $B=0.475~a^{-1}$, fine tuned in order to achieve the ground state saturation for the single $\Omega_{ccc}$ at early Euclidean time $t$. See Fig.~\ref{Fig-M-Omg-tune} for the smearing parameter's determination.
With the fine tuned smearing parameters, the mass of $\Omega_{ccc}$ is determined with high precision as shown in Fig.~\ref{Fig-M-Omg}.

%==========================
%==========================

The projection to spin zero state is performed at both the sink and the source by,
\begin{align}
    \left[\Omega_{ccc}\Omega_{ccc}\right]_0 = \frac{1}{2}&\left(\Omega^{3/2}_{ccc}\Omega^{-3/2}_{ccc} - \Omega^{1/2}_{ccc}\Omega^{-1/2}_{ccc} \right. \nonumber \\
    & + \left.\Omega^{-1/2}_{ccc}\Omega^{1/2}_{ccc} - \Omega^{-3/2}_{ccc}\Omega^{3/2}_{ccc}\right)
\end{align}
with $\Omega^{s_z}_{ccc}$ being the spin $3/2$ interpolating operator with its $z$-component $s_z$.
The $A^+_1$ projection of the cubic group $SO(3,\mathbb Z)$ is also employed as,
\begin{align}
    \mathcal{P}^{A^+_1}R(\vec r, t)=\frac{1}{48}\sum_{E\in SO(3,\mathbb Z)} R(E[\vec r],t).
\end{align}
The Misner's method for the approximate partial wave decomposition on a cubic grid is used for the $S$-wave potential~\cite{Misner:1999ab,Miyamoto2020}.
As for the choice of the sub lattice $\Lambda_\text{sub}$ in Eq.~(\ref{Eq-Sub-Lattice}), we choose $l=8~[a]$ to implement the optimized two-baryon operators at the source.
We confirm that this choice results in a relatively small variance $\sigma^2_{Z_3}$ in Eq.~(\ref{Eq-Tot-Stat-Err})
associated with $Z_3$ noises.  For details, see Appendix~\ref{Sec-err}. 

We use $N_g=120\sim 960$ configurations.
To reduce the statistical fluctuation, the forward and backward propagations are averaged for each configuration, 
the hypercubic symmetry on the lattice (four rotations) is employed, and $N_t=32\sim80$ measurements are performed by shifting the
temporal coordinate of the source position. 
The total number of measurements is $N_g\times N_t\times 2_\text{bf}\times 4_\text{rot}=30720\sim614400$.
The quark propagators are calculated by the \texttt{Bridge++} code~\cite{Bridge} with the PBC for all directions.
The statistical errors are evaluated by the jackknife method,
and the bin-size dependence is confirmed to be small.
Details about the calculation are summarized in Table~\ref{tab-cal}.

%==========================
\begin{table}[htbp]
\caption{Summary of the statistics, where
$N_g$ and $N_t$ are number of configuration and numbers of time slices, respectively, while $N_b$ is the bin size for jackknife analysis, and should be understood as $N_b$ configurations.
The ``compact'' source operator represents the local smeared hexa-quark operator, and 
$ O_0$ or $ O_1$ are optimized two-baryon operators in Eq.~(\ref{Eq-O}) for the ground state or the first excited state, respectively.
The plane-wave operators with relative momentum $|\vec p|=0$ and $|\vec p|=\frac{2\pi}{L}$ are denoted as $\tilde O_0$ and $\tilde O_1$, respectively.
}
\begin{tabular}{cccc}
  \hline\hline
source operator~~~~~~~~~~~~&$N_g$~~~~~~~~~~~~&$N_t$~~~~~~~~~~~~&$N_{b}$
\\
\hline
wall~~~~~~~~~~~~&120~~~~~~~~~~~~&32~~~~~~~~~~~~&20 \\
compact~~~~~~~~~~~~&320~~~~~~~~~~~~&80~~~~~~~~~~~~&40 \\
optimized ($ O_0$)~~~~~~~~~~~~&960~~~~~~~~~~~~&80~~~~~~~~~~~~&160 \\
optimized ($ O_1$)~~~~~~~~~~~~&320~~~~~~~~~~~~&80~~~~~~~~~~~~&40 \\
momentum ($\tilde O_0$)~~~~~~~~~~~~&320~~~~~~~~~~~~&80~~~~~~~~~~~~&80 \\
momentum ($\tilde O_1$)~~~~~~~~~~~~&320~~~~~~~~~~~~&80~~~~~~~~~~~~&80 \\
 \hline\hline
\end{tabular} \label{tab-cal}
\end{table}
%==========================  

%==========================
%==========================

%=================================================
\section{Results}\label{Sec-results}

To demonstrate the efficiency of the optimized operators, we take $\Omega_{ccc}\Omega_{ccc}$ system in $^1S_0$ channel as an typical example for three reasons: (i) the heavy $\Omega_{ccc}$ results in dense spectral of two baryons  
on a large lattice box, necessitating  highly optimized operators for state identification;
(ii) the previous lattice calculation~\cite{Lyu2021} has shown that this system supports a (shallow) bound state, allowing us to test our method on both bound and scattering states; 
and (iii) the system exhibits smaller statistical fluctuations than baryons containing light valence quarks, making optimized operator effects more visible.

We shall start with two commonly used operators: the (extended) wall source constructed using the smearing function in Eq.~(\ref{Eq-Wall}), 
and the (local) compact source constructed using the smearing function in Eq.~(\ref{Eq-Smear-Exp}).
As discussed in Sec.~\ref{Sec-opt-o}, both are not optimized operators in the sense that spatial information between two hadrons is 
approximated by some simple functions.
Nevertheless, these results provide us with initial input for the optimization of two-baryon operators.
The results from optimized operators will be presented subsequently,
followed by a comparison with the conventional variational analysis using
plane-wave operators.

\subsection{The wall operator and the compact operator}

\subsubsection{Hadronic correlation functions}
%==========================
\begin{figure*}[htbp]
  \centering
  \includegraphics[width=8.7cm]{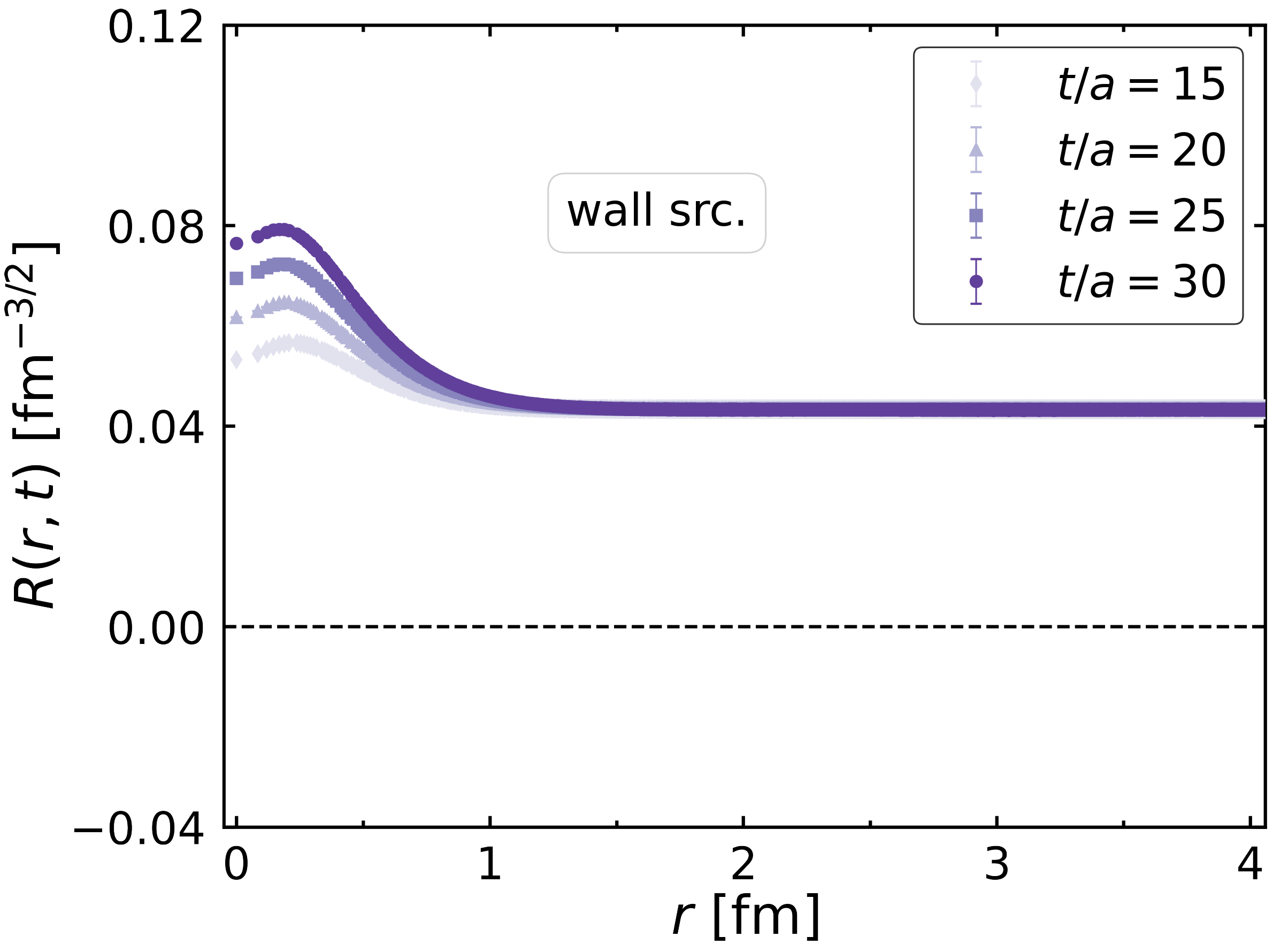}
  \includegraphics[width=8.7cm]{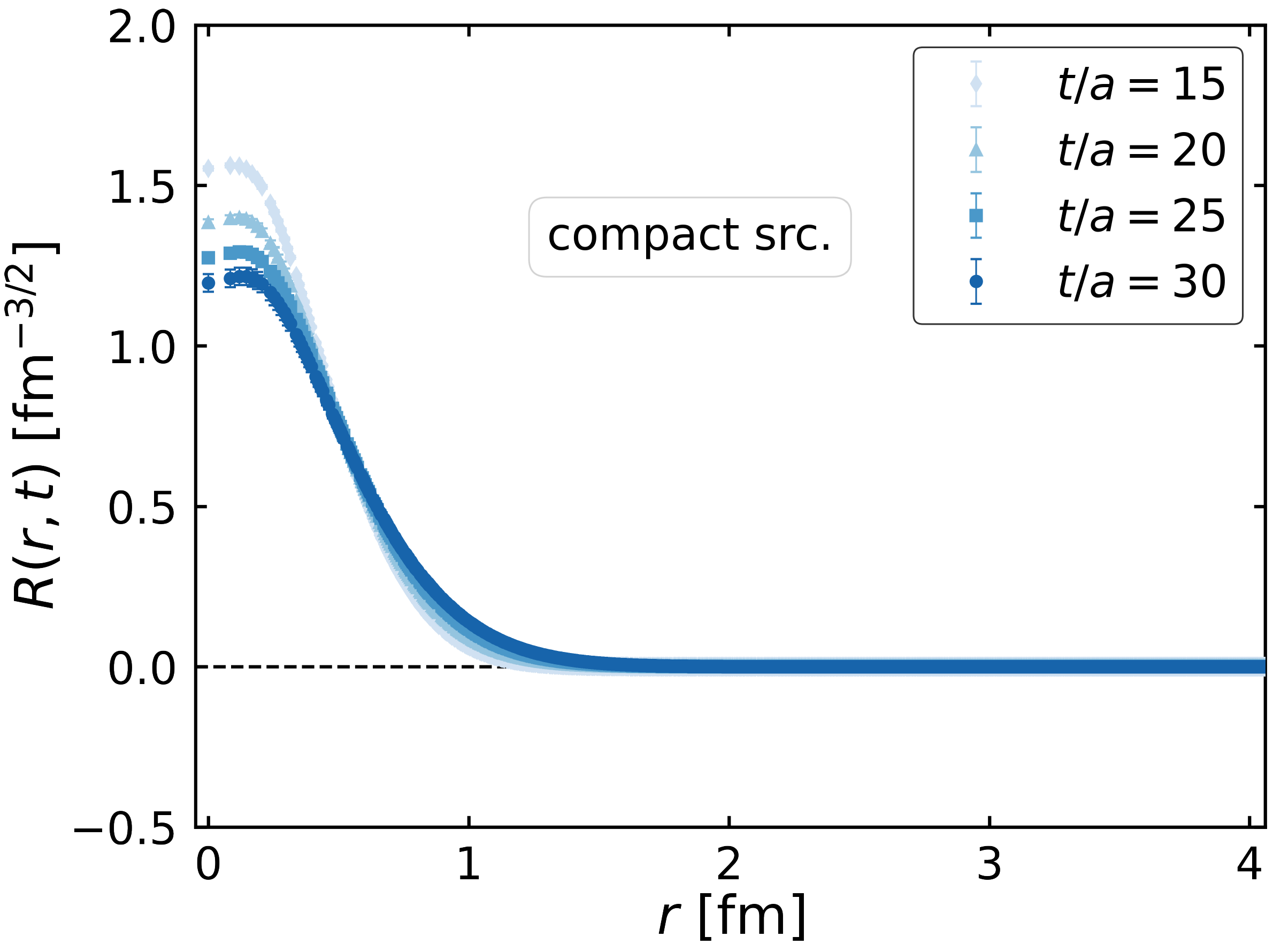}
  \caption{The correlation function $R(\vec r,t)$ calculated with the wall source (left) and the compact source (right) at multiple Euclidean time $t$.  Results are shown with the normalization $\sum\limits_{\vec r\in \Lambda} R^2(\vec r,t) =1$.
  } \label{Fig-R-Wall-Comp}
\end{figure*}
%==========================

%==========================
\begin{figure}[htbp]
  \centering
  \includegraphics[width=8.7cm]{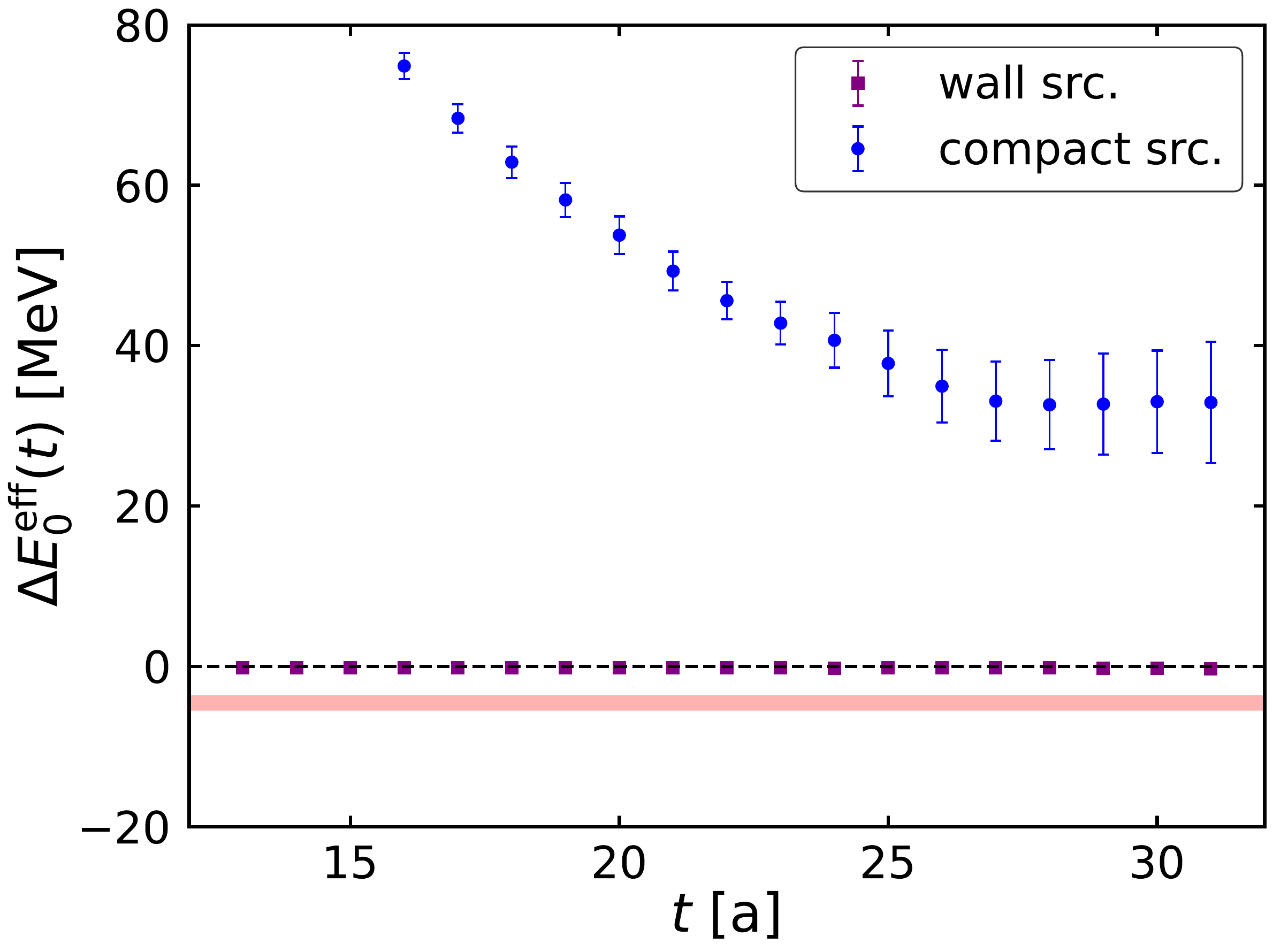}
  \caption{The effective energies extracted from temporal correlation functions in Eqs.~(\ref{Eq-Rt-wall}) and (\ref{Eq-Rt-comp}) calculated using the wall source (purple) and the compact source (blue), respectively.
  The red band represents the genuine ground state energy determined in Fig.~\ref{Fig-Eeff}.
  } \label{Fig-Eeff0-wall-comp}
\end{figure}
%==========================

We first examine the spatial profile of the hadronic correlation functions $R(\vec r, t)$ defined in Eq.~(\ref{Eq-RJ}).
Shown in Fig.~\ref{Fig-R-Wall-Comp} are $R(\vec r, t)$ with the normalization $\sum\limits_{\vec r\in \Lambda} R^2(\vec r,t) =1$ 
at Euclidean time $t/a=15$, $20$, $25$, and $30$ calculated using the wall source and the compact source, respectively.
The spatial profiles exhibit distinct behaviors, and show significant $t$ dependence over a long Euclidean time range of $t/a=15\sim30$,
which indicates that ground state saturation is not achieved within these time slices in either of the cases.
If the ground state saturation were achieved, the spatial profile of $R(\vec r, t)$ would be given by the ground state NBS amplitude $\psi_0(\vec r)$ in Eq.~(\ref{Eq-NBS}), so that it should be independent of source operators used in the calculation and stable against $t$.

The time evolution of $R(\vec r, t)$ in Fig.~\ref{Fig-R-Wall-Comp}
reveals intriguing behaviors.
For the extended wall source, $R(\vec r, t)$ gradually becomes narrower, evolving from an approximately flat/constant profile into a more localized distribution.
In contrast, $R(\vec r, t)$ from the compact source exhibits the opposite trend: the spatial profile broadens over time, evolving from a delta-like configuration into a more extended one. (A similar trend was observed in the $\Xi\Xi$ potential~\cite{Iritani2019PRD}.)
This observation is in line with intuitive expectations, as the extended wall source and the compact source represent two limiting cases, while the true profile of the ground state NBS amplitude $\psi_0(\vec r)$ lies somewhere in between.

The elastic state contamination also manifests itself in the effective energies extracted from temporal correlation functions.
As first pointed out in Ref.~\cite{Iritani2016}, 
achieving ground-state saturation becomes particularly difficult when nearby states are present.
Unoptimized operators, which fail to suppress contaminations from these states in general, may lead to a misidentification of states, likely observing a misleading ``fake plateau'' in the corresponding effective energies.

In Fig.~\ref{Fig-Eeff0-wall-comp}, we show the effective energies 
extracted from respective temporal correlation functions $R(t)$,
\begin{align}
   & R_\text{wall}(t) = \sum_{\vec r\in\Lambda}R_\text{wall}(\vec r, t), \label{Eq-Rt-wall}\\ 
   & R_\text{compact}(t) = \sum_{|\vec r|< r_B }R_\text{compact}(\vec r, t). \label{Eq-Rt-comp}
\end{align}
To make sink operators resemble the source operators, the summation over $\Lambda$ is performed for $R_\text{wall}(\vec r, t)$  
which leads to an extended sink operator. 
By contrast, the summation is restricted to $|\vec r|< r_B$ for $R_\text{compact}(\vec r, t)$ with $r_B=4~[a]$ determined by the smearing parameter in Eq.~(\ref{Eq-Smear-Exp}) resulting in a compact sink operator
\footnote{Here, the effective masses depend on the choice of $r_B$.
Increasing $r_B$ from the current value to $r_B=\sqrt{3}L/2~[a]$, that is when the summation is taken over all $\Lambda$ in Eq.~(\ref{Eq-Rt-comp}),
the resulting plateau gradually shifts from approximately $30$ [MeV] to $-30$ [MeV].}.
The effective energy extracted from Eq.~(\ref{Eq-Rt-wall}) 
displays a plateau near zero, while that from Eq.~(\ref{Eq-Rt-comp})
shows rapid decreasing before reaching a plateau at $t/a=25\sim 30$.
To explain the difference, we note that the extended operators in Eq.~(\ref{Eq-Rt-wall}) represent zero-momentum mode in the free theory limit and therefore is expected to be dominated by low-energy states,
while the compact operators in Eq.~(\ref{Eq-Rt-comp}) receives approximately equal contributions from all elastic scattering states at $t=0$
\footnote{The compact operator with the $\delta$-like function can be written as 
$\frac{1}{V^3}\sum\limits_{\vec{p}\in \tilde{\Lambda} }\tilde{B}(\vec p,t)\tilde{B}(-\vec p,t)$ in the momentum space. See Eq.~(\ref{Eq-O-p}) for a comparison with the optimized operator.}
, and contributions from those high energy states quickly decrease with Euclidean time $t$.
As mentioned above, while the degree of elastic contamination depends on the source operators, the ground state saturation is not achieved in either case. 
As a result, both effective energies in Fig.~\ref{Fig-Eeff0-wall-comp} deviate from the true ground-state energy (determined from the optimized operator later).

\subsubsection{Potentials from the time-dependent HAL QCD method}
%==========================
\begin{figure*}[htbp]
  \centering
  \includegraphics[width=8.7cm]{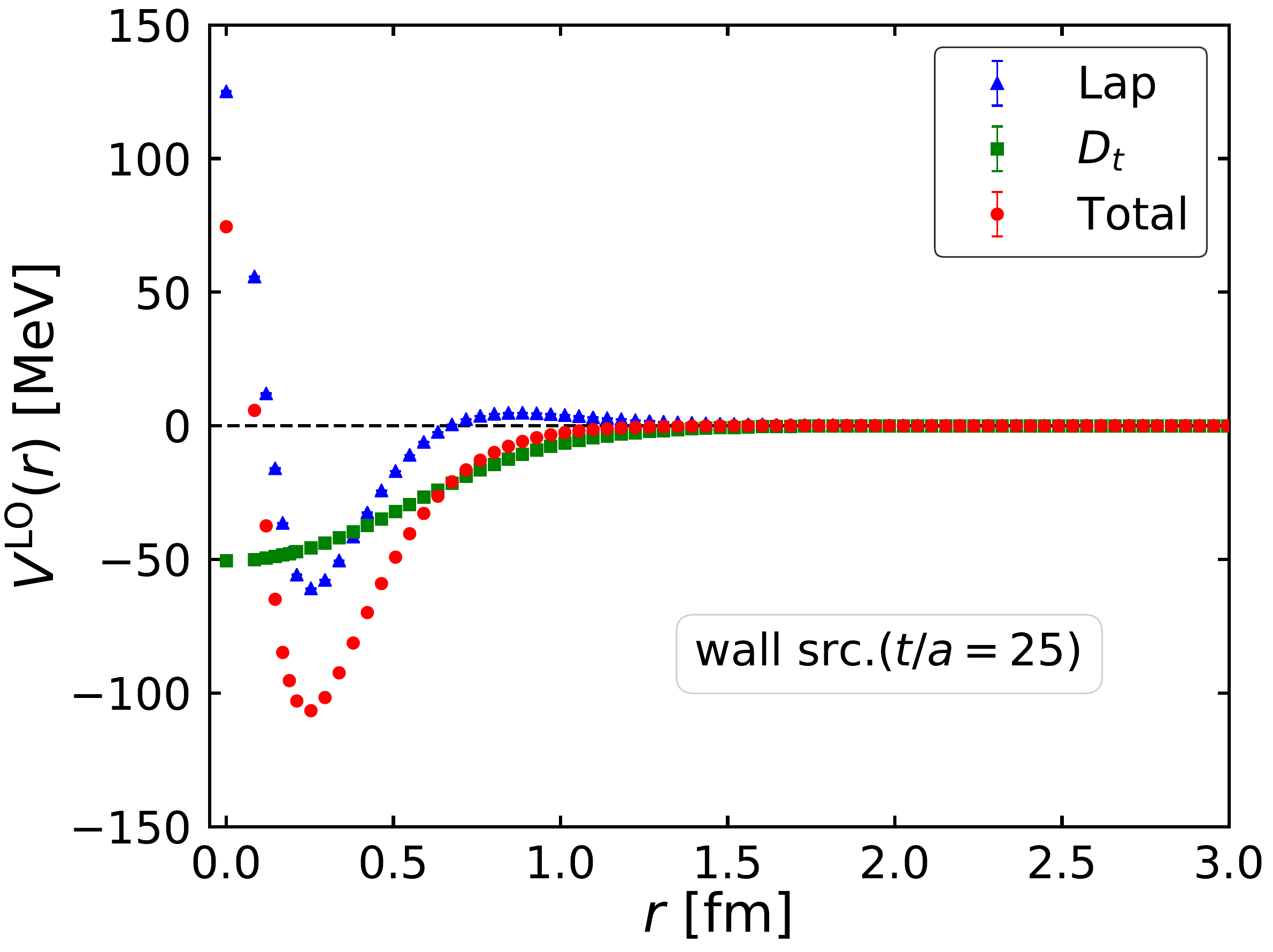}
  \includegraphics[width=8.7cm]{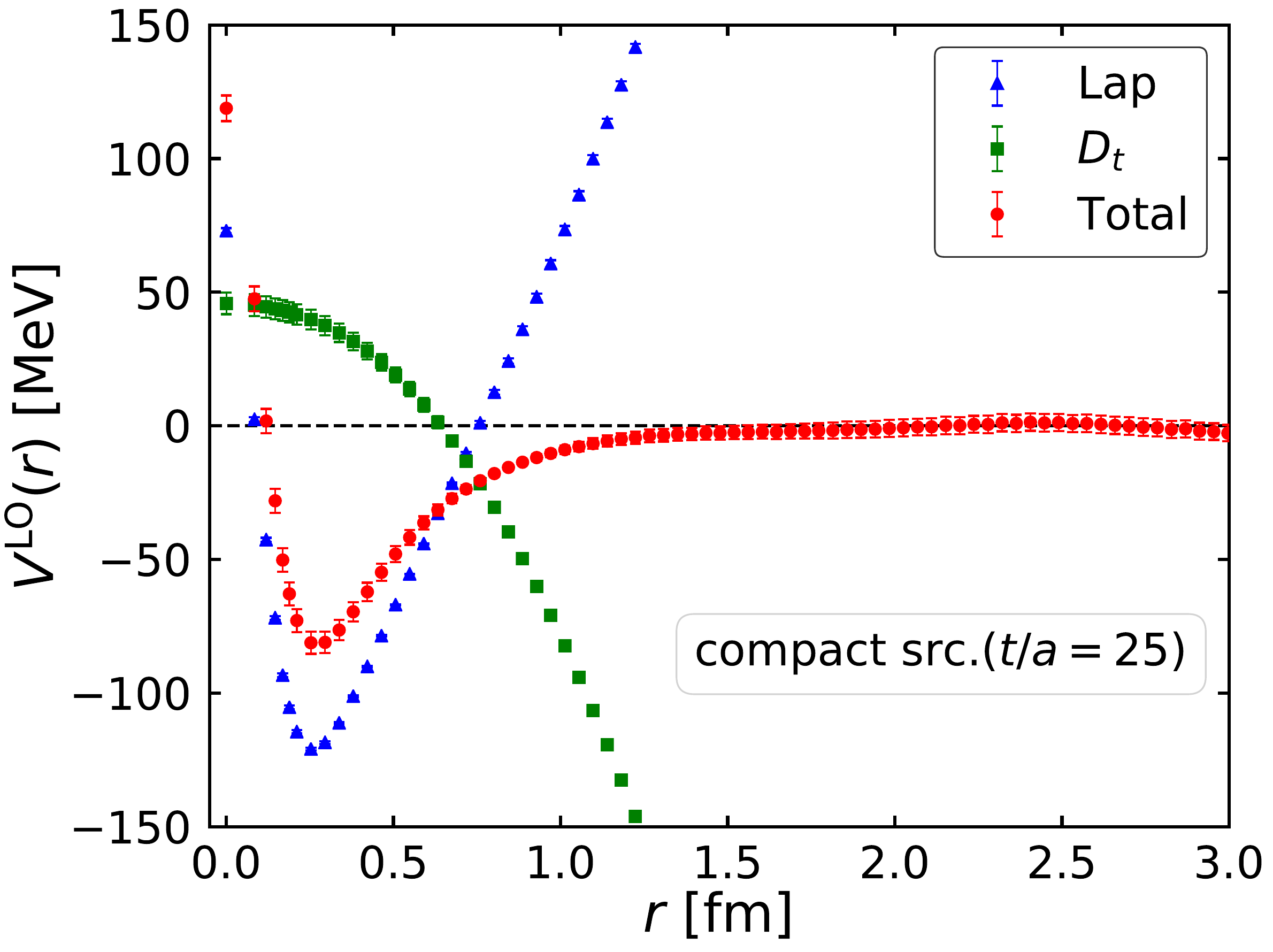}
  \caption{The LO potentials calculated using the wall source (left) and the compact source (right) at Euclidean time $t/a=25$.
  ``Lap'' and``$D_t$'' represent the Laplacian component $R^{-1}(\vec r, t)\frac{\nabla^2}{m_B}R(\vec r, t)$, and the time-derivative component
$R^{-1}\left(\frac{1}{4m_B}\frac{\partial^2}{\partial t^2} -\frac{\partial}{\partial t} \right)R(\vec r, t)$ of the total potential in Eq.~(\ref{Eq-HAL-LOV}).
  } \label{Fig-VLO-Wall-Comp}
\end{figure*}
%==========================

%==========================
\begin{figure}[htbp]
  \centering
  \includegraphics[width=4.2cm]{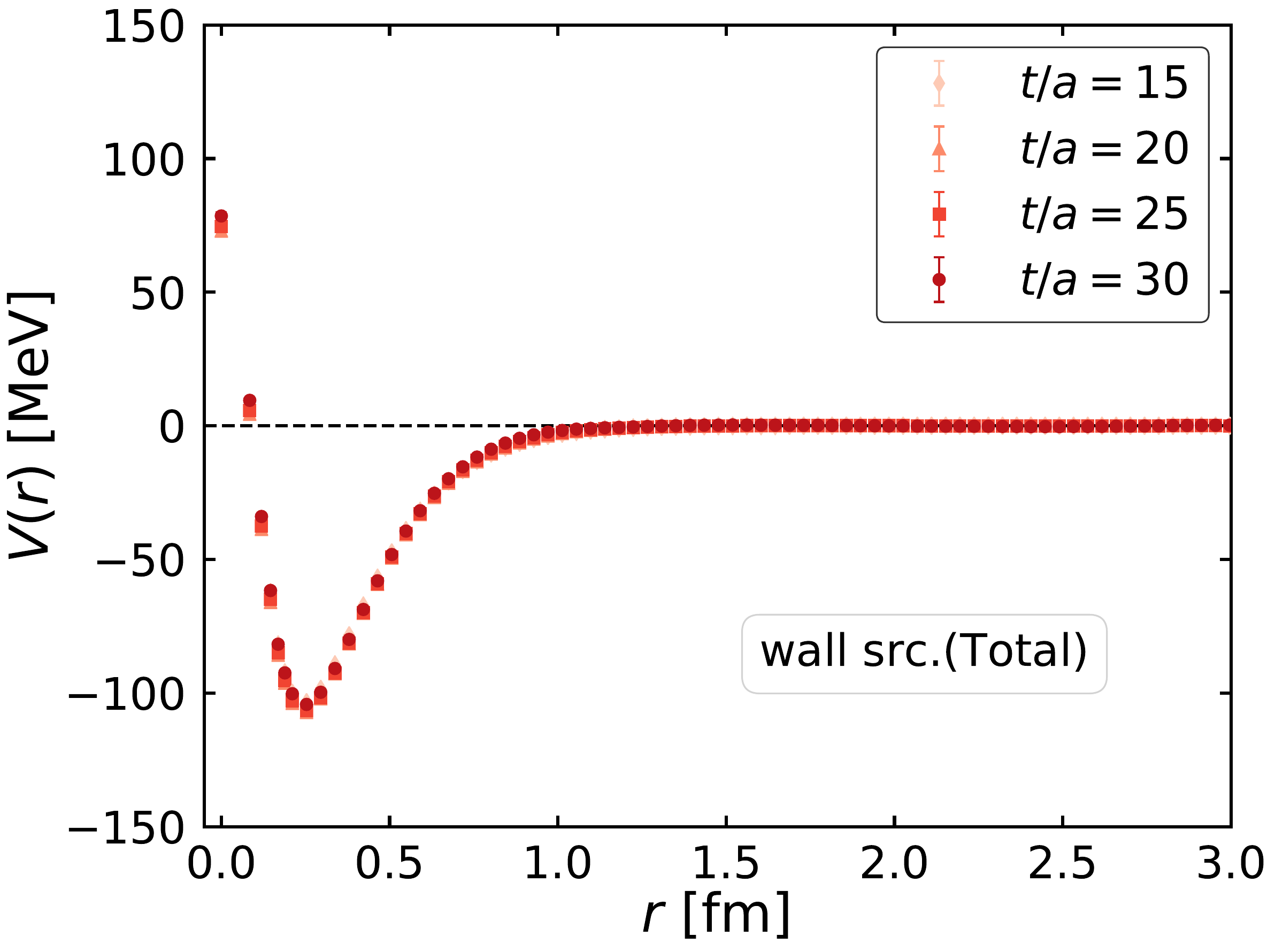}
  \includegraphics[width=4.2cm]{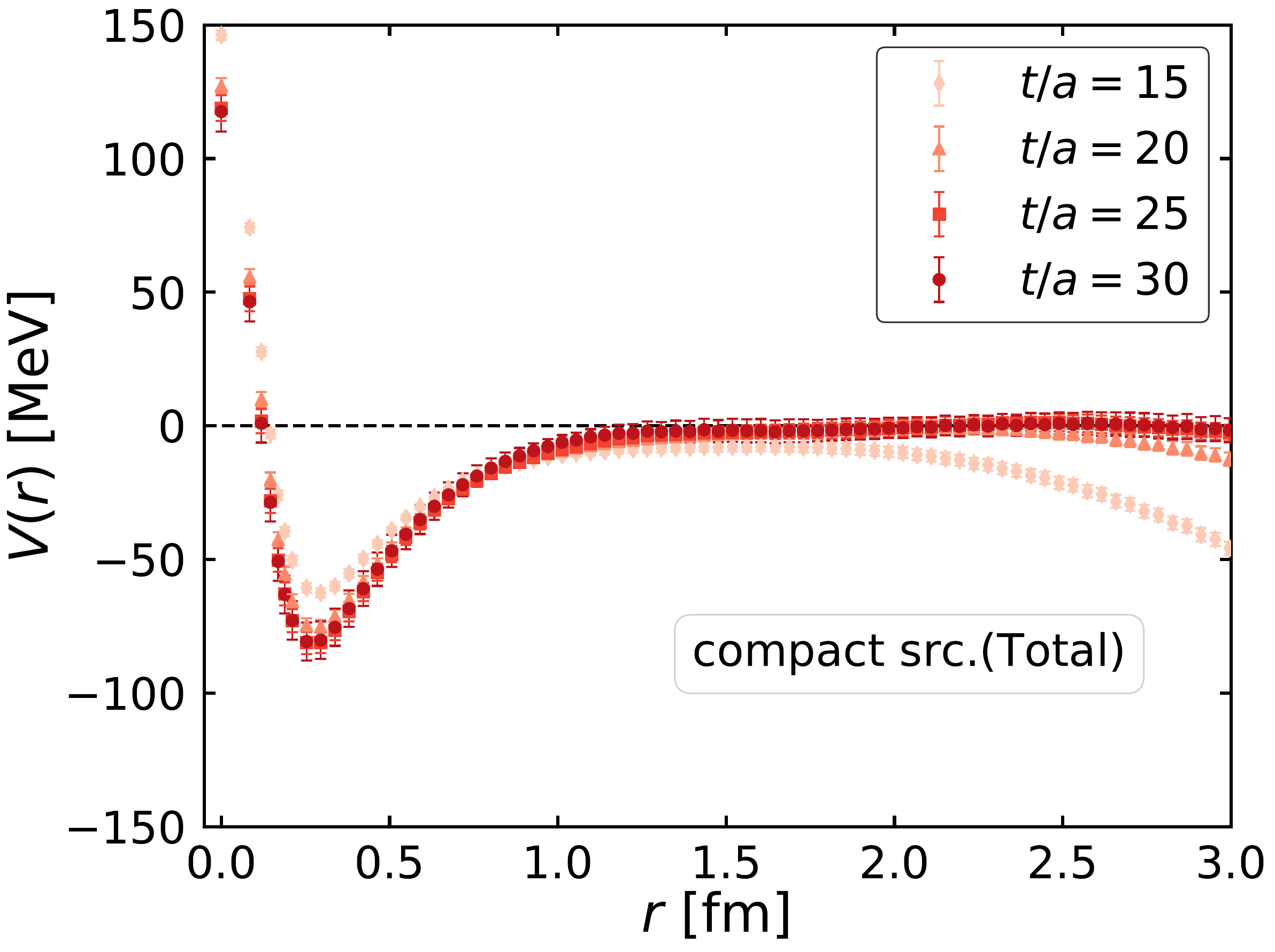}
  \includegraphics[width=4.2cm]{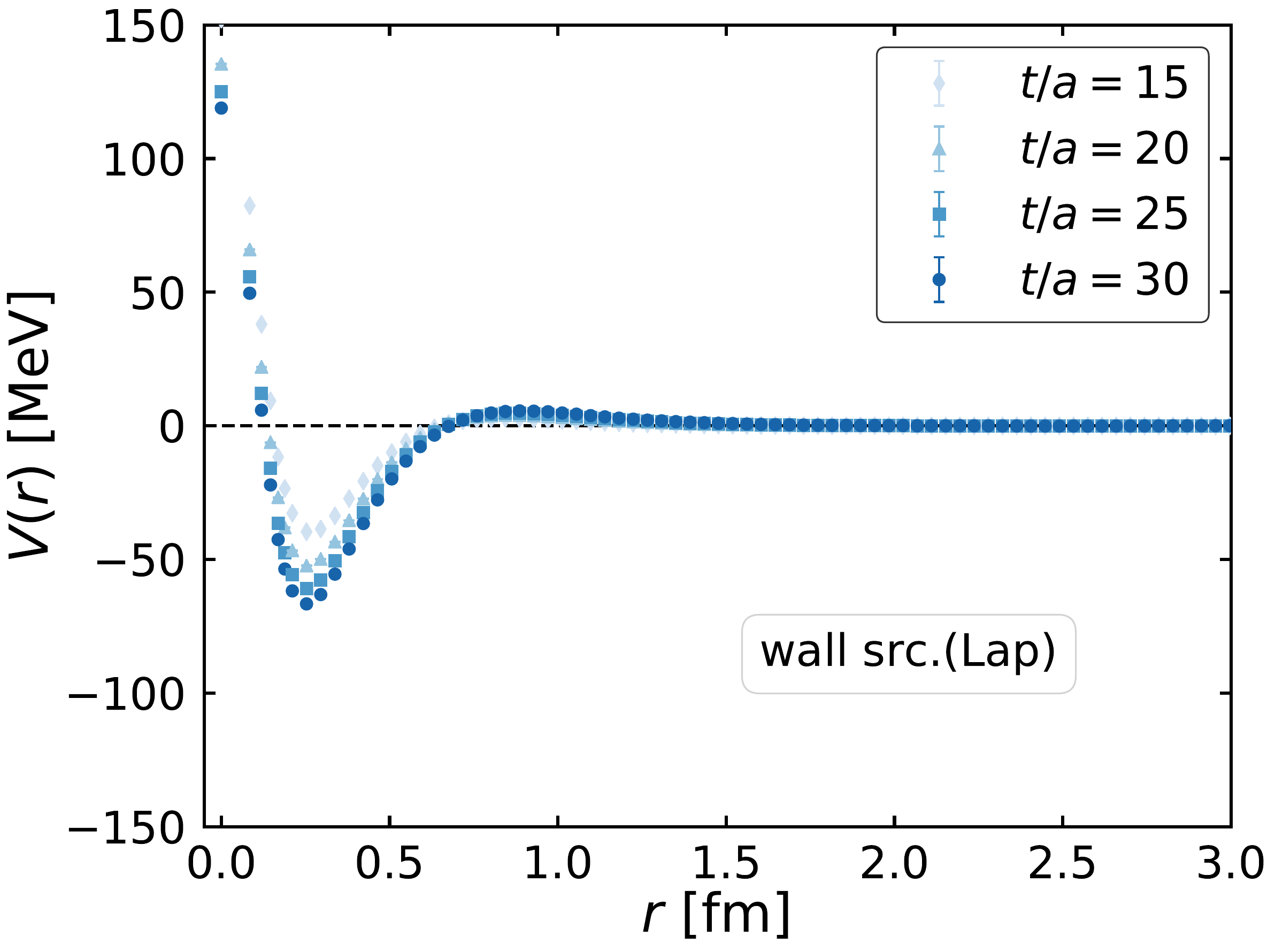}
  \includegraphics[width=4.2cm]{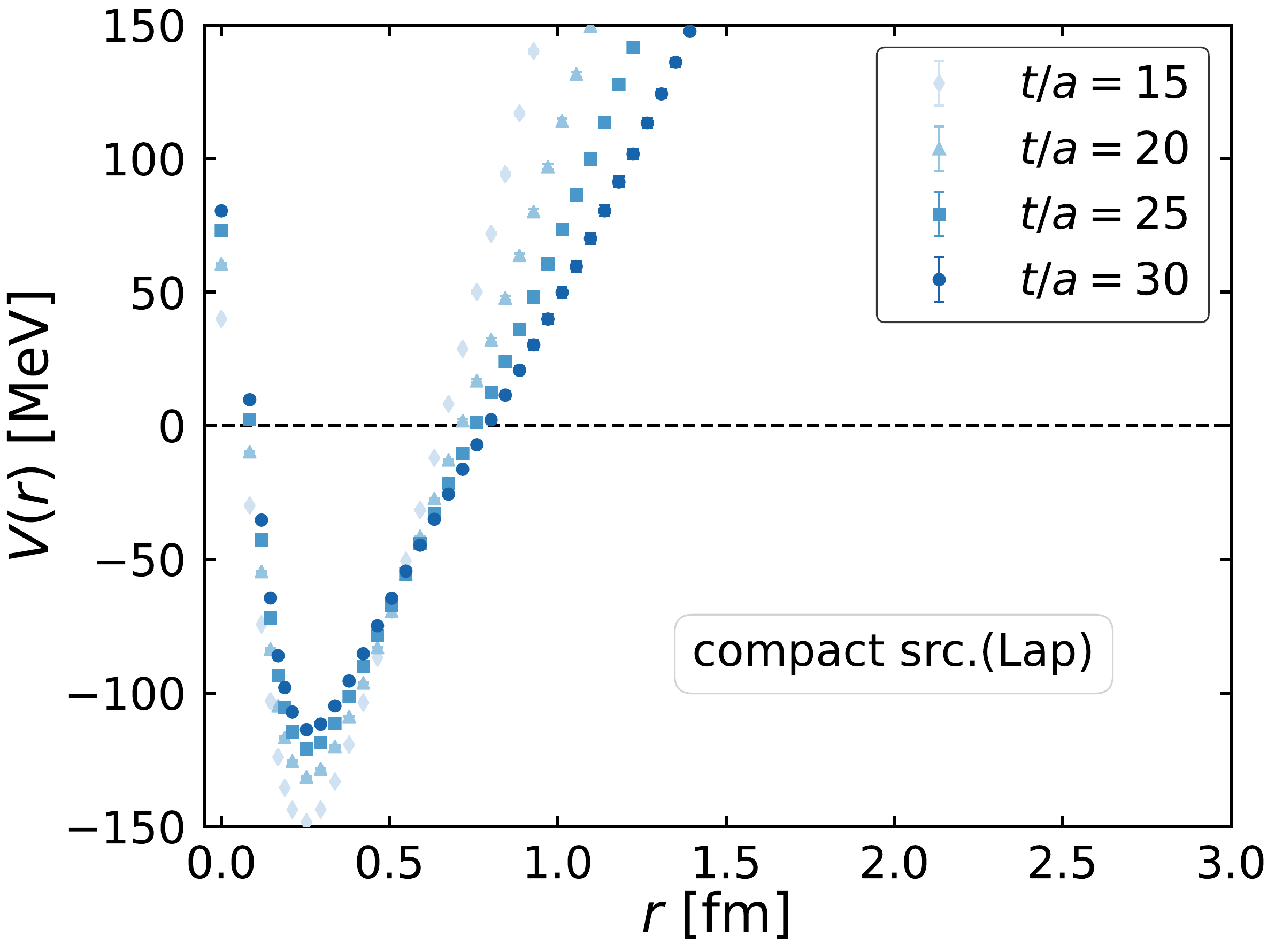}
  \includegraphics[width=4.2cm]{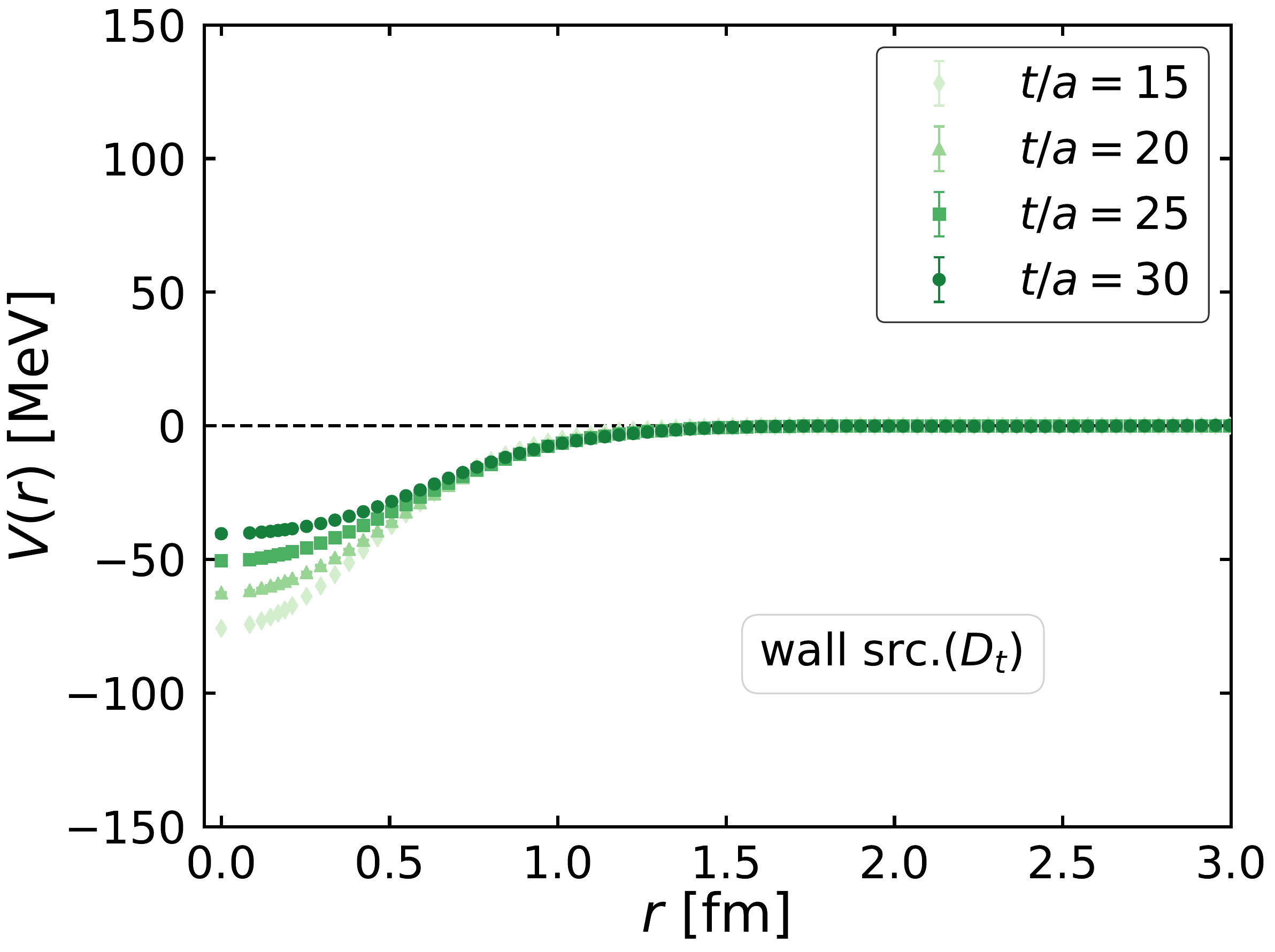}
  \includegraphics[width=4.2cm]{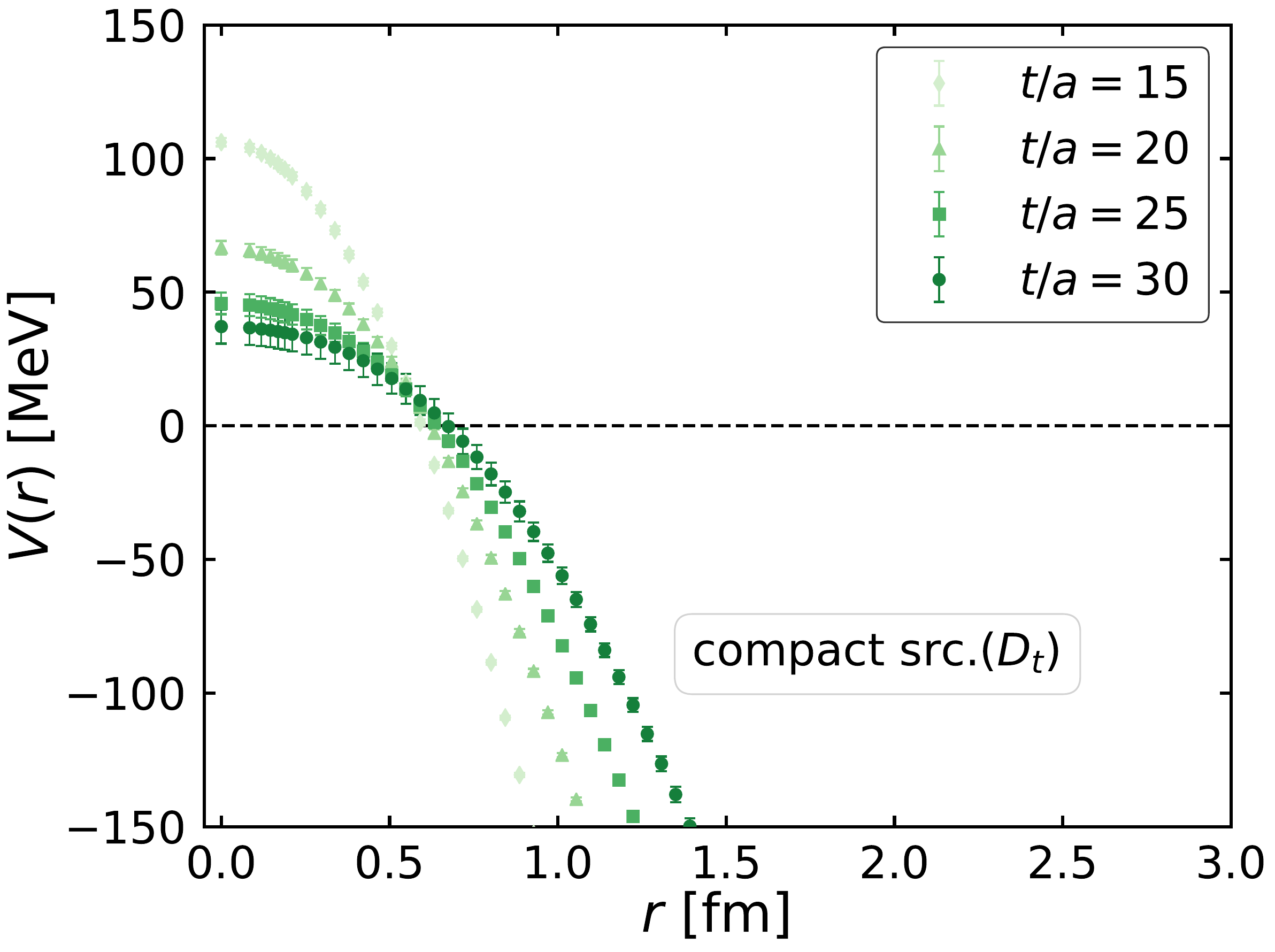}
  \caption{The time dependence of the total potential (top), of its Laplacian component (middle), and of its $D_t$ component (bottom) calculated with the wall source (left) and the compact source (right). 
  } \label{Fig-V-tdep-Wall-Comp}
\end{figure}
%==========================

To overcome the issue of elastic contamination discussed above,
let us perform the time-dependent HAL QCD analysis in this subsection.

Using the $R(\vec r, t)$ shown in Fig.~\ref{Fig-R-Wall-Comp}, we extract the LO potentials using Eq.~(\ref{Eq-HAL-LOV}).
Shown in Fig.~\ref{Fig-VLO-Wall-Comp} is a comparison of the LO potential calculated using the wall source and that using the compact source at Euclidean time $t/a=25$.
The total potential is shown together with each of its Laplacian component $R^{-1}(\vec r, t)\frac{\nabla^2}{m_B}R(\vec r, t)$ (denoted by ``Lap''), and its time-derivative term
$R^{-1}\left(\frac{1}{4m_B}\frac{\partial^2}{\partial t^2} -\frac{\partial}{\partial t} \right)R(\vec r, t)$ (denoted by ``$D_t$'').
We find that the total potentials from both sources are quantitatively agree with each other except for small deviations at short range ($r\lesssim0.3$~fm), unlike the huge difference observed in effective energies in Fig.~\ref{Fig-Eeff0-wall-comp}.
However, the contributions from the Laplacian term and the $D_t$ term to the total potential differ significantly in two cases.
For the wall source, the Laplacian term is similar to the total potential in shape while the $D_t$ term contributes a relative small portion, and both terms vanish at long distance.
In the case of the compact source, however, the Laplacian term
strongly increases with $|\vec r|$ while the $D_t$ term decreases sharply.
The cancellation between these two terms results in a total potential that is quite similar to the one obtained from the wall source.

To further disentangle how elastic states appear in potentials extracted from the time-dependent HAL QCD method, we show in Fig.~\ref{Fig-V-tdep-Wall-Comp} 
a comparison of potentials (the total potential, the Laplacian term, and the $D_t$ term) from each source at multiple Euclidean time $t$.
We observe that the total potentials calculated using the wall source are almost unchanged with respect to $t$,
while those obtained with the compact source exhibit sizable $t$-dependence at $t=15$ and $20$ before stabilizing.
By contrast, the Laplacian term and $D_t$ term for each source
show  significant $t$ dependence, particularly pronounced in the case of the compact source.

The above observations lead to the following conclusions.
\begin{enumerate}
    \item[(i).] The nearly identical potentials from two sources and their  weak $t$ dependence suggest that different combinations of elastic scattering states are governed by the same underlying (nonlocal) potential,
    which is well approximated by the LO potential extracted from the time-dependent HAL QCD method.  
    \item[(ii).] The time-dependent HAL QCD method's ability of utilizing both temporal and spatial information of elastic scattering states is crucial to extract correct information without disentangling each elastic state.
    In fact, the stability of the total potentials arises from the cancellation of the $t$-dependent contributions between the Laplacian term and the $D_t$ term.
    \item[(iii).] The wall source results are dominated by low energy states, whereas the compact source results receive large contributions from higher energy states, as evidenced by the larger $D_t$ term observed for the compact source.
    This implies the potential calculated using the wall source provides better description to interactions at low energy, and the difference between potentials from two sources are relevant to higher energy scatterings. 
\end{enumerate}

\subsubsection{Eigen functions on a finite box}

Using the LO potential $V^\text{LO}$ calculated with the wall source, we solve the eigen equation in Eq.~(\ref{Eq-FV}) on the a three dimensional discrete finite box under PBC with $a$ and $L$ being same as our lattice setup.

\begin{figure*}[htbp]
  \centering
  \includegraphics[width=8.7cm]{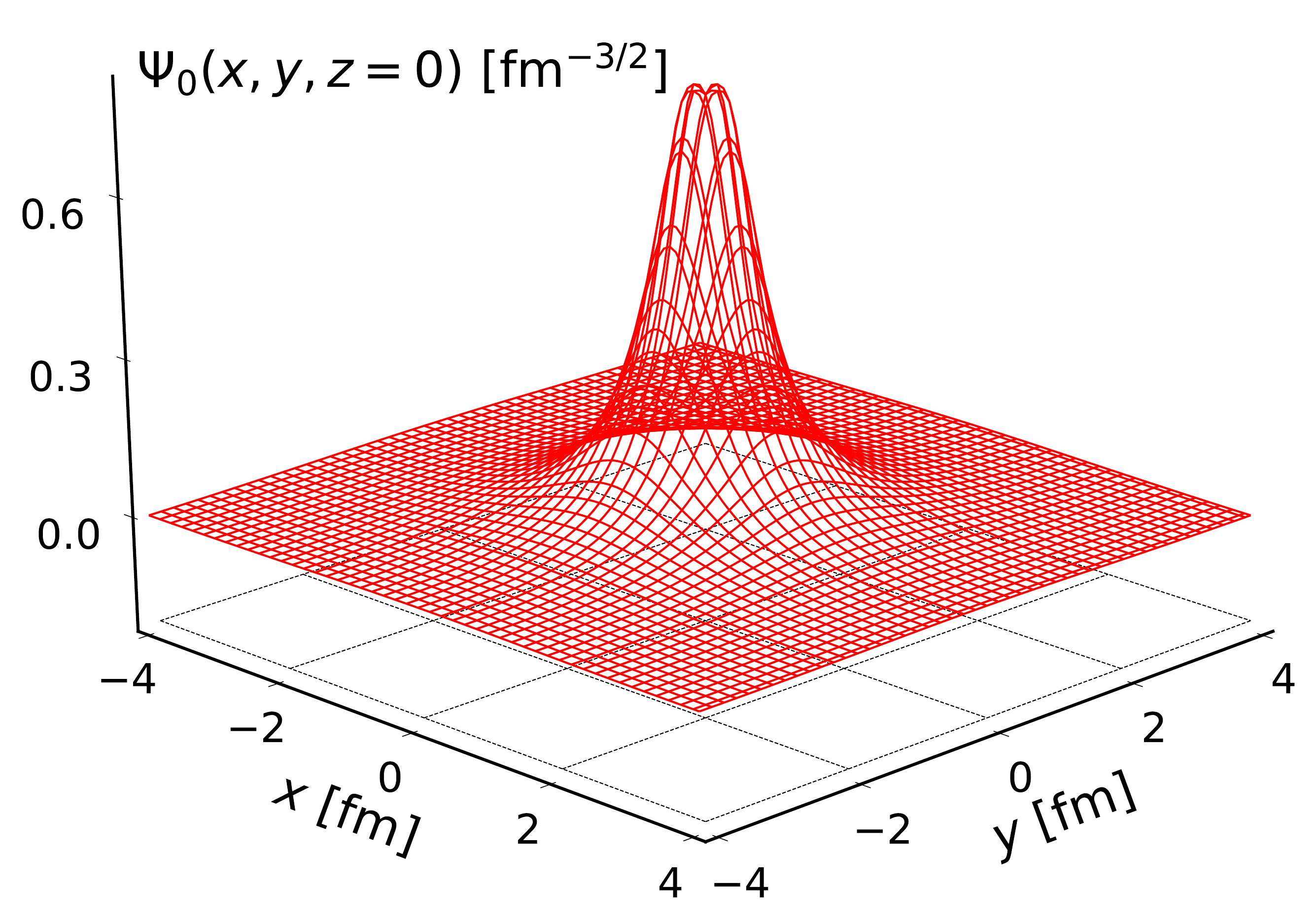}
  \includegraphics[width=8.7cm]{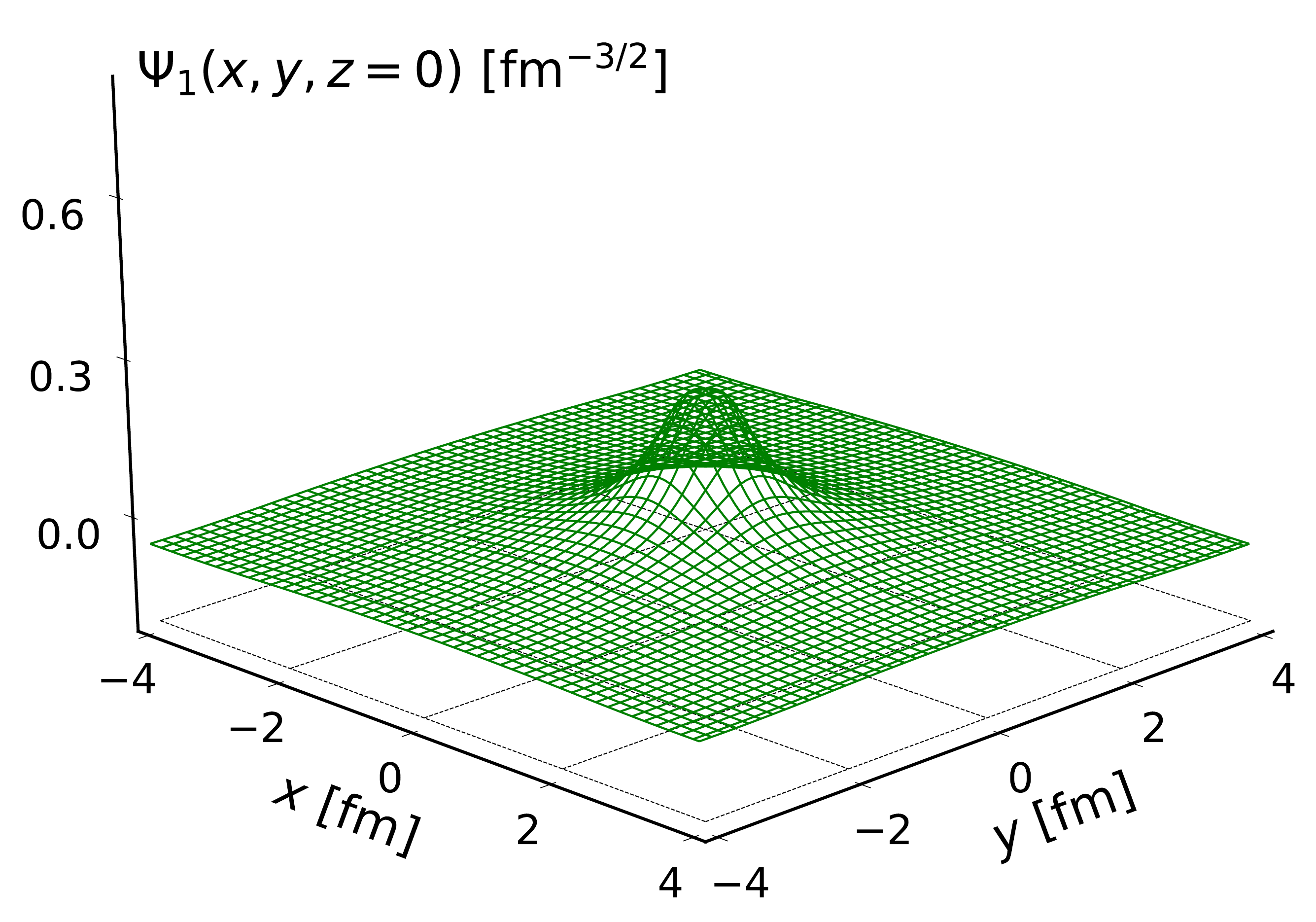}
  \caption{The dual functions $\Psi_{0,1}(\vec r)$ for the ground state (left) and the first excited state (right) constructed using the LO potential $V^\text{LO}$ of the wall source, 
  with the normalization condition in Eq.~(\ref{Eq-norm}).
  } \label{Fig-Psi_01_3d}
\end{figure*}

\begin{figure}[htbp]
  \centering
  \includegraphics[width=8.7cm]{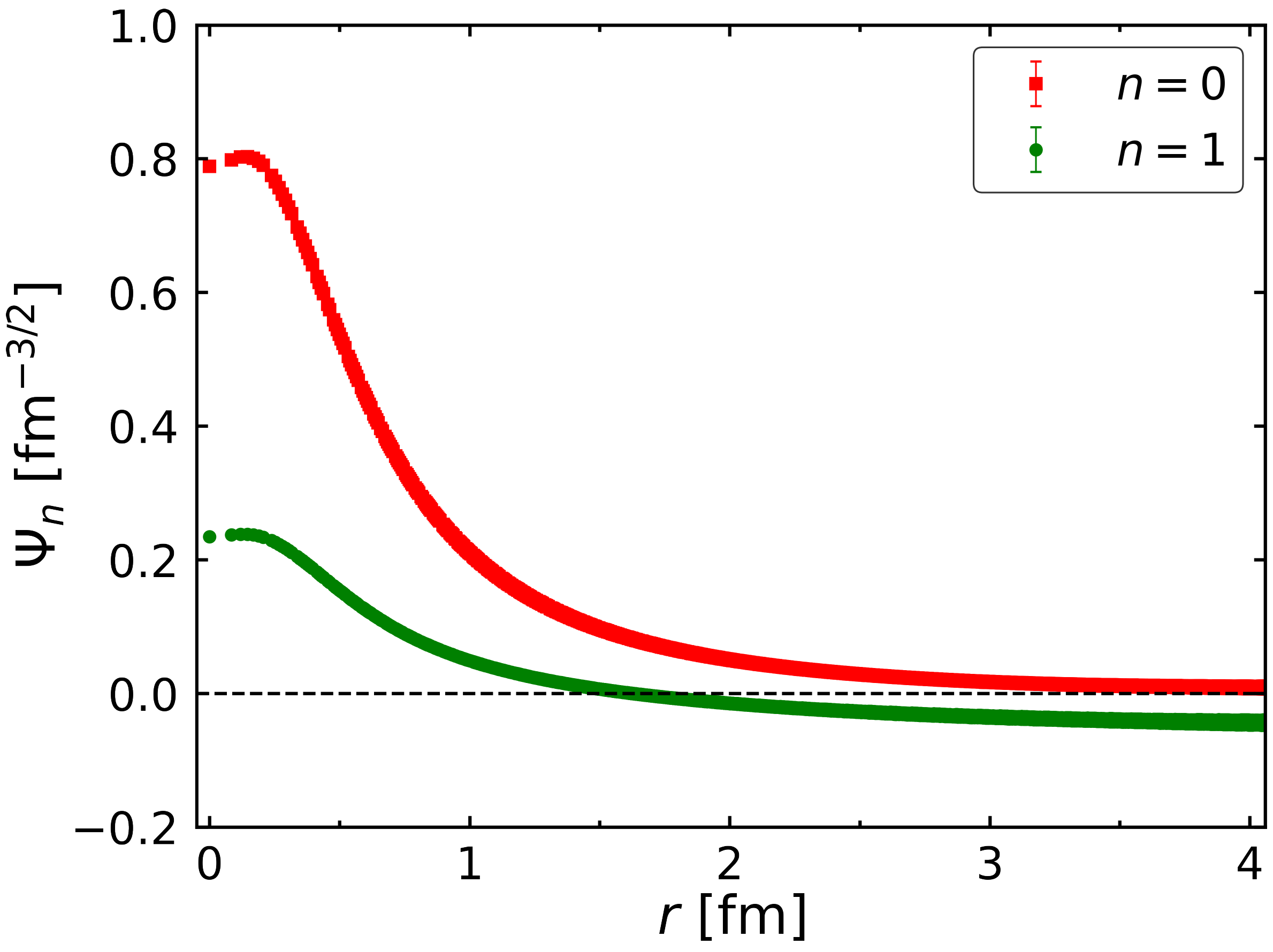}
  \caption{The radial projection of the dual functions $\Psi_{0,1}(\vec r)$ in Fig.~\ref{Fig-Psi_01_3d}.
  } \label{Fig-Psi_01_1d}
\end{figure}

The obtained eigen energies defined in Eq.~(\ref{Eq-FV-E})  for the ground state and the first excited state are given by,
\begin{align}
    & \varepsilon_0=-4.6(4)~\text{MeV}, \label{Eq-E0}\\
    & \varepsilon_1=0.6(1) ~\text{MeV}. \label{Eq-E1}
\end{align}
The obtained eigen functions provide good approximation to underling NBS amplitudes
\footnote{
These eigen functions are orthogonal to each other, as they are
eigen modes with different eigen values of a Hermitian operator.
Using them as initial approximation of the underlying NBS amplitudes
results in the dual functions $\Psi^{(0)}_n(\vec r)=\psi^{(0)}_n(\vec r)$.
}, and can be used to construct dual functions
in Eq.~(\ref{Eq-APsi}). 
Shown in Fig.~\ref{Fig-Psi_01_3d} are resulting $\Psi_{0}(\vec r)$ and $\Psi_{1}(\vec r)$ in the $A^+_1$ representation with a normalization 
\begin{align}
     \sum_{\vec r\in \Lambda} |\Psi_{0,1} (\vec r)|^2 =1. \label{Eq-norm}
\end{align}
Their radial profiles are shown in  Fig.~\ref{Fig-Psi_01_1d}.
We find $\Psi_{0}(\vec r)$ is very localized, consistent with typical bound state wavefunctions, while $\Psi_{1}(\vec r)$
is extended and has one node as expected from the quantization condition given by the PBC.

%==========================
%==========================

%==========================
%==========================

\subsection{Optimized operators}

%==========================
\begin{figure*}[htbp]
  \centering
  \includegraphics[width=8.7cm]{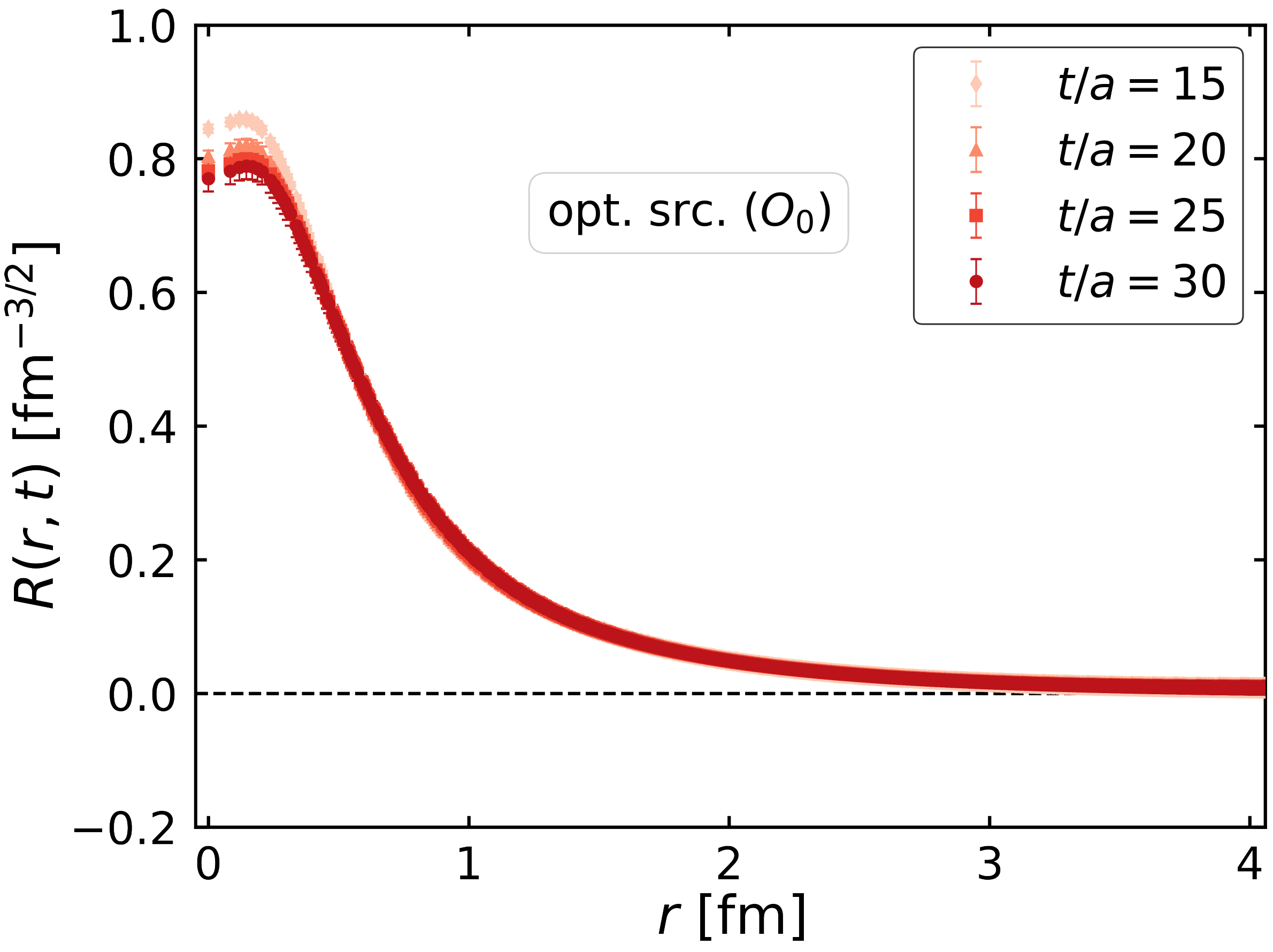}
  \includegraphics[width=8.7cm]{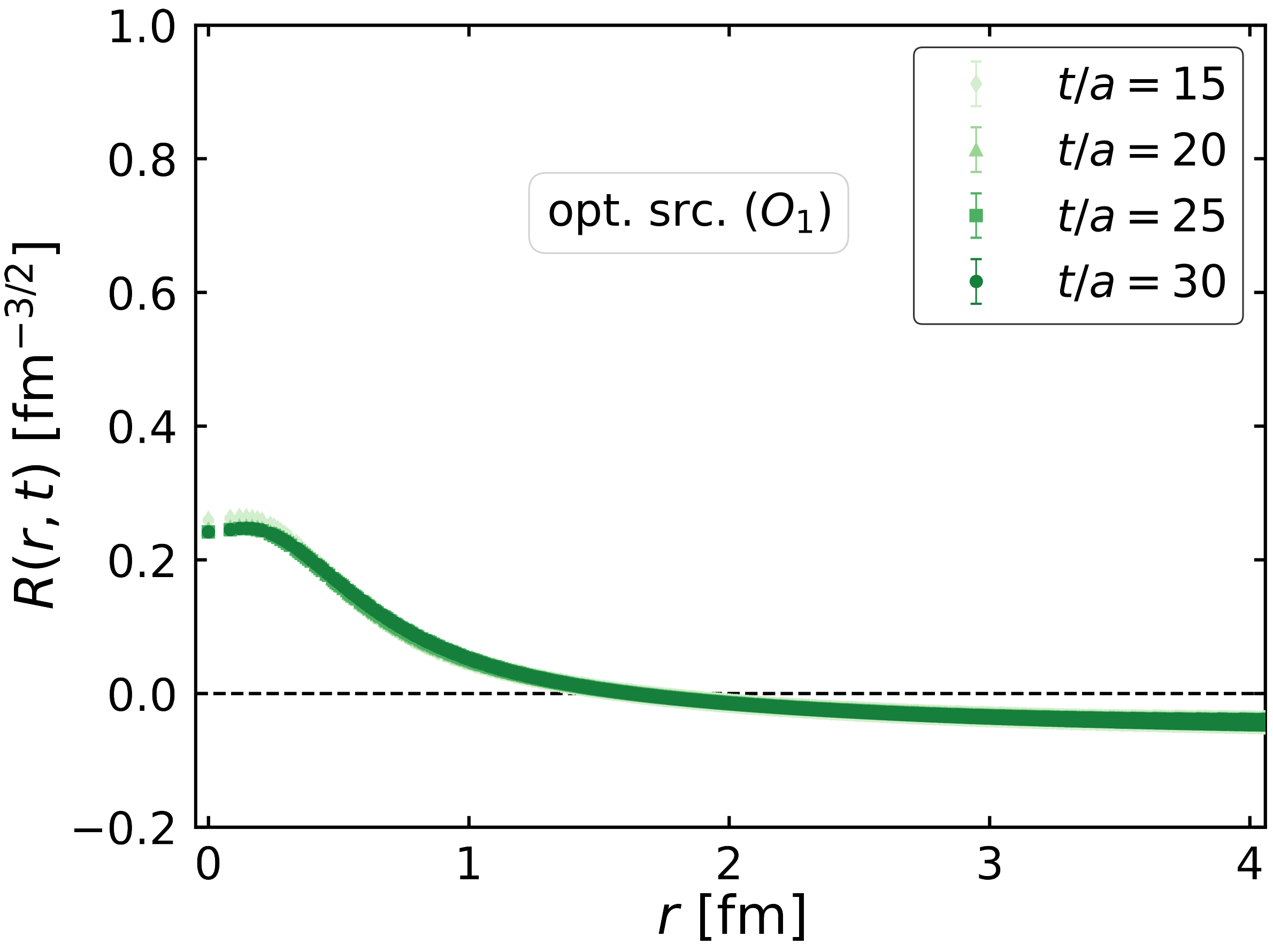}
  \caption{The $R(\vec r,t)$ correlation function calculated with the optimized two-baryon sources for the ground state (left) and the first excited state (right) at multiple Euclidean time $t$.  Results are shown with the normalization $\sum\limits_{\vec r\in \Lambda} R^2(\vec r,t) =1$ .
  } \label{Fig-R-opt01}
\end{figure*}
%==========================

%==========================
\begin{figure}[htbp]
  \centering
  \includegraphics[width=8.7cm]{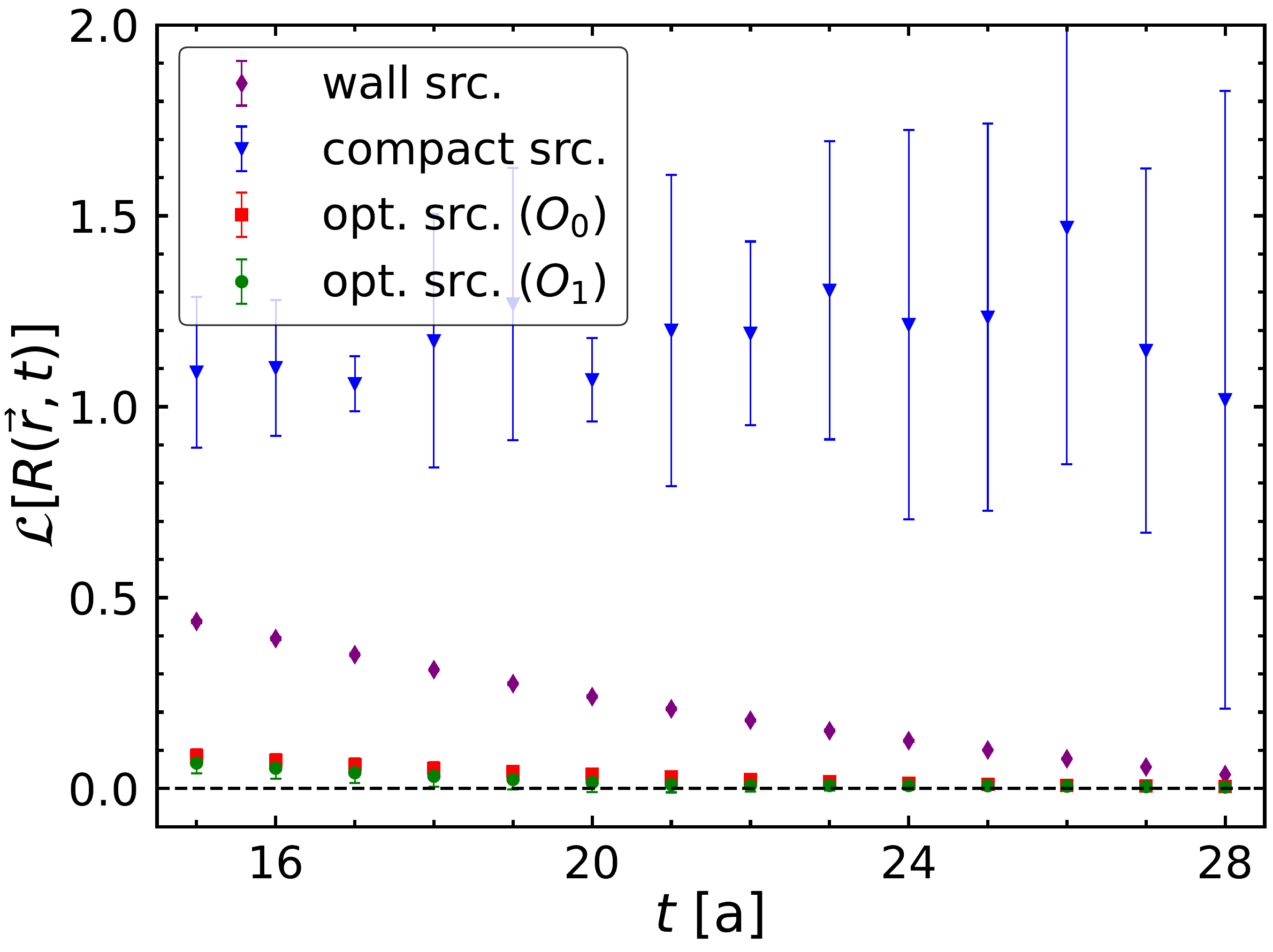}
  \caption{The residue factor $\mathcal{L}[R(\vec r,t)]$ defined in Eq.~(\ref{Eq-Res-R}) for quantifying  stability of the spatial profile of $R(\vec r,t)$ against Euclidean time $t$.
  Results calculated using the wall source, the compact source, and the optimized sources for the ground state ($O_0$) and the first excited state ($O_1$) are presented together for comparison.
  } \label{Fig-Res-R}
\end{figure}
%==========================

\subsubsection{Hadronic correlation functions}
Using the dual functions in Fig.~\ref{Fig-Psi_01_3d}, we 
construct optimized two-baryon operators defined in Eq.~(\ref{Eq-O}) and use them as source operators according to Sec.~\ref{Sec-Smear-A}
to compute hadronic correlation functions given in Eq.~(\ref{Eq-Rr}).

The hadronic correlation function $R_0(\vec r,t)$ for the ground state and $R_1(\vec r,t)$ for the first excited state are shown in Fig.~\ref{Fig-R-opt01} with the normalization $\sum\limits_{\vec r\in \Lambda} R^2(\vec r,t) =1$
at $t/a=15$, $20$, $25$, and $30$.
Unlike the case of the wall/compact sources where clear $t$ dependence is observed in correlation functions in Fig.~\ref{Fig-R-Wall-Comp},
$R_{0,1}(\vec r, t)$ here display very stable spatial profiles over a long Euclidean time period $t/a=15\sim30$.
To further quantify how stable the spatial profile of $R(\vec r, t)$ is with respect to $t$, we define a residue factor,
\begin{align}
\mathcal{L}[R(\vec r,t)]=\frac1V\sum\limits_{\vec r\in\Lambda}\left|\frac{R(\vec r, t)/R(\vec 0, t) }{R(\vec r, t_\text{f})/R(\vec 0, t_\text{f})} - 1\right|, \label{Eq-Res-R}
\end{align}
with $t_\text{f}/a=30$. 
The residue factor $\mathcal{L}[R(\vec r,t)]$ vanishes when the single state domination in $R(\vec r, t)$ is achieved. 
Shown in Fig.~\ref{Fig-Res-R} are corresponding residue factors for $R_{0}(\vec r, t)$ and $R_{1}(\vec r, t)$, together with those from the wall source and the compact source for comparison.
Small residue factors are observed for $R_{0,1}(\vec r,t)$,
which indicates $R_{0}(\vec r,t)$ and $R_{1}(\vec r,t)$ are dominated by the ground state and the first excited state, respectively.
These observations lead to a conclusion that the iterative process for operator optimization proposed in Sec.~\ref{Sec-Dual-Func-B} is converged at this step.

The spatial profiles of $R_{0}(\vec r,t)$ and $R_{1}(\vec r,t)$ closely match with the dual functions $\Psi_{0}(\vec r)$ and $\Psi_{1}(\vec r)$ in Fig.~\ref{Fig-Psi_01_1d}, respectively.
This agreement suggests that the underlying NBS amplitudes $\psi_0(\vec r)$ and $\psi_1(\vec r)$ are nearly orthogonal to each other.
This observed near-orthogonality also explains the rapid convergence in the iterative optimization process:
the initial eigen functions $\psi^{(0)}_{0,1}(\vec r)$, derived from the wall source LO potential,
preserve their orthogonal relationship and thus provide a good approximation to the genuine NBS amplitudes.
In general cases where NBS amplitudes are merely linearly independent (but not orthogonal), however,  more iterations would be required for convergence.
Typically in such cases, the resulting converged dual functions $\Psi_n(\vec r)$ display spatial profiles different from the converged correlation functions $R_{n}(\vec r,t)$ (proportional to the NBS amplitude), and $\{\Psi_n(\vec r)\}$ would not be orthogonal to each other
\footnote{We expect, however, that orthogonality of $\{\Psi_n(\vec r)\}$ is recovered in the infinite volume limit as long as the interaction is short-ranged.}.

\begin{figure}[htbp]
  \centering
  \includegraphics[width=8.7cm]{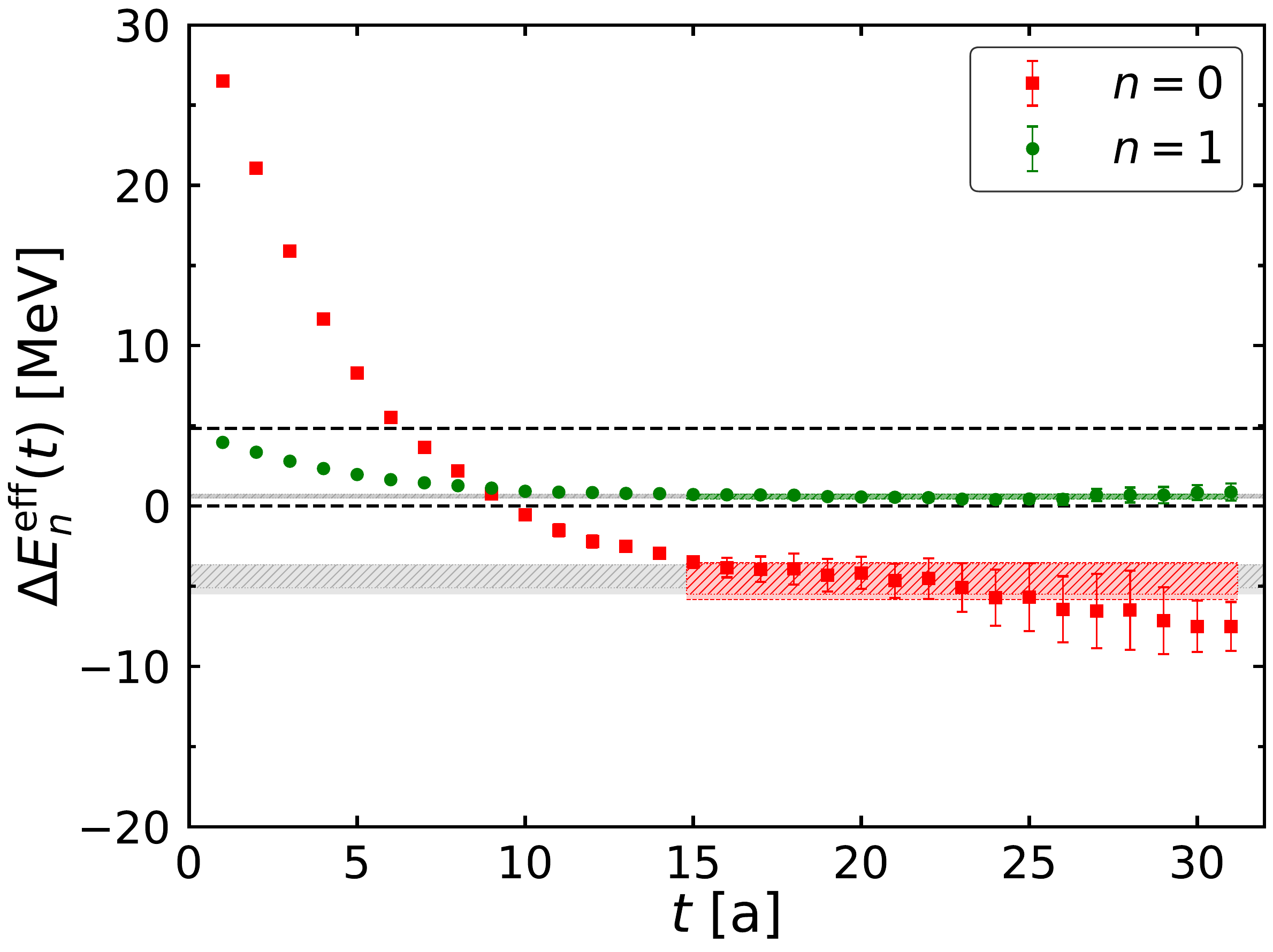}
  \caption{The effective energies $\Delta E^{\text{eff}}_n(t)=\ln\left[\frac{R_n(t)}{R_n(t+1)}\right]$ calculated with optimized two-baryon operators for the ground state ($n=0$) and the first excited state ($n=1$), respectively.
  Results from a single-state fit to $R_{0,1}(t)$ are shown by the red and green bands. 
  The energies from solving 3D Schr\"odinger equation using the HAL QCD potentials are shown by the gray bands.
  The inner bands represent the statistical errors, while the outer bands are obtained by adding the systematic uncertainties in quadrature.
  Exact values are provided in Eqs.~(\ref{Eq_E0_fit})-(\ref{Eq_E1_HAL_fit}).
  Non-interacting spectra are shown by black dashed lines for reference.
  } \label{Fig-Eeff}
\end{figure}

Having hadronic correlation functions $R_{1,0}(\vec r, t)$ 
with least nearby state contamination, we should be able to identify each state in their spectra.
Shown in Fig.~\ref{Fig-Eeff} are the effective energies $\Delta E^{\text{eff}}_{0,1}(t)$, derived from $R_{0,1}(t)$ in Eq.~(\ref{Eq-Rt}) obtained by further applying the optimized operators at the sink. 
Effective energies $\Delta E^{\text{eff}}_{0,1}(t)$ are found to exhibit stable plateau against a long period of $t$.
A single-state fit to $R(t)$ leads to,
\begin{align}
    & \Delta E_0= -4.6(9)\left(^{+4}_{-9}\right)~\text{[MeV]}, \label{Eq_E0_fit}\\
    & \Delta E_1= 0.6(2)\left(^{+0}_{-0}\right)~\text{[MeV]}, \label{Eq_E1_fit}
\end{align}
which are consistent with the energies obtained from solving the 3D Schr\"odinger equation using the HAL QCD potentials (to be shown in the next subsubsection),
\begin{align}
    & \Delta E^\text{HAL}_0= -4.4(7)\left(^{+3}_{-9}\right)~\text{[MeV]}, \label{Eq_E0_HAL_fit}\\
    & \Delta E^\text{HAL}_1= 0.6(1)\left(^{+0}_{-1}\right)~\text{[MeV]}. \label{Eq_E1_HAL_fit}
\end{align}
In both determinations for $\Delta E_{0,1}$, the centra value and statistical error are obtained by a fit for $t/a\in[15,32]$, while the systematic uncertainties are estimated by comparing results from different fit windows $t/a\in[20,32]$ and $t/a\in[15,25]$.
These results demonstrate our optimized operators clearly disentangle the ground state and the first excited state, both lying around $2m_{\Omega_{ccc}} \simeq 9700$ MeV with the energy gap as narrow as $\sim 5$ MeV.
Contrary to the naive expectation, 
$\Delta E_0^{\text{eff}}(t)$ in Fig.~\ref{Fig-Eeff} shows large statistical errors than $\Delta E_1^{\text{eff}}(t)$. 
Our interpretation is as follows. The ground state wavefunction is more localized than the first excited one. Hence,
the number of space points contributing to the signal in Eq.~\eqref{Eq-sp-src} is effectively less for the ground state than for the first excited state.

%==========================
%==========================

%==========================
\begin{figure*}[htbp]
  \centering
  \includegraphics[width=8.7cm]{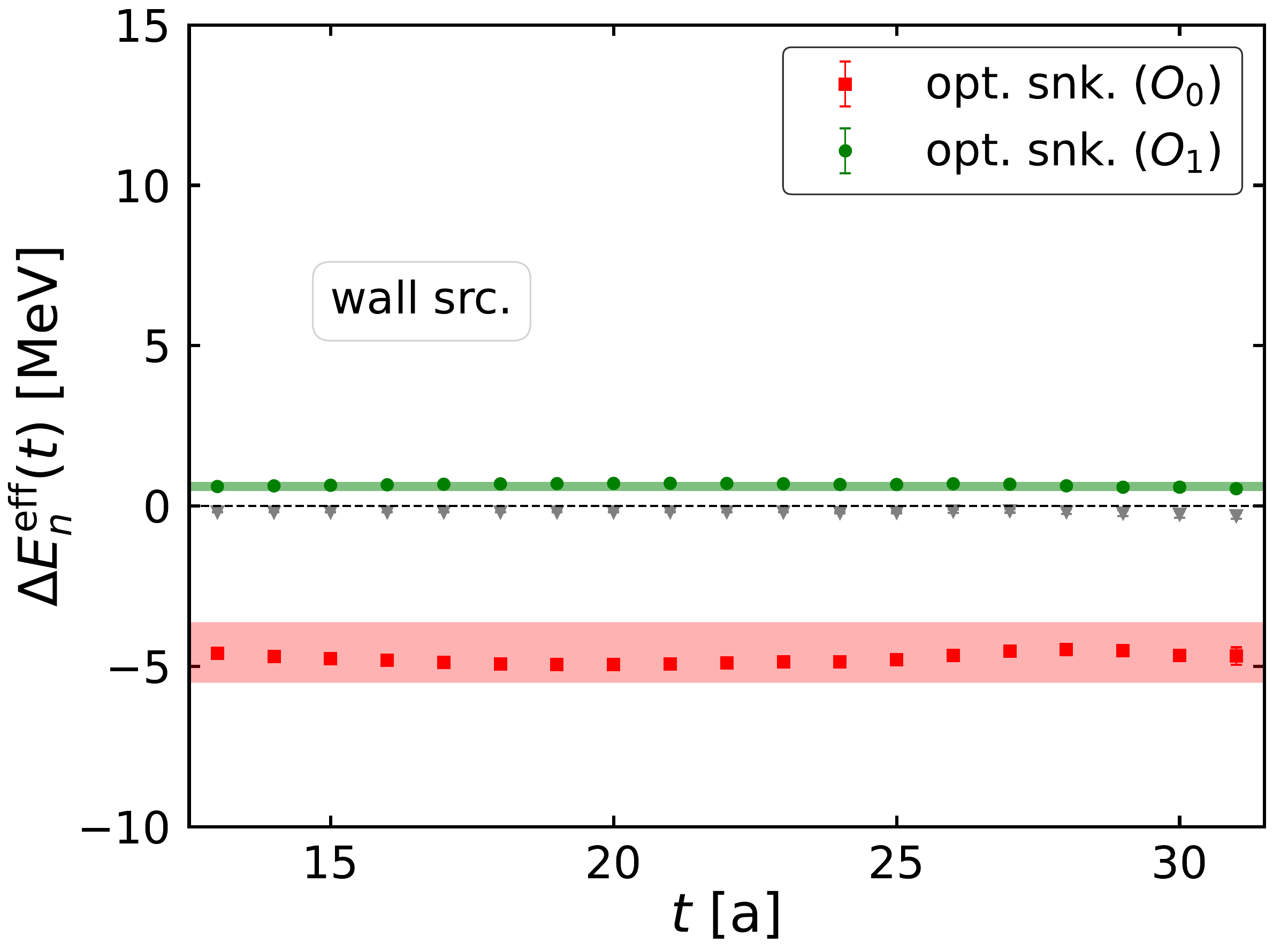}
  \includegraphics[width=8.7cm]{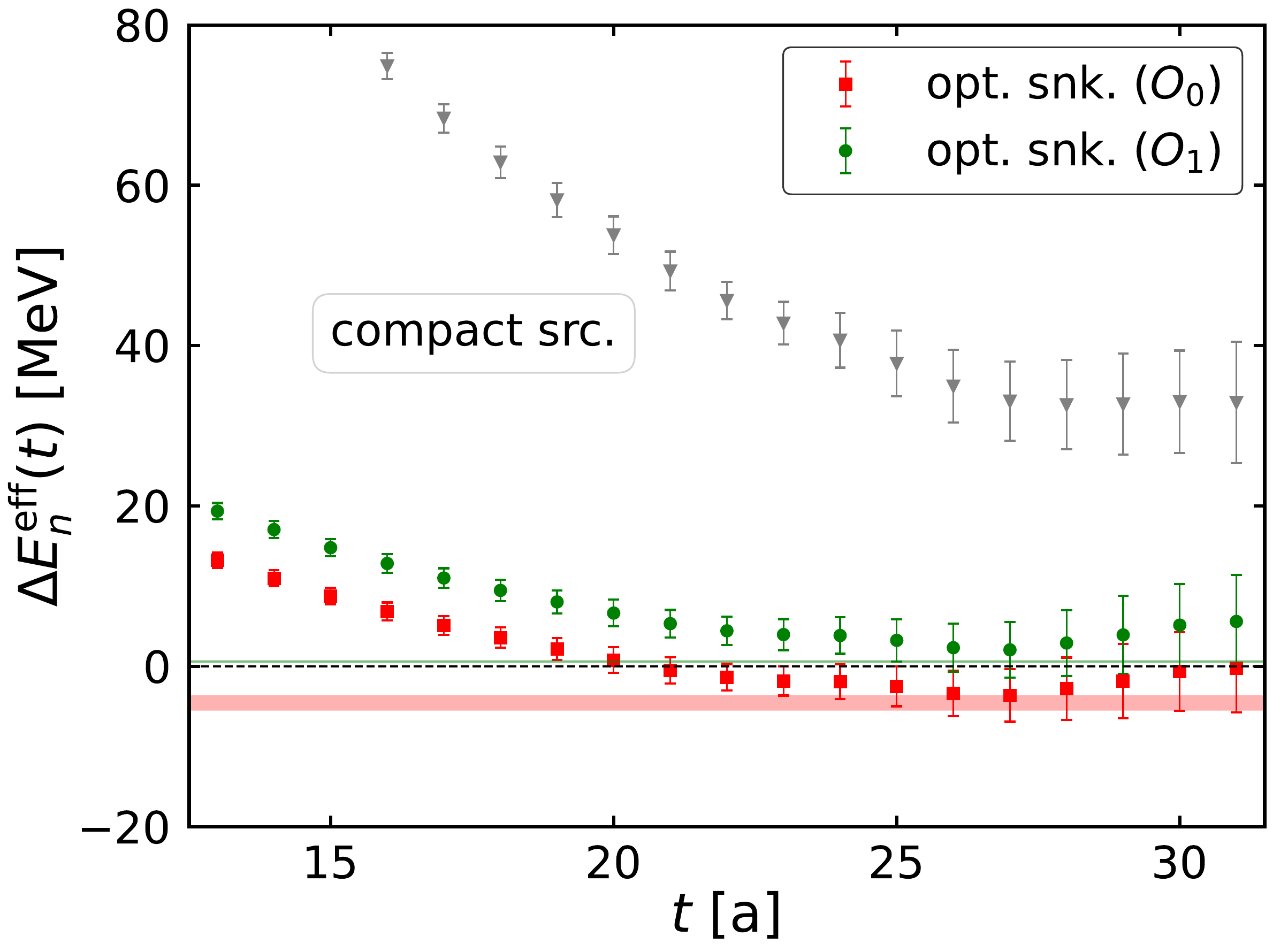}
  \caption{The effective energies for the ground state (red symbols) and the first excited state (green symbols) derived by the temporal correlation functions
  with the optimized sinks operators and the wall (left)/compact (right) sources defined in Eq.~(\ref{Eq-R-O-J}).
  The red and green bands are the genuine $\Delta E_0$ and  $\Delta E_1$ obtained from the plateaux in Fig.~\ref{Fig-Eeff}, respectively.
  Results in Fig.~\ref{Fig-Eeff0-wall-comp} are also shown by gray symbols for comparison.
  } \label{Fig-Eeff01-wall-comp}
\end{figure*}
%==========================

To further explore the ability of optimized operators $ O_{0,1}$, we apply them as sink operators in the cases of the wall source and the compact source, namely, we compute the following temporal correlation functions:
\begin{align}
    R^J_{0,1}(t) &= \langle  O_{0,1}(t) \bar J(0)\rangle, \nonumber \\
         &= \frac{1}{V}\sum_{\vec r\in\Lambda} \Psi^*_{0,1}(\vec r) R_J(\vec r, t), \label{Eq-R-O-J}
\end{align}
where $\bar J$ being the wall/compact sources, and $R_J(\vec r, t)$ are the hadronic correlation functions shown in Fig.~\ref{Fig-R-Wall-Comp}.
Using them, we can derive the effective energies for the ground state and the first excited state, which are shown in Fig.~\ref{Fig-Eeff01-wall-comp}.
We find that obtained effective energies match well with the genuine $\Delta E_{0,1}$, though those from the compact source contain large uncertainties.

\subsubsection{Potentials from the time-dependent HAL QCD method}
\begin{figure*}[htbp]
  \centering
  \includegraphics[width=8.7cm]{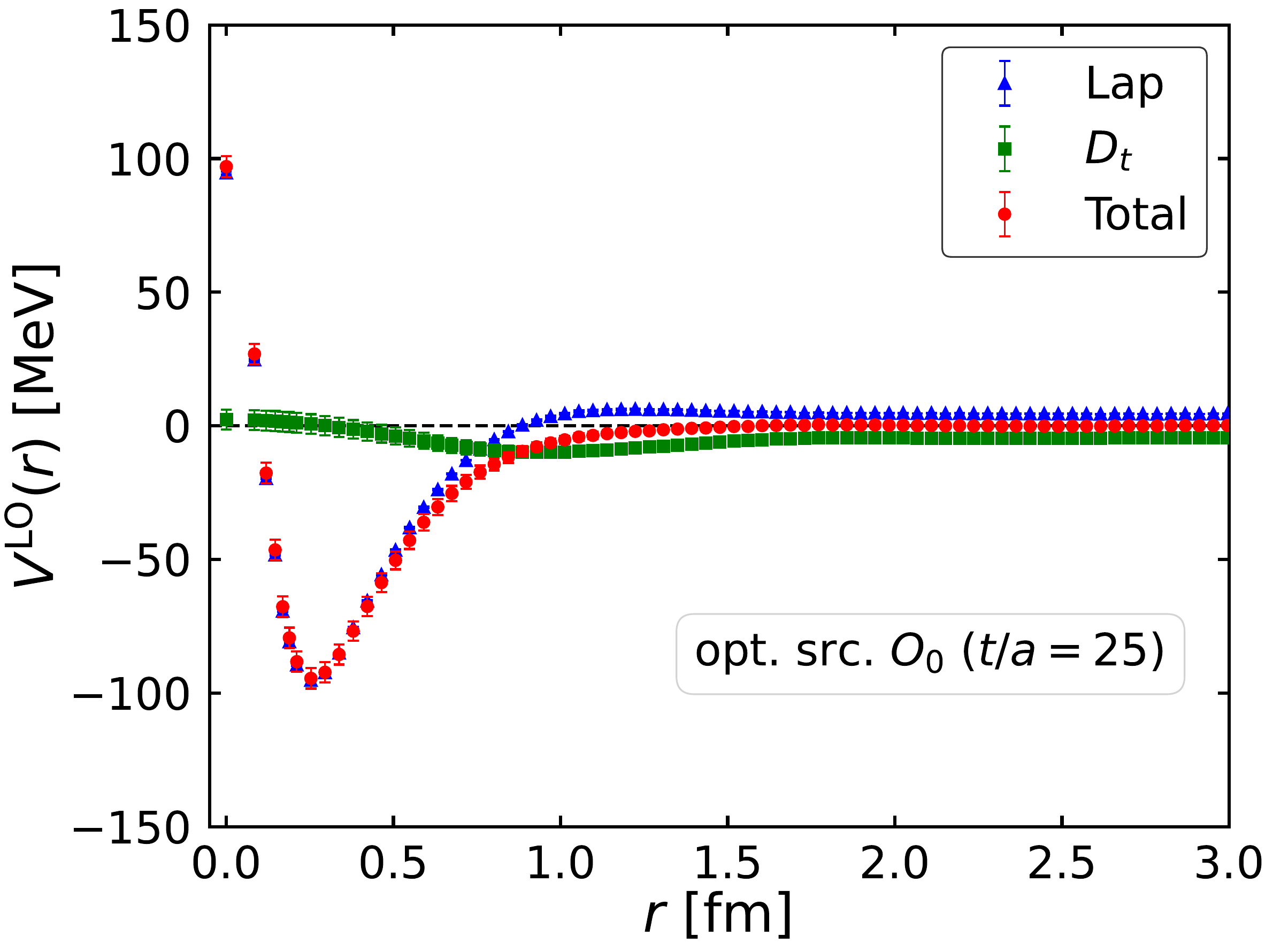}
  \includegraphics[width=8.7cm]{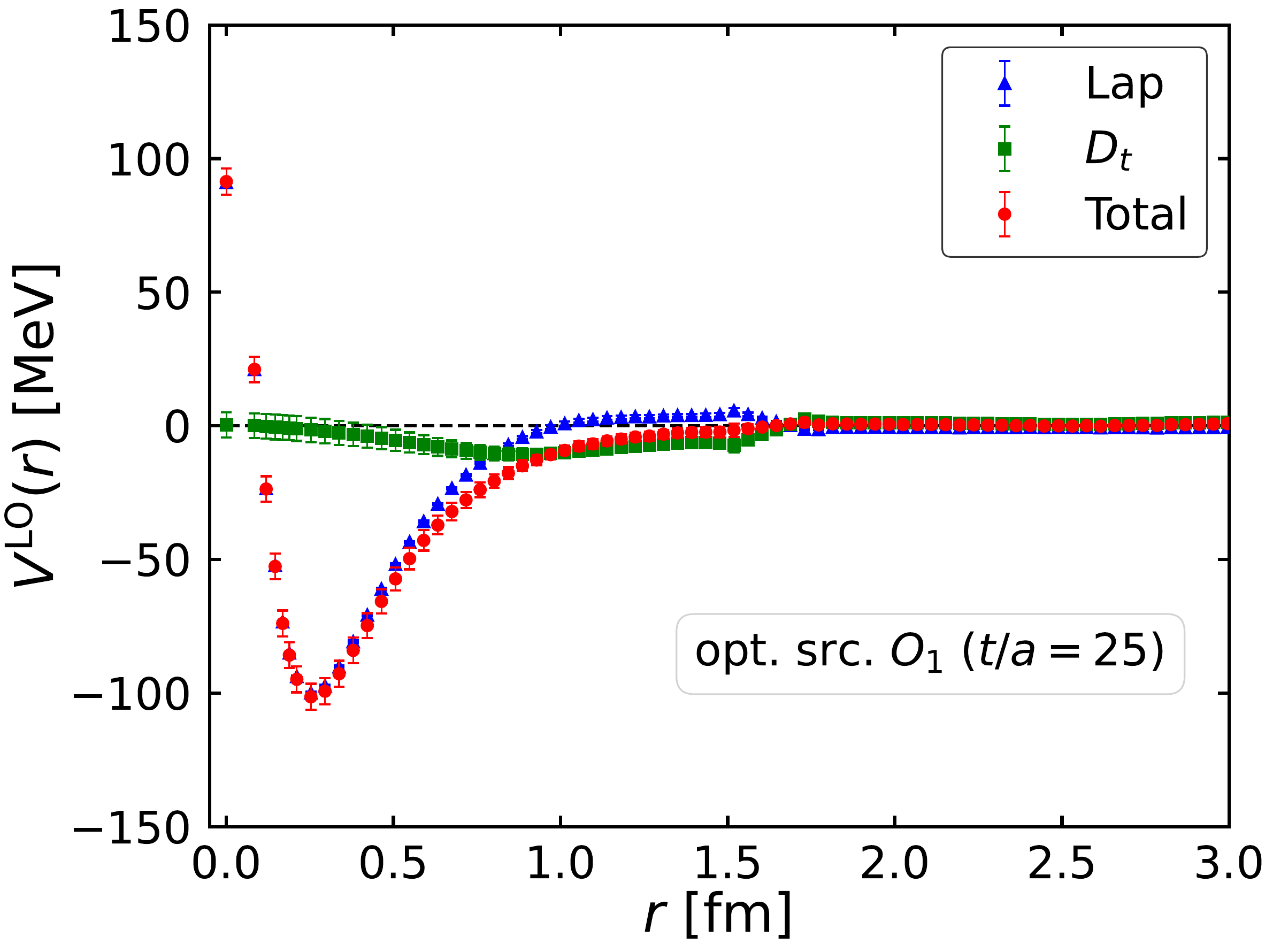}
  \caption{The LO potentials calculated using the optimized sources $ O_0$ for the ground state (left) and $ O_1$ for the first excited state (right) at Euclidean time $t/a=25$.
  ``Lap'' and``$D_t$'' represent the Laplacian component $R^{-1}(\vec r, t)\frac{\nabla^2}{m_B}R(\vec r, t)$, and the time-derivative component
$R^{-1}\left(\frac{1}{4m_B}\frac{\partial^2}{\partial t^2} -\frac{\partial}{\partial t} \right)R(\vec r, t)$ of the total potential in Eq.~(\ref{Eq-HAL-LOV}).
  } \label{Fig-VLO-opt01}
\end{figure*}

\begin{figure}[htbp]
  \centering
  \includegraphics[width=4.2cm]{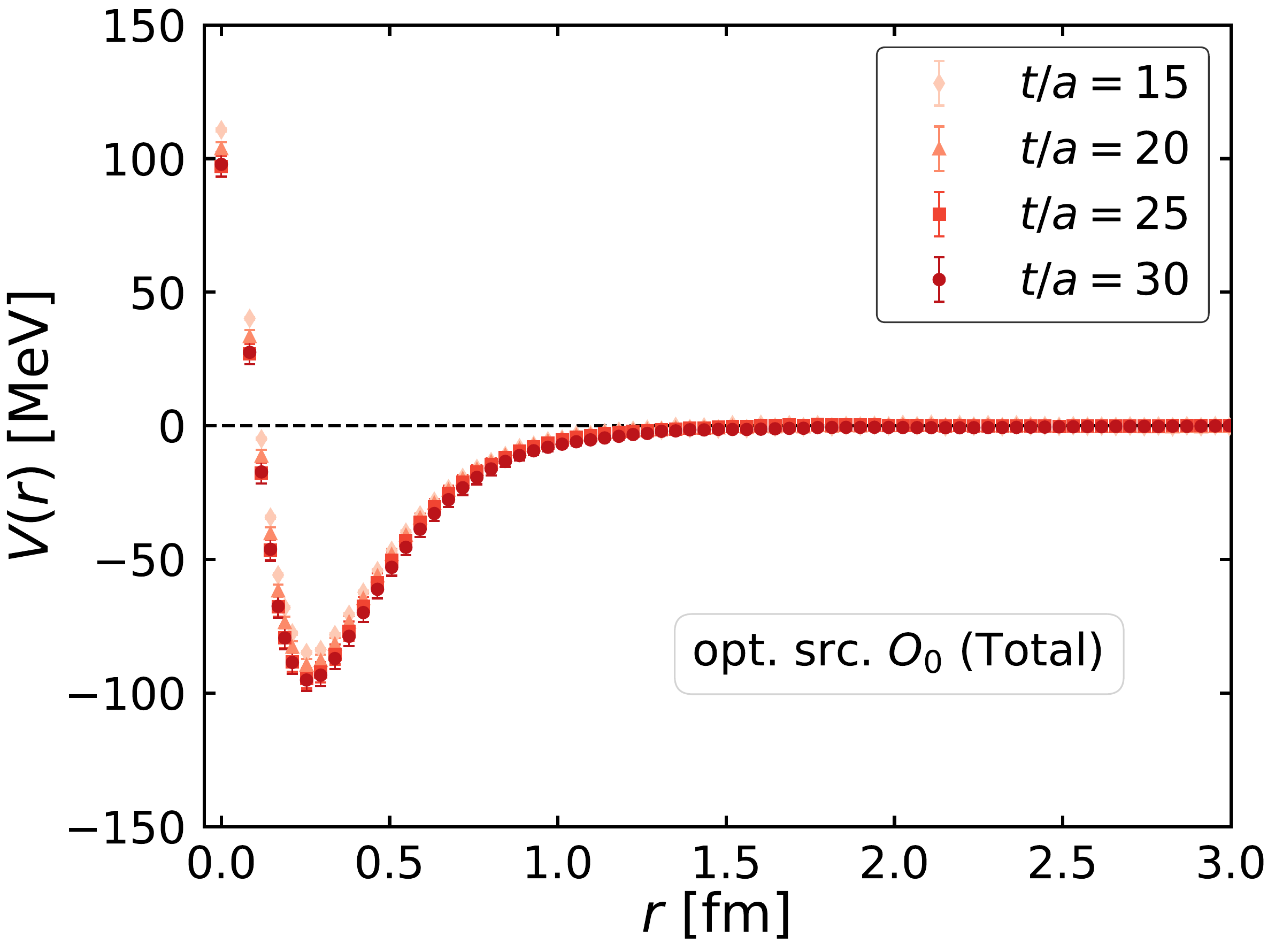}
  \includegraphics[width=4.2cm]{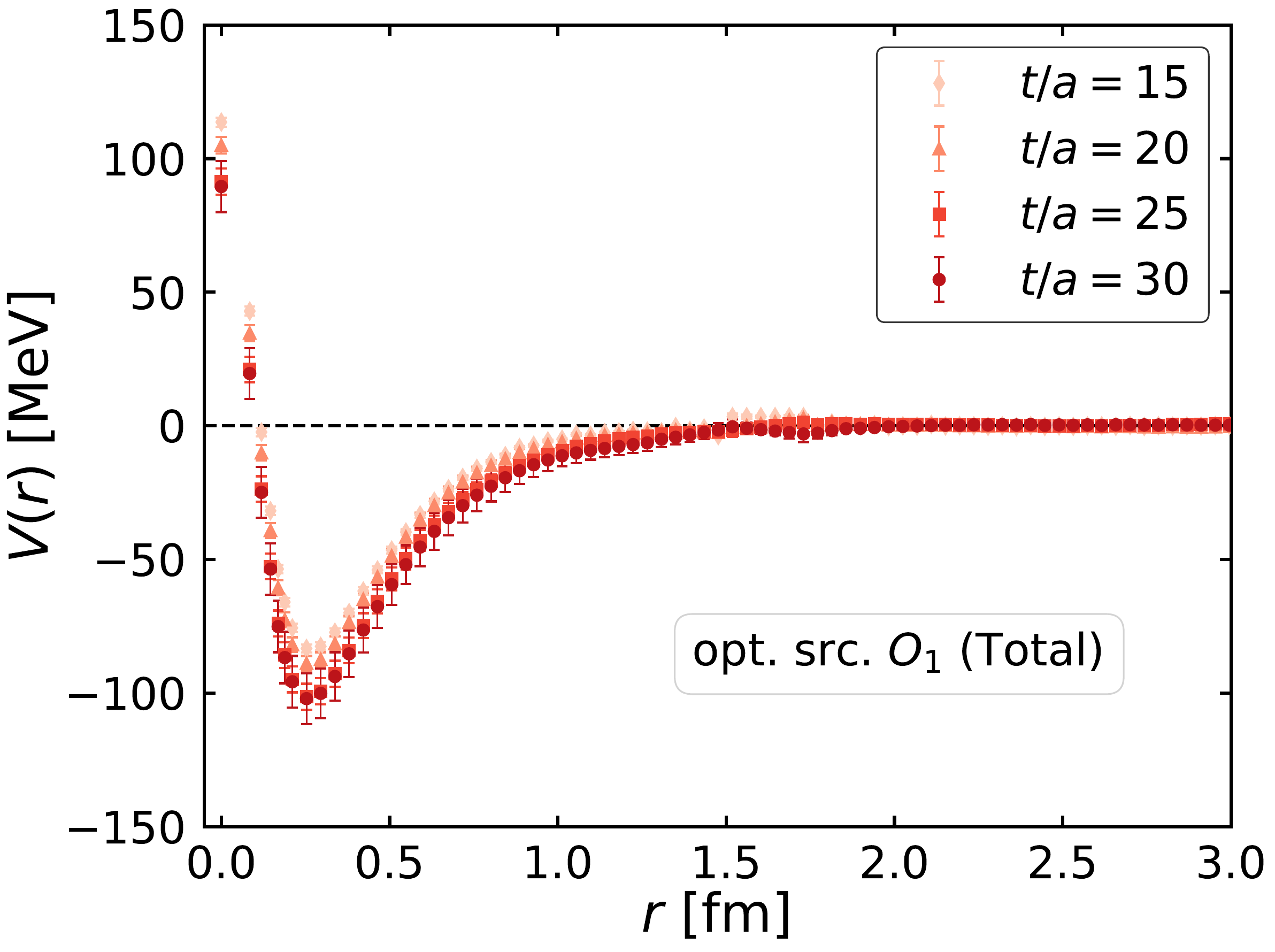}
  \includegraphics[width=4.2cm]{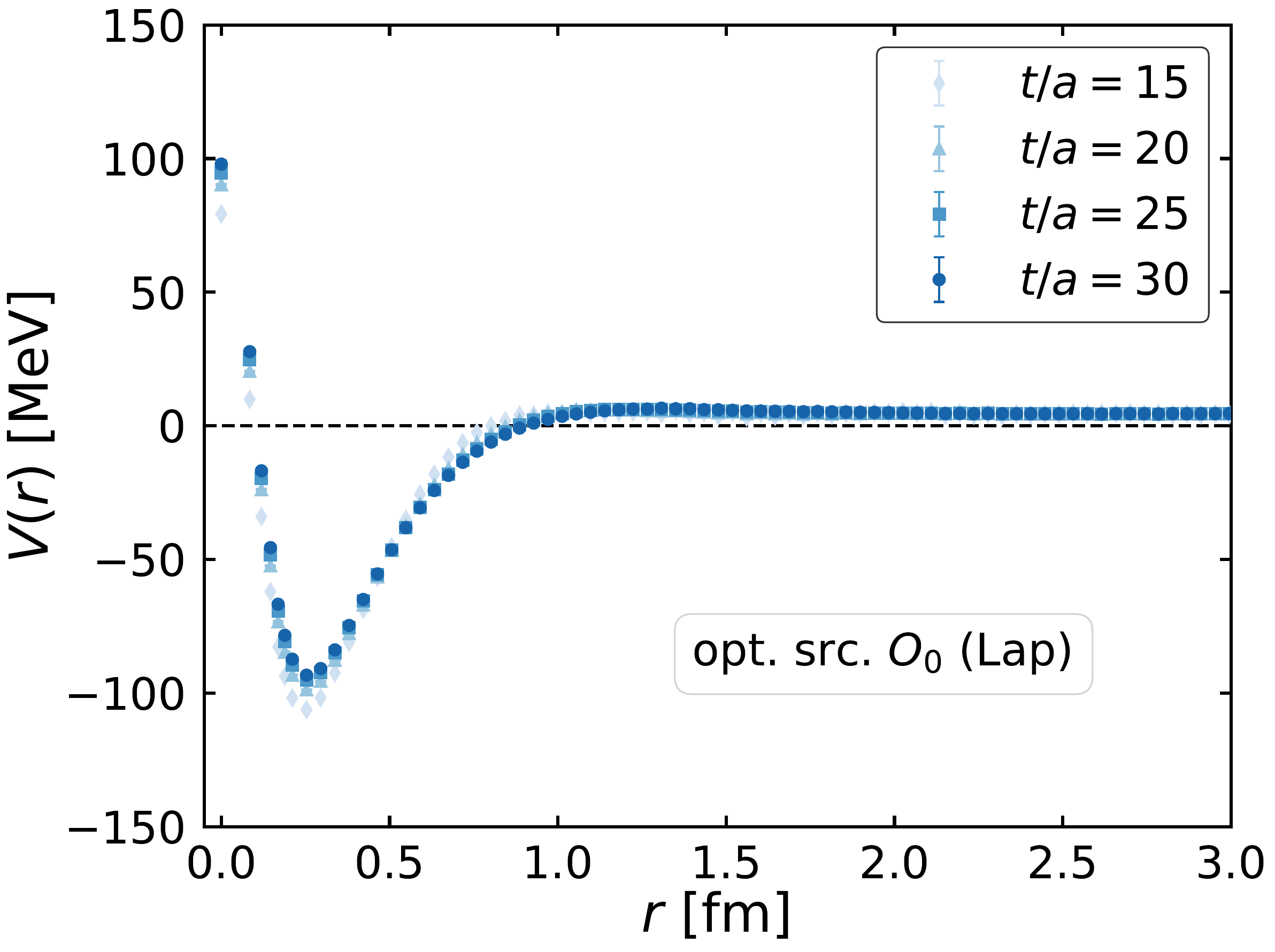}
  \includegraphics[width=4.2cm]{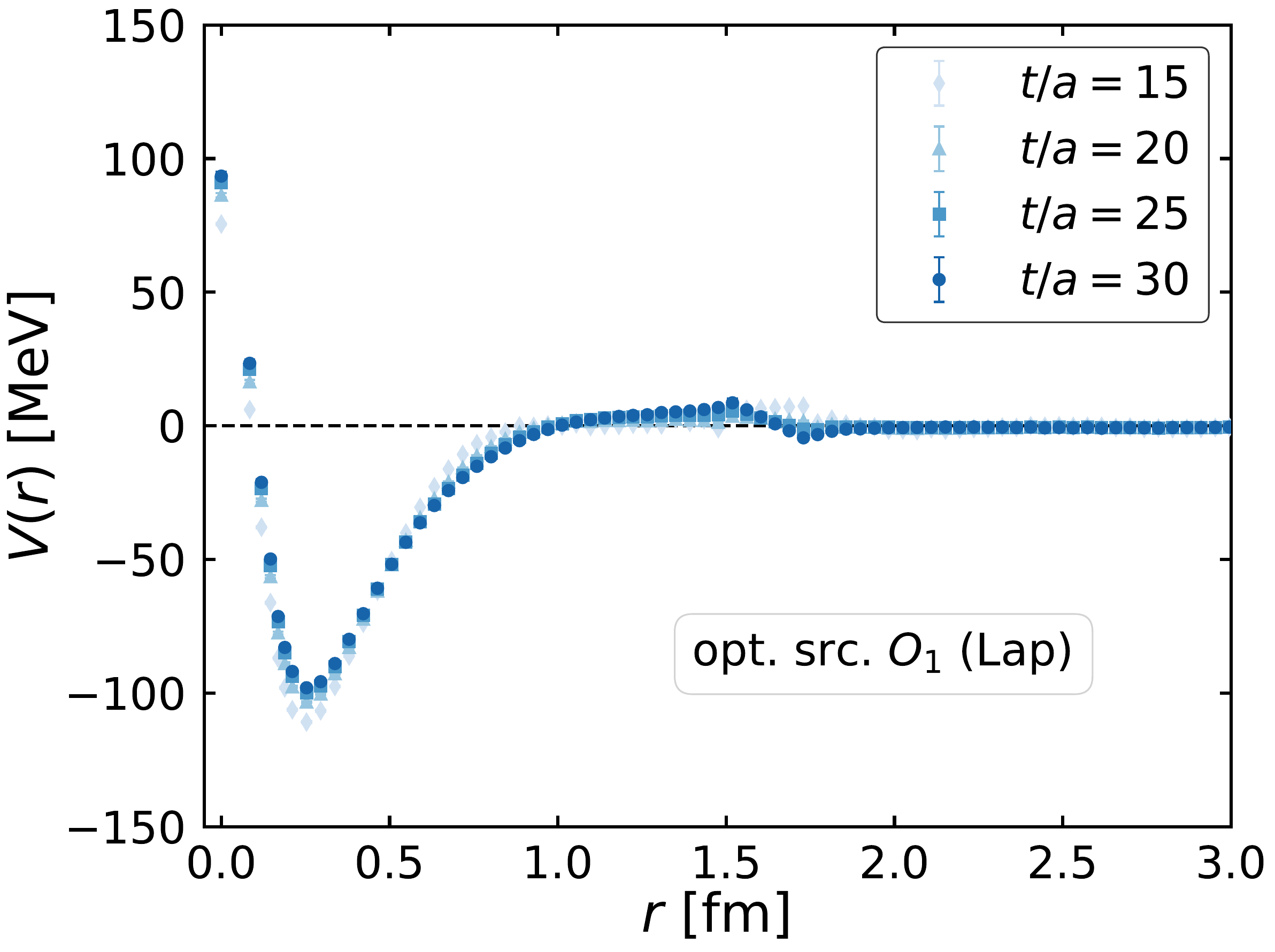}
  \includegraphics[width=4.2cm]{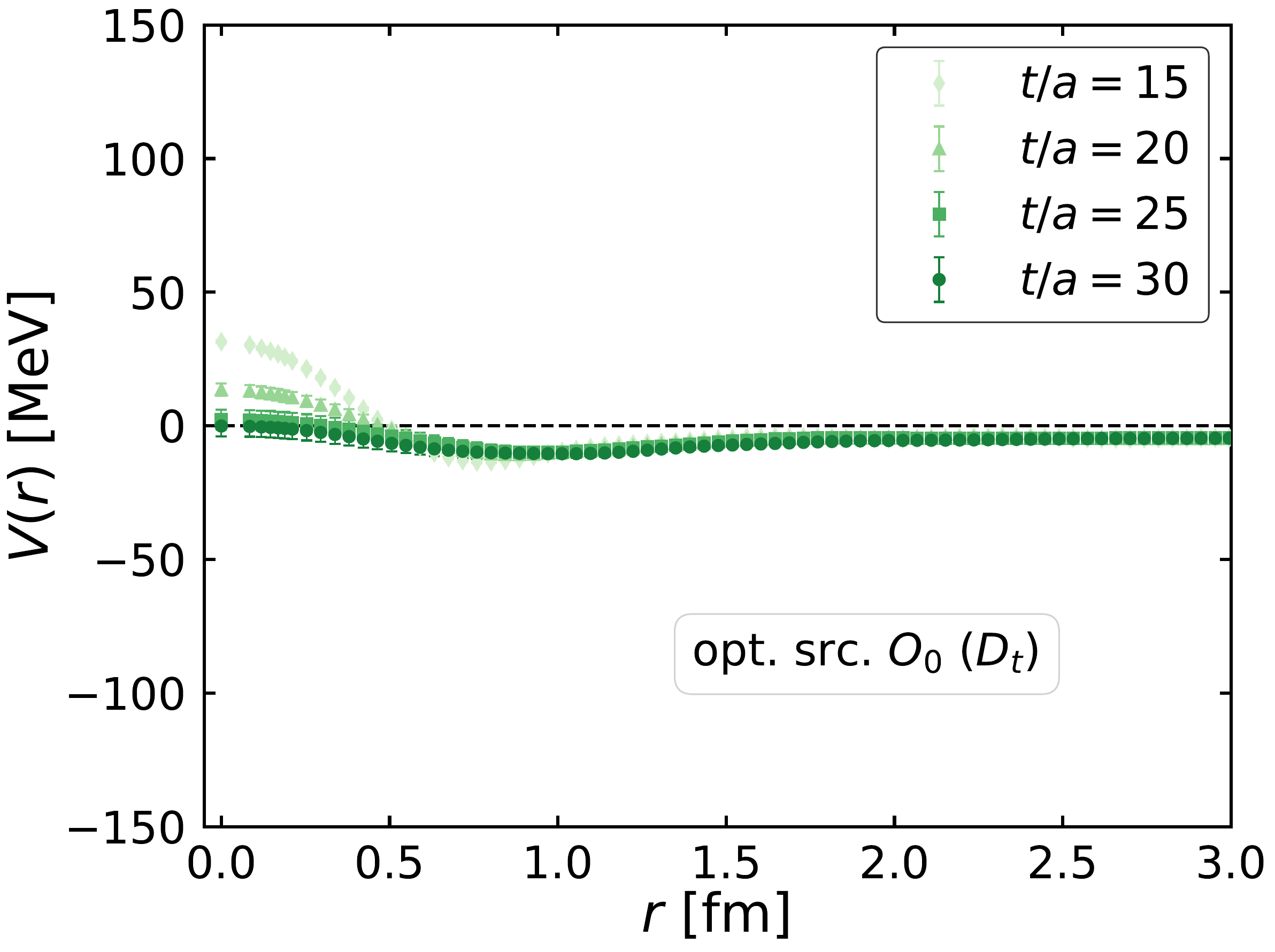}
  \includegraphics[width=4.2cm]{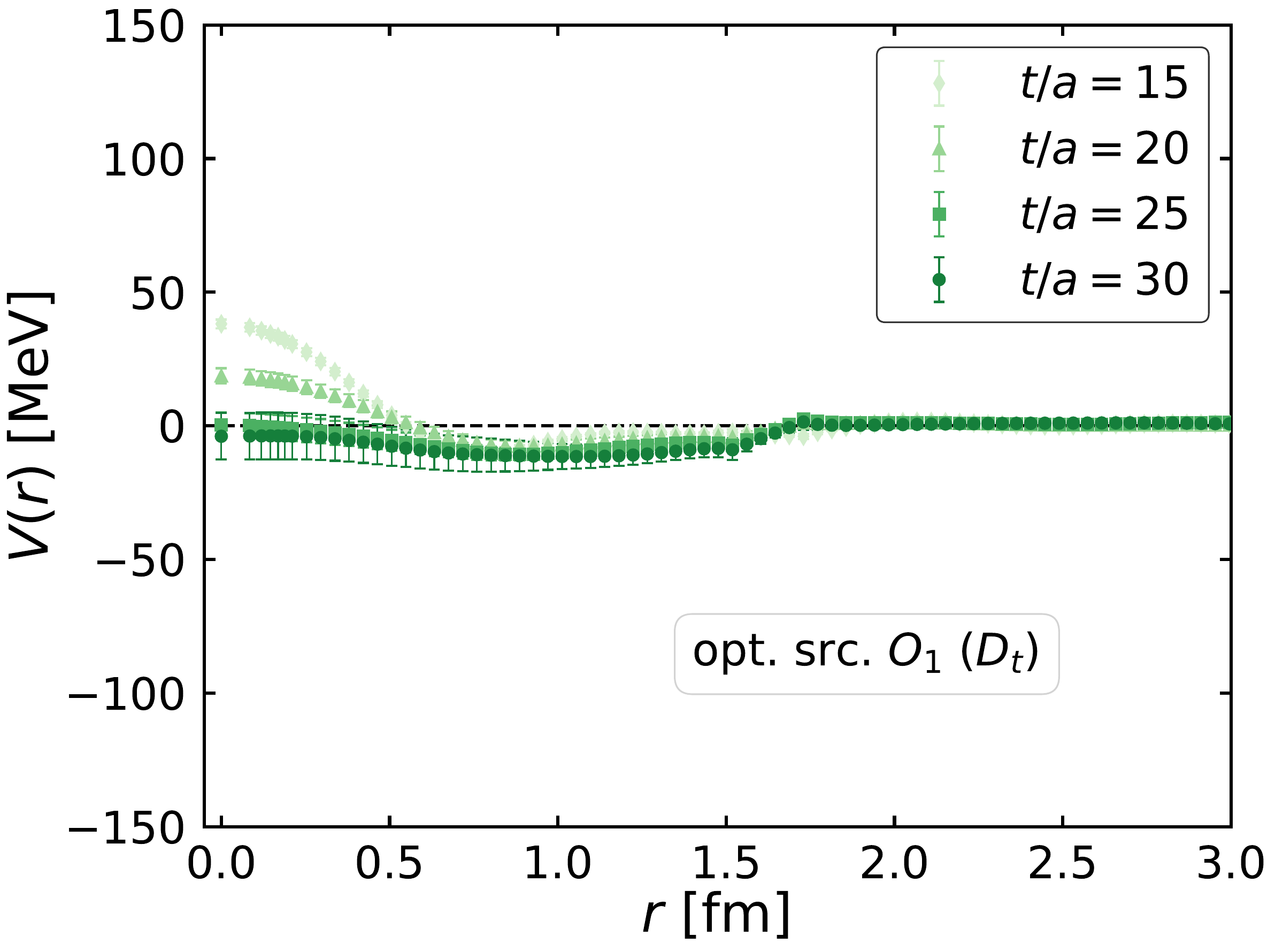}
  \caption{The time dependence of the total potential (top), of its Laplacian component (middle), and of its $D_t$ component (bottom) calculated with the optimized sources $O_0$ for the ground state (left) and $O_1$ for the first excited state (right). 
  } \label{Fig-V-tdep-opt01}
\end{figure}

Let us now perform the time-dependent HAL QCD analysis using the $R_{0,1}(\vec r, t)$ in Fig.~\ref{Fig-R-opt01}.
In Fig.~\ref{Fig-VLO-opt01}, we show  the LO potentials extracted at Euclidean time $t/a=25$ by using Eq.~(\ref{Eq-HAL-LOV}).
The total potentials in both cases
%calculated using $R_{0}(\vec r, t)$ and $R_{1}(\vec r, t)$ 
are dominated by the Laplacian terms, with small contributions from the $D_t$ terms,
which show weak $r$-dependence and become nearly constant at long range.
%For results from both $R_0(\vec r,t)$ and $R_1(\vec r,t)$, 
As shown in Fig.~\ref{Fig-V-tdep-opt01},
the time dependence of the total potentials,  of the Laplacian terms, and of the $D_t$ terms all are weak.
These observations are natural consequences of the dominance of a single state in 
 $R_0(\vec r,t)$ or $R_1(\vec r,t)$, namely the ground state or the first excited state, respectively.

\begin{figure}[htbp]
  \centering
  \includegraphics[width=8.7cm]{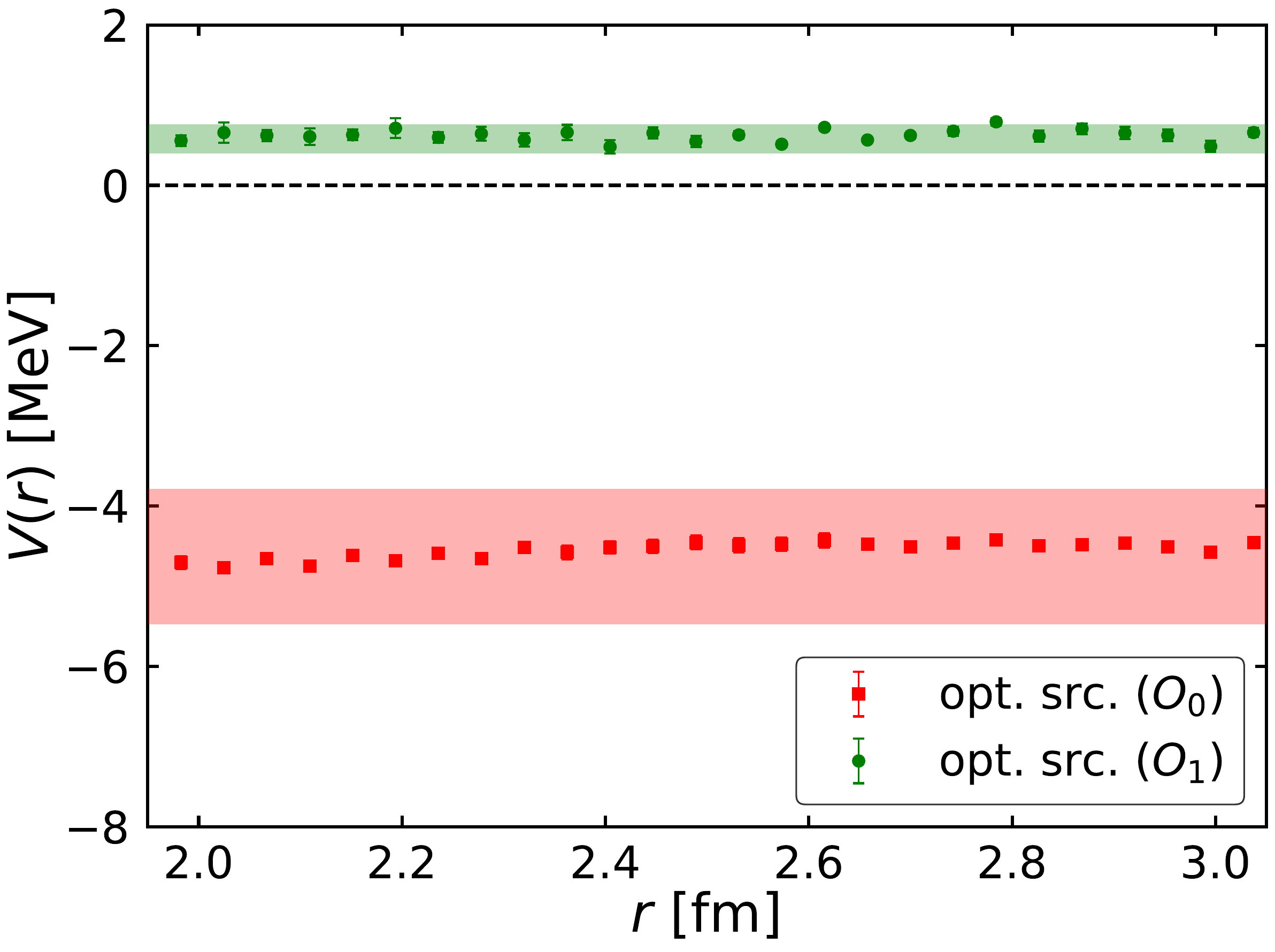}
  \caption{ Numerical confirmation of Eq.~(\ref{Eq-Helmholtz}).
  The red and green symbols are $\frac{-\nabla^2R_{0,1}(\vec r,t)}{m_B R_{0,1}(\vec r,t)}$ calculated using the optimized sources $O_0$  and $O_1$  at Euclidean time $t/a=25$, respectively.
  The red and green bands are $\frac{p^2_{0,1}}{m_B}$ with $p_{0,1}$ being the finite volume momenta from the effective energies.
  } \label{Fig-VLap-opt01}
\end{figure}

We have made a further test using the Laplacian terms.
Fig.~\ref{Fig-VLap-opt01} compares long range behaviors of the Laplacian terms (multiplied by $-1$) with $\frac{p^2_{0,1}}{m_B}$,  
where $p_{0,1}$ are the finite volume momentum for the ground state and the first excited state extracted from the plateaux of the effective energies in Fig.~\ref{Fig-Eeff}. 
In the asymptotic region, the NBS amplitudes satisfy the Helmholtz equation,
\begin{align}
    \left(\frac{\nabla^2}{m_B} + \frac{p_n^2}{m_B}\right)\psi_n(\vec r) = 0, \label{Eq-Helmholtz}
\end{align}
if the volume is large enough.
The consistency between the long-range plateaux of the Laplacian terms and $\frac{p^2_{0,1}}{m_B}$ in  Fig.~\ref{Fig-VLap-opt01} 
means that (i) the volume is large enough to have the asymptotic region; and (ii) $R_{0,1}(\vec r,t)$ is dominated by a single state, whose finite volume momentum is $p_{0,1}$.

In lattice calculations of two-body systems, it is generally difficult to evaluate the finite-volume effects on the obtained spectrum. A qualitative assessment can be made by checking whether the dimensionless parameter $m_\pi L$ is sufficiently large, or by examining physical observables extracted from spectra at multiple volumes to verify their expected behavior. 
These are all indirect and empirical assessments.
We emphasize that the above test provides direct evidence that the finite volume effect on the spectrum is well under control.

%==========================
%==========================

%==========================
%==========================

%==========================
%==========================

\subsection{Comparison with variational analysis using plane-wave operators}
The exceptionally large overlap of the optimized operators with targeted states lies in the 
nontrivial spatial wavefunctions incorporated in their constructions.
To demonstrate this point more explicitly, here we performed a  variational analysis using plane-wave operators as a comparison.

\begin{figure}[hbpt]
    \centering
    \includegraphics[width=8.7cm]{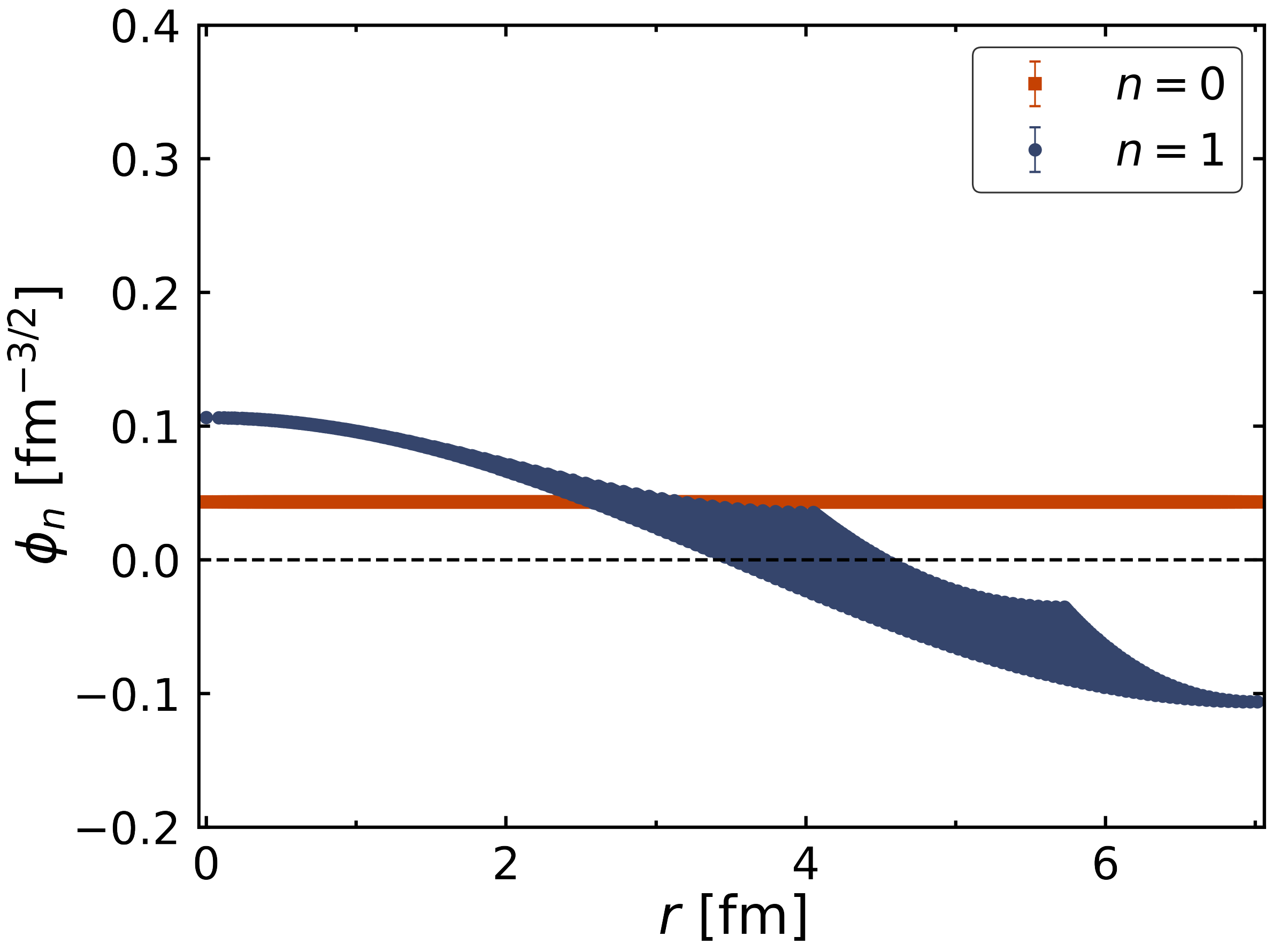}
    \caption{The free wavefunctions $\phi_{0,1}(r)$ in the $A_1^+$ irrep derived from the Eq.~(\ref{Eq_free_H}).}
    \label{Fig_2B_psi_p01}
\end{figure}

Within the current framework, the plane-wave operator is just a special case in Eq.~\ref{Eq-O}, where the realistic wavefunction is replaced by a plane wave $\phi_{\vec p}(\vec r)$,
\begin{align}
\tilde O(\vec p, t) & =\tilde B(\vec p,t)\tilde B(-\vec p,t) \nonumber\\
  & =\frac{1}{V^2}\sum_{\vec x,\vec r\in\Lambda} B(\vec x+\vec r,t)B(\vec x,t)\phi^*_{\vec p}(\vec r),\label{Eq_O_free_p}\\
H_0\phi_{\vec p}(\vec r) &=\frac{\vec p^2}{m_B}\phi_{\vec p}(\vec r),\label{Eq_free_H}
\end{align}
where $\tilde B(\vec p)$ is the Fourier transform of $B(\vec r)$,
the free Hamiltonian is defined as $H_0=-\nabla^2/m_B$, and the (discretized) relative momentum $\vec p$ is given by the PBC, $\vec p=\frac{2\pi}{L}\vec n$ with $\vec n\in \mathbb{Z}^3$. 
Operator $\tilde O(\vec p, t)$ in different irreps can be constructed by projecting $\phi_{\vec p}(\vec r)$ appropriately. 
Shown in Fig.~\ref{Fig_2B_psi_p01} are the first two free wavefunctions $\phi_{0,1}(r)$ projected to the $A^+_1$ irrep, which are obtained by solving Eq.~(\ref{Eq_free_H}) on a 3D box with the same geometry as our lattice and PBC. 
For later convenience, we denote operators constructed using $\phi_{0,1}(r)$ as $\tilde O_{0,1}(t)$.

\subsubsection{Symmetric source and sink operators}
\begin{figure*}[hbpt]
    \centering
\includegraphics[width=8.7cm]{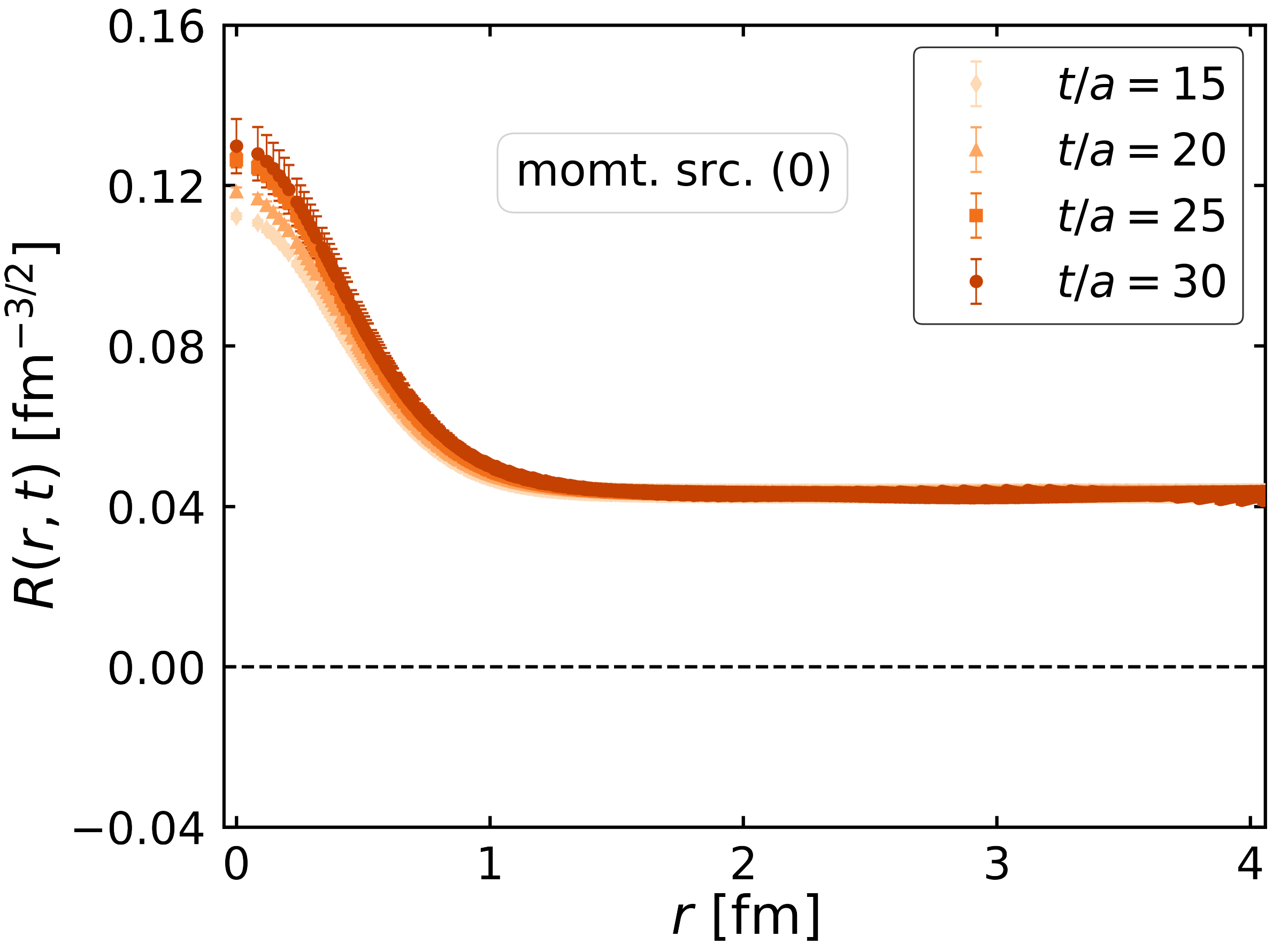}
 \includegraphics[width=8.7cm]{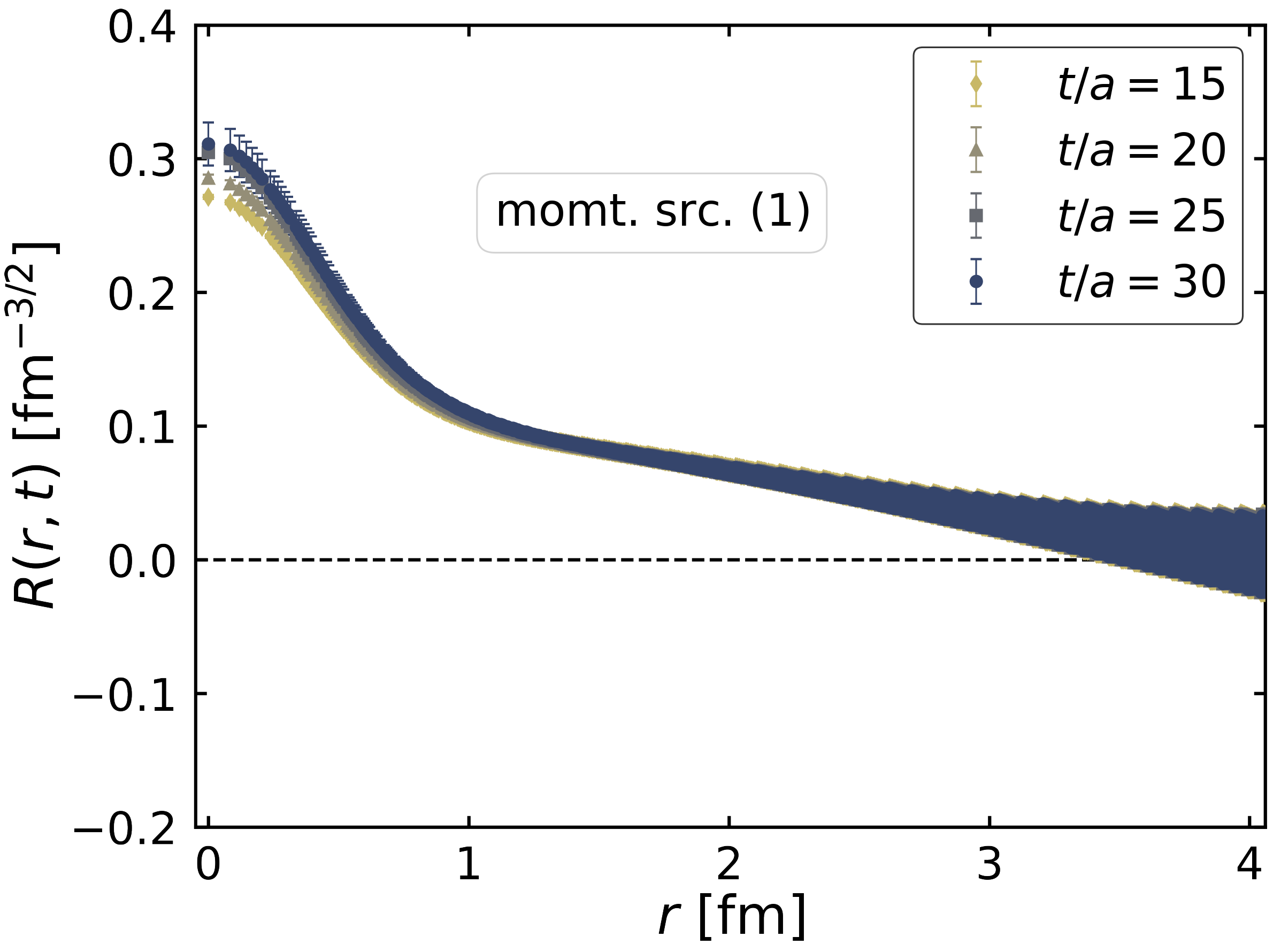}
    \caption{The correlation function $R(\vec r,t)$ in the $A_1^+$ irrep calculated using the plane-wave operators at source with relative momentum $p=0$ (left) and $p=1$ (right) in the unit of $2\pi/L$. The smeared $B$ operator are adopted at both the source and sink. 
    Results are shown with the normalization $\sum\limits_{\vec r\in \Lambda}R^2(\vec r, t)=1$.}
    \label{Fig_R_smr_p01}
\end{figure*}

According to the newly proposed calculating strategy using $Z_3$ noise and sparse field summation in Eqs.~(\ref{Eq-Smear-1}) and (\ref{Eq-Smear-2}), we implemented
the plane-wave operators $\tilde O_{0,1}(t)$ at source,
and obtained the following correlation functions, 
\begin{align}
& F_{0,1}(\vec r,t) = \frac{1}{V}\sum_{\vec x\in\Lambda}\langle B(\vec x +\vec r,t)B(\vec x ,t)\tilde O^\dagger_{0,1}(0) \rangle, \label{Eq_2B_F} \\
& R_{0,1}(\vec r,t) = \frac{F_{0,1}(\vec r,t) } {C_B(t)\cdot C_B(t)}.
\end{align}
In contrast to the asymmetric choice in the previous calculations (point-like $B$ at the sink, smeared $B$ at the source), smeared $B$ operators are adopted at both the source and sink. 
This {\it symmetric} setup ensures a {\it Hermitian} correlation matrix for the subsequent analysis.
In Fig.~\ref{Fig_R_smr_p01}, we show the spatial profile of $R(\vec r, t)$ over a long range of Euclidean time, which is normalized with the condition $\sum\limits_{\vec r\in \Lambda}R^2(\vec r, t)=1$ (Note, $R(\vec r,t)$ and $F(\vec r,t)$ are same under this normalization condition).
Both correlation functions have not yet achieved single-state dominance, as evidenced by their clear time-dependence.
We also note that the correlation function from the zero-momentum source in Fig.~\ref{Fig_R_smr_p01} (left)
resembles the one from the wall source in Fig.~\ref{Fig-R-Wall-Comp} (left).    

\begin{figure*}[hbpt]
    \centering
\includegraphics[width=8.7cm]{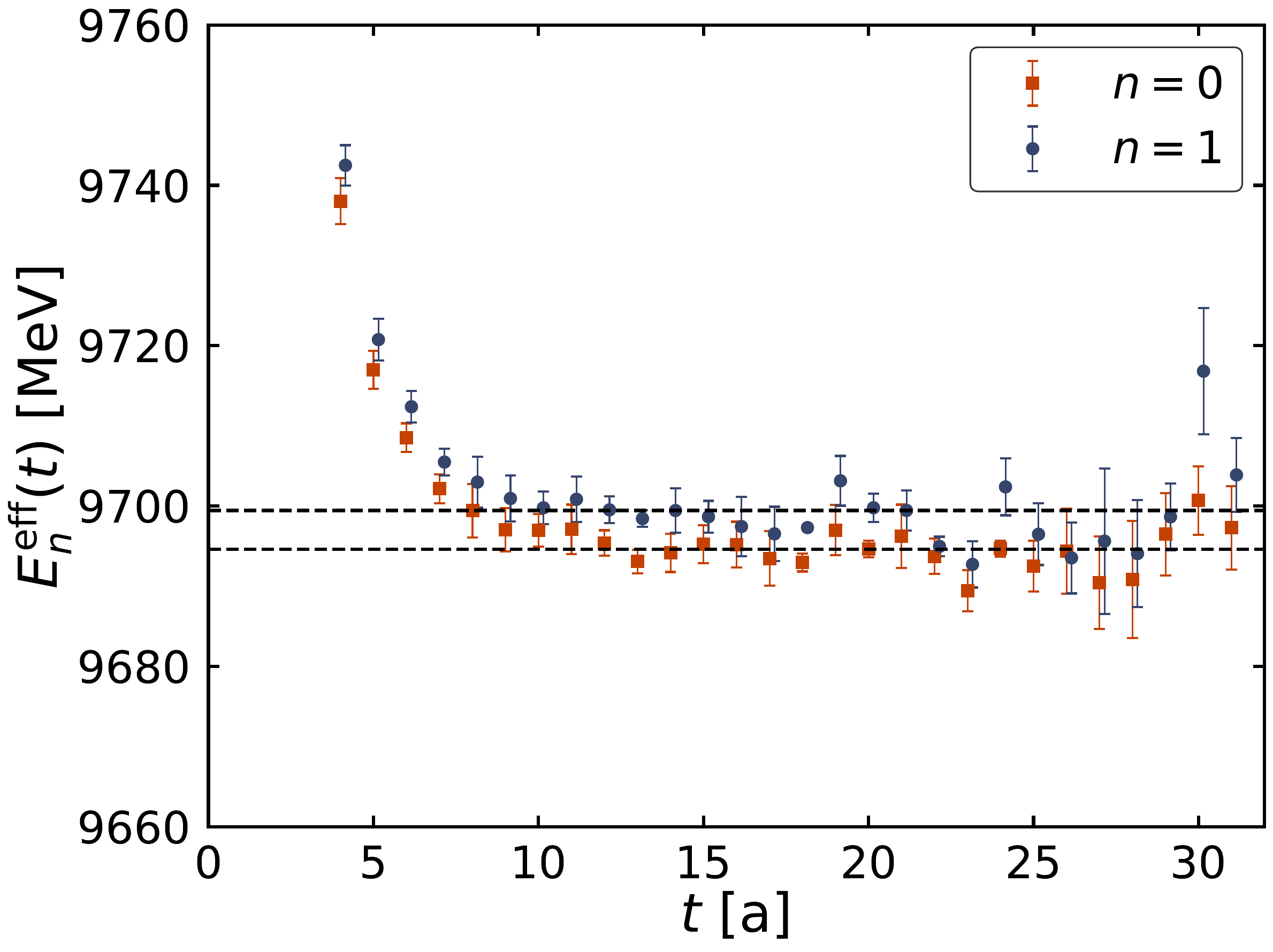}    \includegraphics[width=8.7cm]{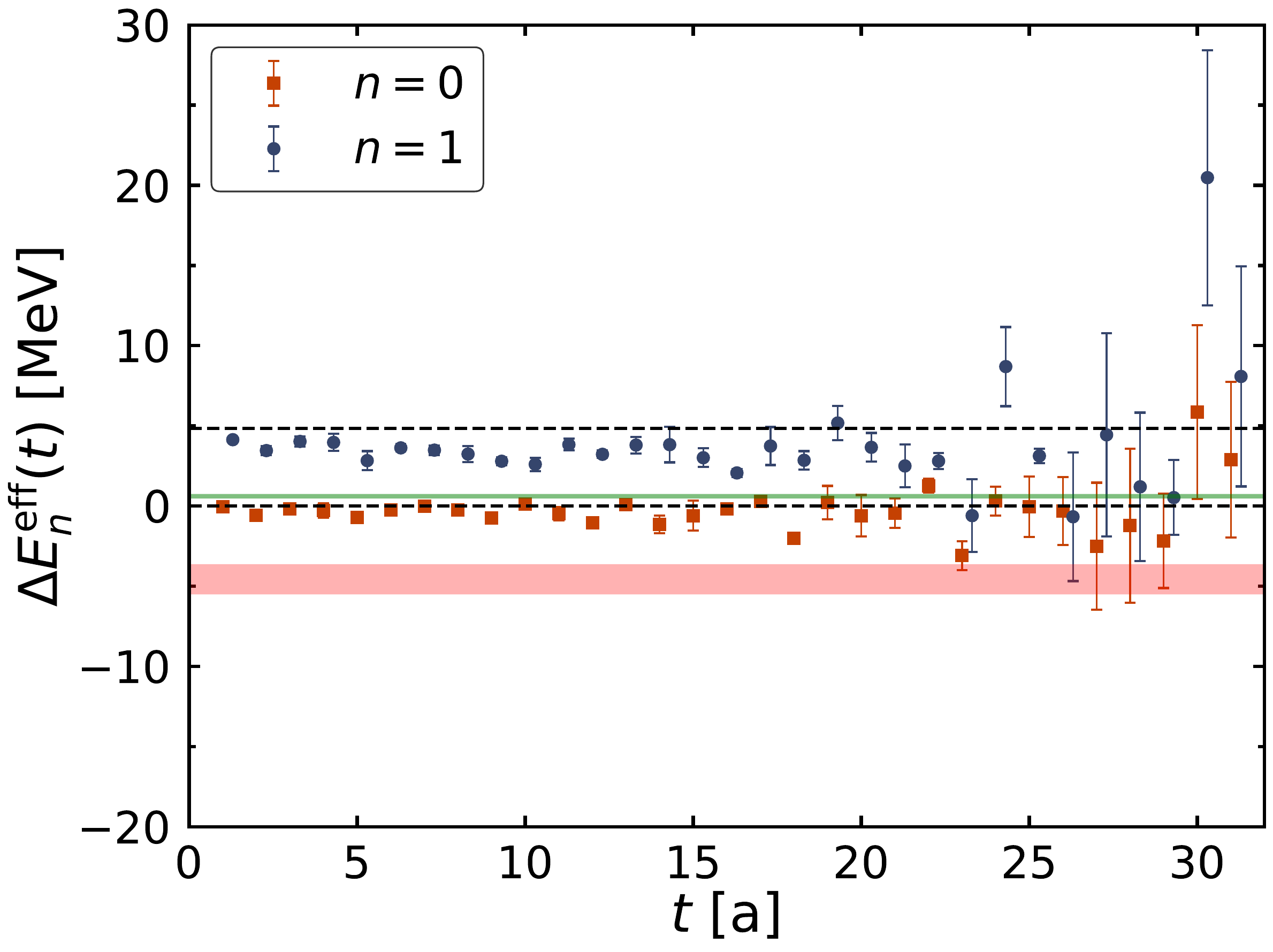}
\caption{The two-baryon effective energy $E^\text{eff}_n(t)$ and its difference with respect to the two-baryon threshold $\Delta E^\text{eff}_n(t)$ defined in Eqs.~(\ref{Eq_E}) and (\ref{Eq_DE}). 
Dashed lines are noninteracting energy levels, while bands are genuine energy levels determined from our wavefunction-optimized operators. 
}\label{Fig_2B_E_p01}
\end{figure*}

To proceed with the variational analysis, let us define the following $2\times2$ {\it Hermitian} matrix of temporal  correlation function
using $F_{0,1}(\vec r, t)$ in Eq.~(\ref{Eq_2B_F}),
\begin{align}\label{Eq_C_mtx}
    C_{mn}(t)=\langle \tilde O_m(t) \tilde O^\dagger_n(0) \rangle=\frac{1}{V}\sum_{\vec r\in\Lambda} \phi^*_m(\vec r) F_n(\vec r, t),
\end{align}
where $m,n\in[0, 1]$, and then solve the following generalized eigen value problem (GEVP),
\begin{align}\label{Eq_GEVP}
    C(\tau_D)v_n=\lambda_n C(\tau_0)v_n.
\end{align}
The obtained eigenvectors form a rotation matrix with each column being a eigenvector, $V=[v_0,v_1]$, which can be used to diagonalize the $C(t)$ matrix as,
\begin{align}\label{Eq_rot_C}
    \tilde C(t)=V^\dagger C(t)V.
\end{align}
We chose $\tau_0/a=12$ and $\tau_D/a=20$ to construct the rotation matrix $V$, and confirmed that the results shown below have very little dependence on the choice of 
$\tau_0$ and $\tau_D$.
The effective energy $E^\text{eff}(t)$ and its difference with respect to the two-baryon threshold $\Delta E^\text{eff}(t)$ can be obtained using the diagonal element $\tilde C_{nn}(t)$ as,
\begin{align}
&E^\text{eff}_n(t)=\ln\left[\frac{\tilde C_{nn}(t)}{\tilde C_{nn} (t+1)}\right],\label{Eq_E}\\
&\Delta E^\text{eff}_n(t)=\ln\left[\frac{\tilde R_{nn}(t)}{\tilde R_{nn}(t+1)}\right], \label{Eq_DE}
\end{align}
where $\tilde R_{nn}(t)$ is defined as $\tilde R_{nn}(t)=\frac{\tilde C_{nn}(t)}{C_B(t)C_B(t)}$.
As shown in Fig.~\ref{Fig_2B_E_p01} (left), the extracted effective energy $E^\text{eff}_{0,1}$ exhibits a monotonic decrease.
In the initial phase ($t/a\lesssim 10$), 
it falls rapidly toward the noninteracting energy level (dashed lines). 
This time range is consistent with the saturation of the single-baryon ground state shown in Fig.~\ref{Fig-M-Omg}, which indicates that 
inelastic states contamination in $\tilde C(t)$ has been significantly suppressed over this time region, leaving elastic states to dominate thereafter.
After this initial drop, the decrease slows significantly, with no visible change over an extended time range.
Fig.~\ref{Fig_2B_E_p01} (right) further reveals a clear discrepancy between $\Delta E^\text{eff}_{0,1}$ and the genuine energy levels (shown by bands) for $t/a \lesssim 25$.
At larger time slices, the data are dominated by statistical uncertainties, precluding any definitive statement.

Based on results shown above, we conclude that conventional variational analysis using plane-wave operators $\tilde O_0$ and $\tilde O_1$ is insufficient to  strongly couple to the first two lowest states, and therefore fails to achieve single-state dominance over a substantial range of Euclidean time. 

\subsubsection{Asymmetric source and sink operators}

In the above study, we used smeared $B$ operator at both source and sink sides.
Empirical evidence, however, indicates that adopting a point-like $B$ operator at the sink (while keeping the smeared source) yields a clearer statistical signal, 
which motivated the use of a point-like sink in our main results.
To ensure a fair comparison with those results and 
to better guide the interpretation of data at larger $t$ in Fig.~\ref{Fig_2B_E_p01}, we have performed additional calculations using point-like $B$ operators at the sink while keeping all other setup same as before.
The drawback of using asymmetric source and sink operators is that the resulting correlation matrix is no longer strictly Hermitian. 
Consequently, $E^\text{eff}_n(t)$ does not show monotonic falloff.
Keeping these limitations in mind, we proceed with the analysis as described above.

\begin{figure*}[hbpt]
    \centering
\includegraphics[width=8.7cm]{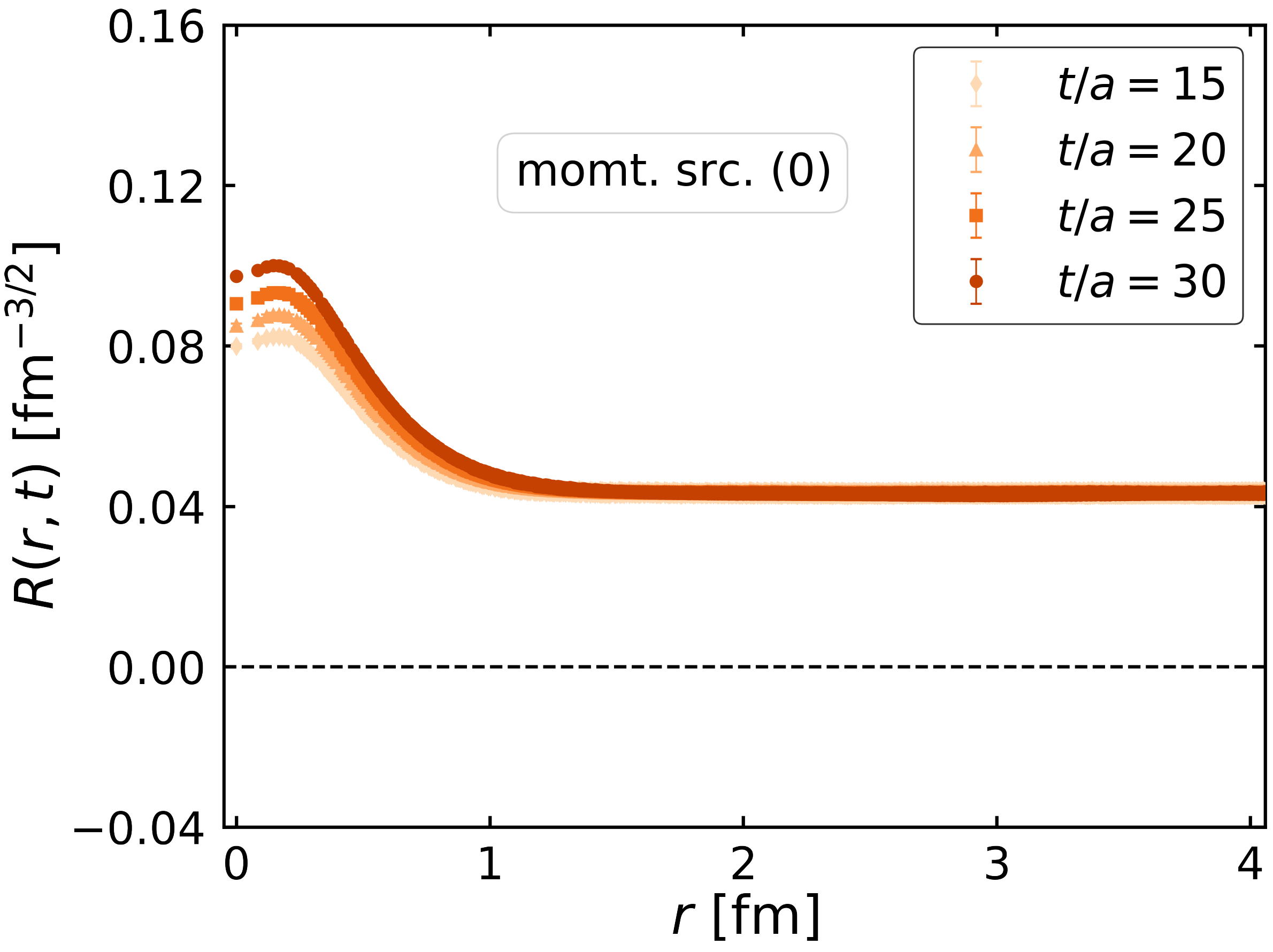} \includegraphics[width=8.7cm]{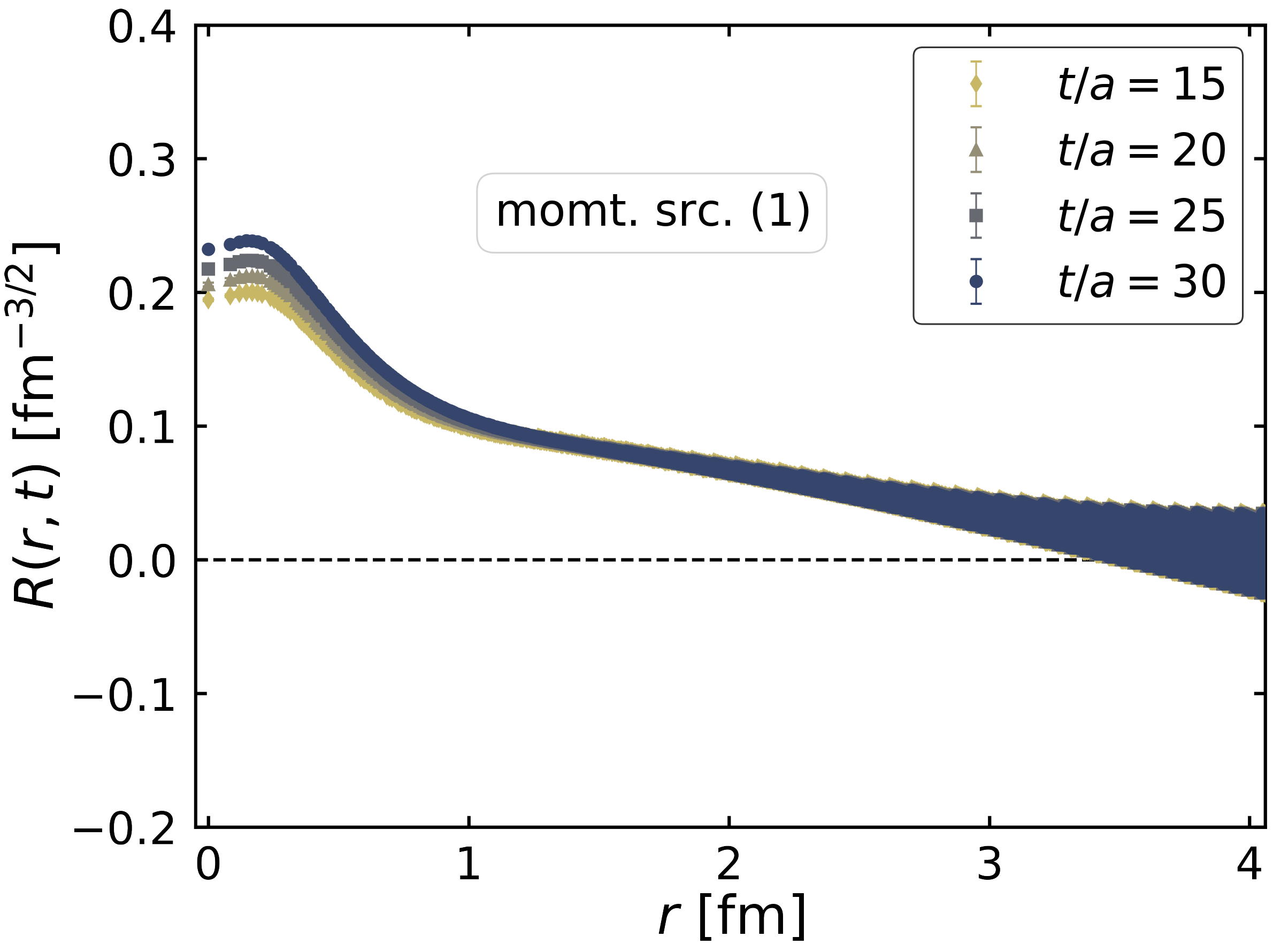}
    \caption{Same as Fig.~\ref{Fig_R_smr_p01}, but calculated using point-like $B$ operator at the sink.}
    \label{Fig_R_smr_p01_pt}
\end{figure*}

\begin{figure*}[hbpt]
    \centering
\includegraphics[width=8.7cm]{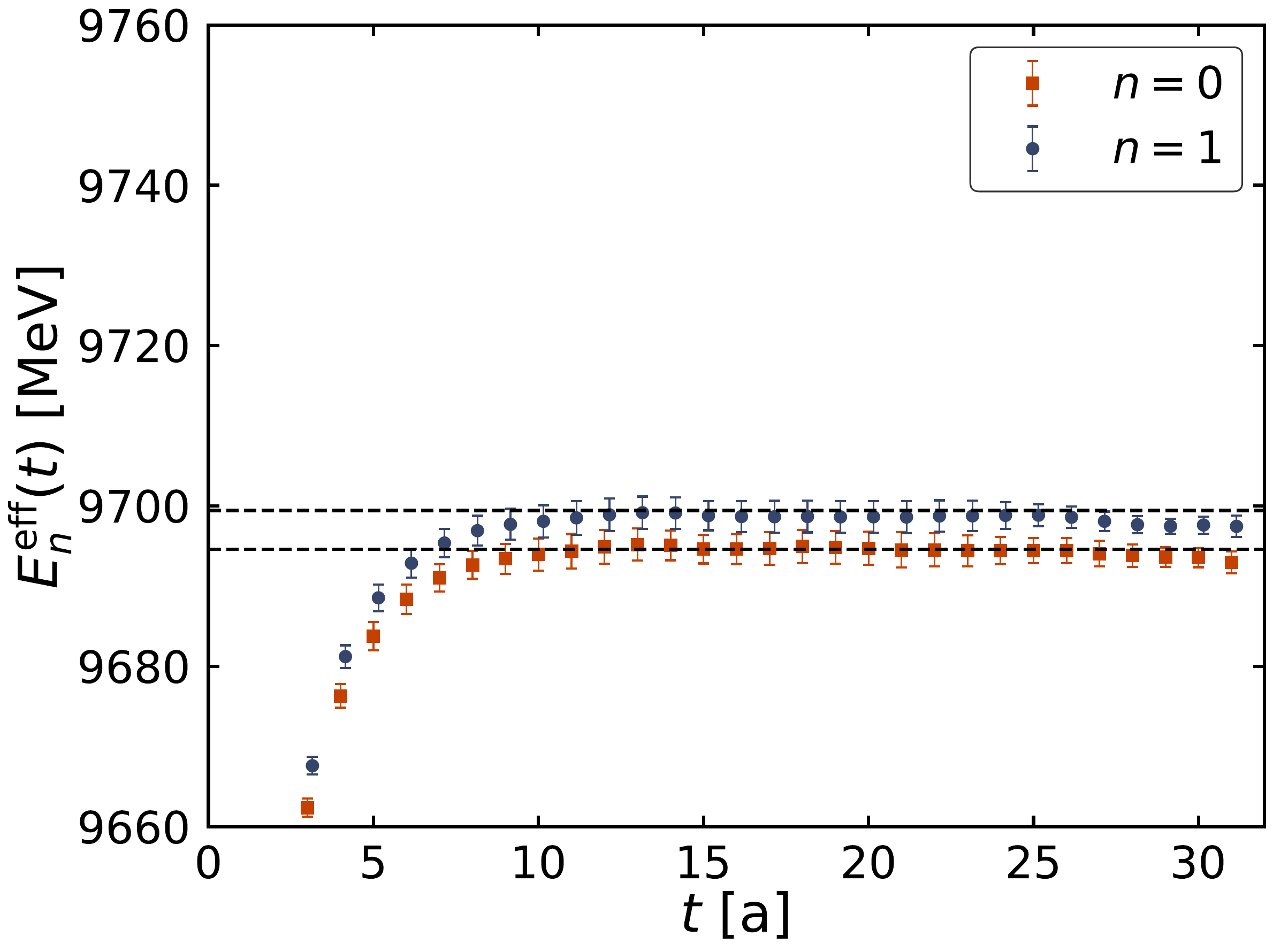}    \includegraphics[width=8.7cm]{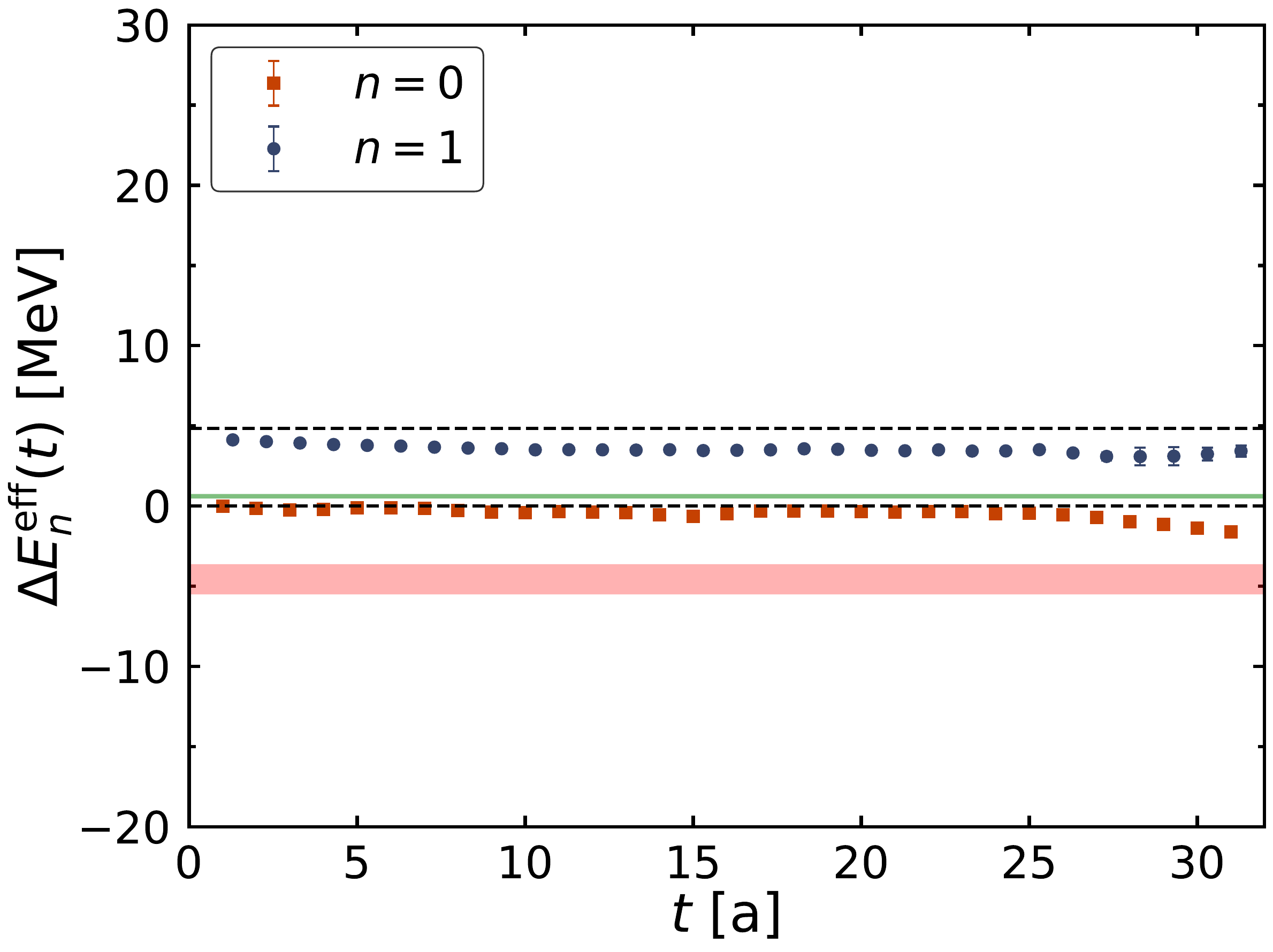}
\caption{Same as Fig.~\ref{Fig_2B_E_p01}, but calculated using point-like $B$ operator at the sink.}
    \label{Fig_2B_E_p01_pt}
\end{figure*}

The obtained $R(\vec r, t)$ correlation functions are shown in Fig.~\ref{Fig_R_smr_p01_pt},
which exhibit very similar profiles to those in Fig.~\ref{Fig_R_smr_p01}.
Visible difference only appears at short distances $r\lesssim 0.3$ fm, which is consistent with the size of single $\Omega_{ccc}$, $\sqrt{\langle r^2\rangle}\simeq 0.28$ fm as calculated in Ref.~\cite{Can2015}.
Then, we define correlation matrix as Eq.~(\ref{Eq_C_mtx}) and 
perform GEVP analysis\footnote{Such an analysis of diagonalizing  a non-Hermitian correlation matrix is analogous to the study in Ref.~\cite{Francis2019}.} as Eqs.~(\ref{Eq_GEVP}) and (\ref{Eq_rot_C}).
The resulting $E^\text{eff}_{0,1}(t)$ and $\Delta E^\text{eff}_{0,1}(t)$ are shown in Fig.~\ref{Fig_2B_E_p01_pt}, which clearly shows that the $E^\text{eff}_{0,1}(t)$ and $\Delta E^\text{eff}_{0,1}(t)$ descend very slowly and are still far from the genuine energy levels (bands) even when $t/a=31$.
These results together with results shown in Fig.~\ref{Fig_2B_E_p01} further strengthen our conclusion that 
plane-wave operators $\tilde O_0$ and $\tilde O_1$ are not enough to form optimal combinations achieving large overlap to the ground and first excited states.

To improve the overlap to targeted states, the number of operator basis used in GEVP must be increased.
Since increasing the number of source operators requires additional computation, 
we leave such additional computations to future work.
Instead, we estimate the required number of operators by enlarging the sink basis while keeping the source operators fixed,
which can be carried out based on our current results.
 

\begin{figure*}[hbpt]
    \centering
\includegraphics[width=8.7cm]{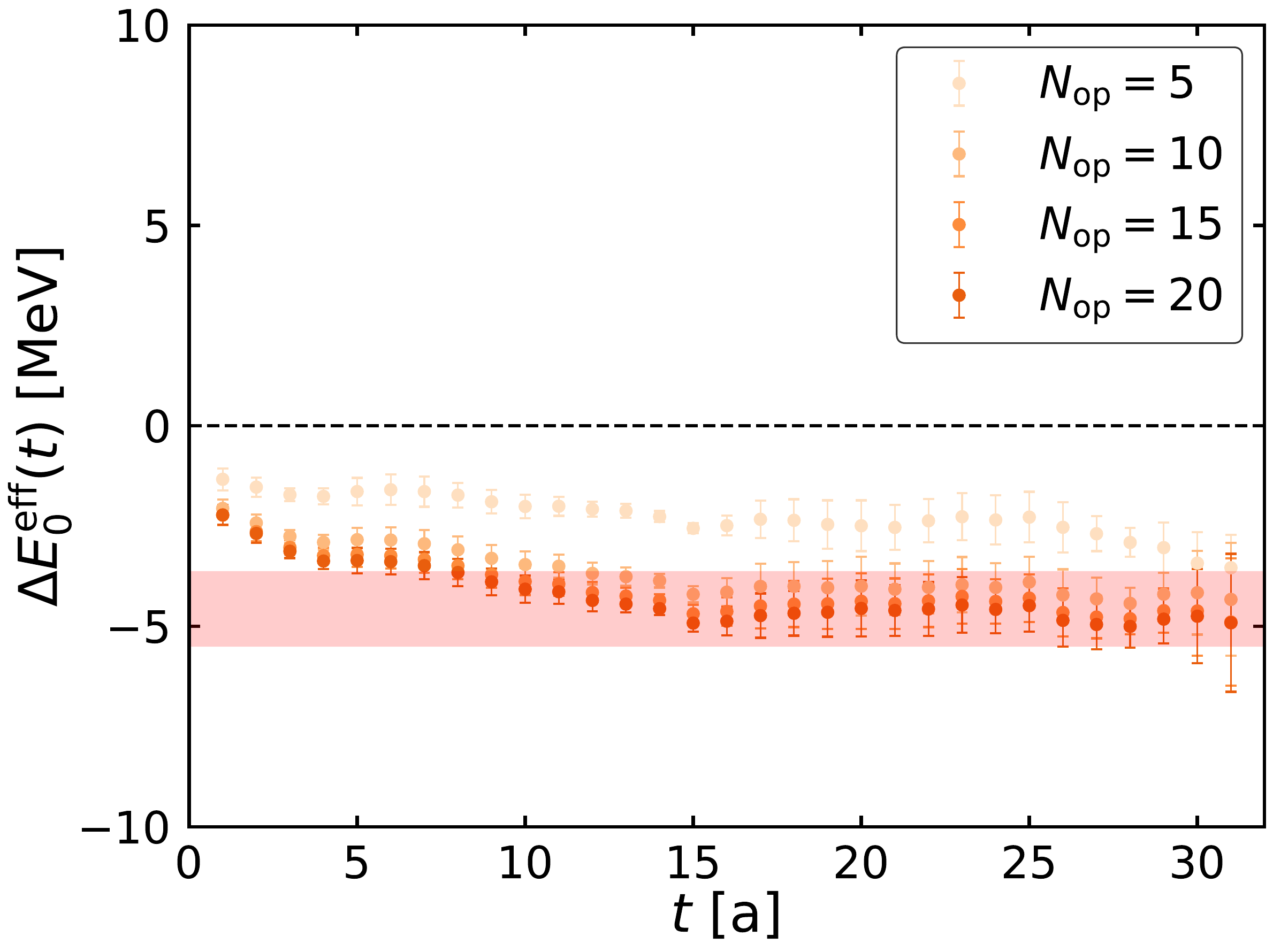}
\includegraphics[width=8.7cm]{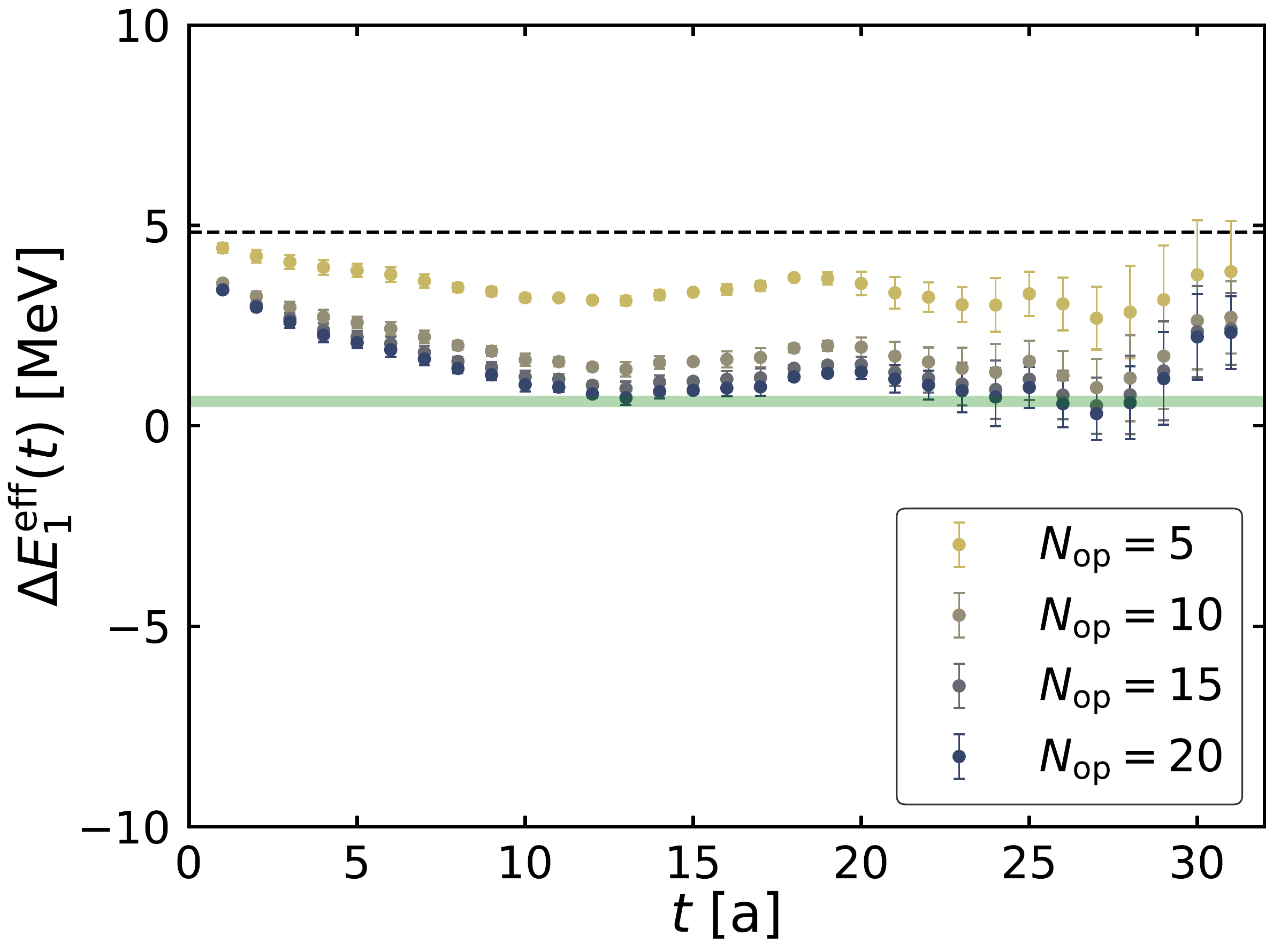}
    \caption{The effective energy $\Delta E^\text{eff}(t)$ for the ground state (left) and first excited state (right) derived from correlation functions in Eqs.~(\ref{Eq_C_eff_O0}) and (\ref{Eq_C_eff_O1})
    using the sink operator $\mathcal O^\text{eff}_{0,1}(t; N_\text{op}-1)$ in Eq.~(\ref{Eq_eff_O}), where the first $N_\text{op}$ plane-wave operators are explicitly included in the sink operator construction.
    The bands are the genuine energies determined from our wavefunction-optimized operators. }
    \label{Fig_2B_DE_vary_n}
\end{figure*}

To this end, we define effective operators $\mathcal O^\text{eff}_{0,1}(t;n)$ for the ground and first excited states, 
which are constructed as a weighted combination of the
first $n+1$ plane-wave operators in Eq.~(\ref{Eq_O_free_p}) projected to the $A_1^+$ irrep.
The weights are given by the optimized wavefunctions in momentum space $\tilde{\Psi}_{0,1}(\vec p)$, obtained from the Fourier transform of $\Psi_{0,1}(r)$ shown in Fig.~\ref{Fig-Psi_01_1d}. Thus explicitly, we have
\begin{align}\label{Eq_eff_O}
    \mathcal O^\text{eff}_{0,1}(t;n) = \sum_{|\vec p|\leq p_n} \tilde O(\vec p, t) \tilde{\Psi}^*_{0,1}(\vec p), 
\end{align}
where $p_n$ is the relative momentum corresponding to the $n+1$-th eigenvalue in Eq.~(\ref{Eq_free_H}), and the summation is restricted to all momenta within the sphere with radius $p_n$.
For instance, $\mathcal O^\text{eff}(t; n=1)$ contains 7 momenta within $p_{1}=\frac{2\pi}{L}$:
$\{(0,0,0), (0,0,\pm1),
(0,\pm1,0), (\pm1,0,0)\}\frac{2\pi}{L}$.
Using such operators at the sink, we calculate the following correlation functions,
\begin{align}
 &  \mathcal C_0(t;n)=\langle \mathcal O^\text{eff}_0(t;n)\tilde O^\dagger_0(0)\rangle,  \label{Eq_C_eff_O0}\\
& \mathcal C_1(t;n)=\langle \mathcal O^\text{eff}_1(t;n)\tilde O^\dagger_1(0)\rangle \label{Eq_C_eff_O1}. 
\end{align}
The corresponding effective energy  $\Delta E^\text{eff}_{0,1}(t)$ can be derived 
in a similar manner as Eq.~(\ref{Eq_DE}) using $\mathcal C_{0,1}(t;n)$.
As shown in Fig.~\ref{Fig_2B_DE_vary_n},
$\Delta E^\text{eff}_{0,1}(t)$ 
converge to the respective true energies at earlier $t$
as the number of operator ($N_\text{op}$) used at sink increases.
Similar results are obtained even if source operators in
Eqs.~(\ref{Eq_C_eff_O0}) and~(\ref{Eq_C_eff_O1}) are
replaced with the combinations of $\tilde O^\dagger_0$ and $\tilde O^\dagger_1$ identified in the previous GEVP analysis.
Results in Fig.~\ref{Fig_2B_DE_vary_n} also indicate that, in order to match the exceptionally large overlap of our wavefunction-optimized operator with the targeted state, 
which achieve single-state dominance as early as $t/a\simeq 15$, the number of operator basis in GEVP should satisfy $N_\text{op}\gtrsim 10$ (corresponding to 147 momentum modes or more).
Given the dense spectrum of the system, with a typical level spacing $\varepsilon=E_{n+1}-E_n\simeq5$~MeV,
a set of $10$ or more plane-wave operators is indeed needed to cover an energy region up to $50$ MeV or higher above the threshold. 

Several remarks concerning the conventional variational approach based on plane-wave operators and our wavefunction-based operator optimization are list here.
\begin{enumerate}
    \item[(i)] While plane-wave operators are a simple and widely used choice, they are not intrinsically superior to operators constructed using other inter-hadron wavefunctions, except in the weakly interacting limit.  
    From this perspective, our wavefunction-optimized operator is a natural generalization of the plane-wave approach.
    It reduces to plane-wave operators in the weak-interaction limit, while allowing the incorporation of more realistic spatial wavefunctions in nontrivial interacting systems.

    \item[(ii)] Our wavefunction-optimized operators and the variational analysis are mutually informative. On the one hand, operators constructed using realistic wavefunctions provide a superior variational basis than the plane waves for variational studies. 
    On the other hand, the optimal linear combinations of basis functions a la plane waves identified by variational studies can provide a better initial guess for constructing more realistic wavefunctions.
    
    \item[(iii)] In practice, our calculation strategy also offers important computational advantages. In conventional calculations, plane-wave operators are typically constructed using advanced techniques such as distillation. 
    For lattice volumes with $La \lesssim 4$ fm, 
    the typical number of quark propagators required is $O(100)$ (spatial only), scaling approximately as $L^{3}$.
    This rapidly becomes computationally demanding for larger volumes, such as the $La \simeq 8.1$~fm lattice used in the present work.\\
    In contrast, for an $N \times N$ correlation matrix with $N\sim O(1-10)$ in most cases, our calculation strategy requires only $N+1$ quark propagators, one computed using Eq.~(\ref{Eq-Smear-1}) and the remaining $N$ using Eq.~(\ref{Eq-Smear-2}).
    Importantly, this cost does not scale with $L$.  
    This feature is crucial for enabling the variational analysis presented in this subsection on large-volume lattices.
\end{enumerate}

Using the $R(\vec r, t)$ correlation functions in Figs.~\ref{Fig_R_smr_p01} and \ref{Fig_R_smr_p01_pt}, we also investigated the sink-operator dependence of HAL QCD potential in Appendix~\ref{Sec-sink-opr}.

\subsection{Analysis of scattering properties}

\subsubsection{Results from LO potentials}
\begin{figure*}[htbp]
  \centering
  \includegraphics[width=8.7cm]{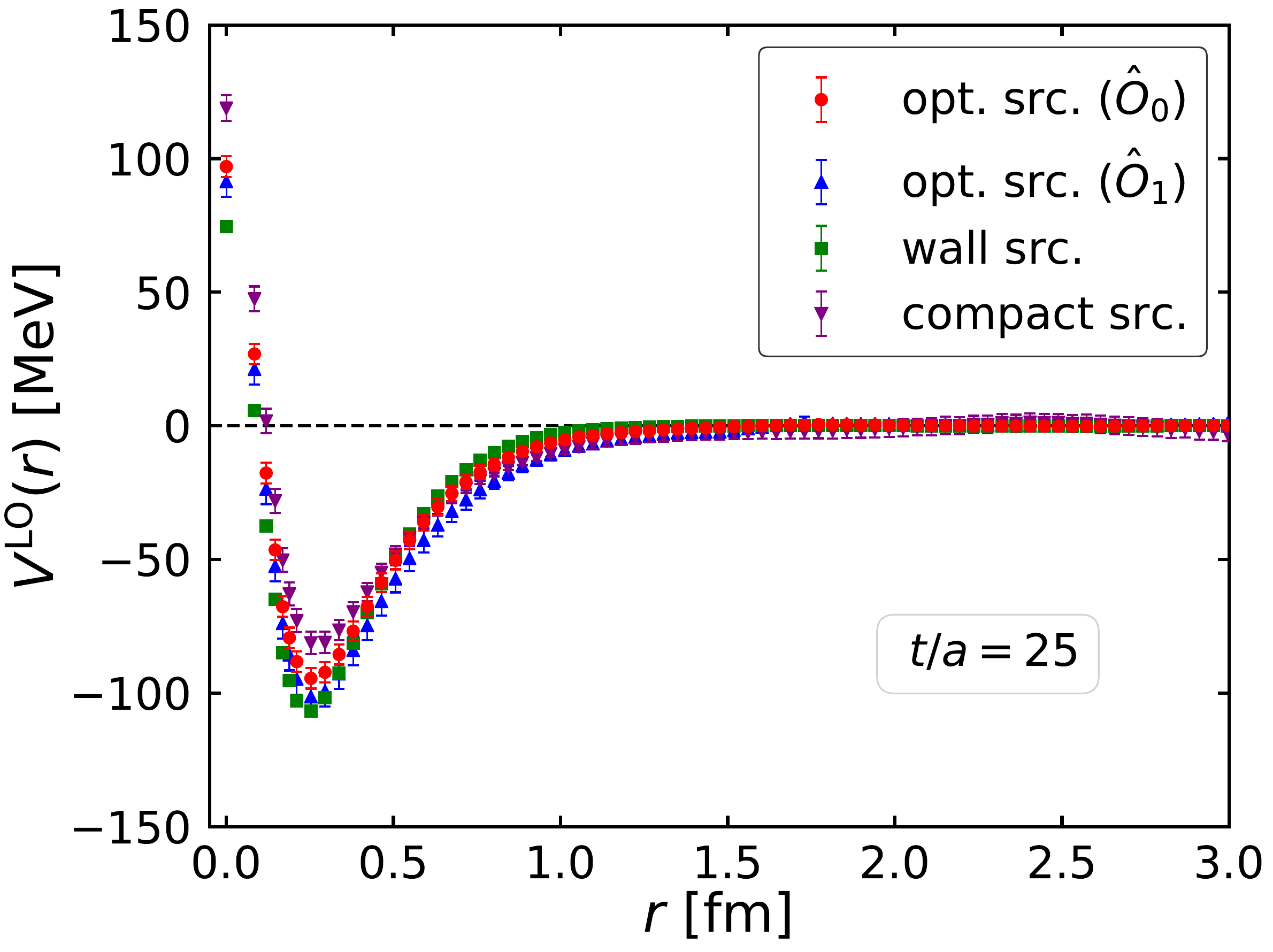}
  \includegraphics[width=8.7cm]{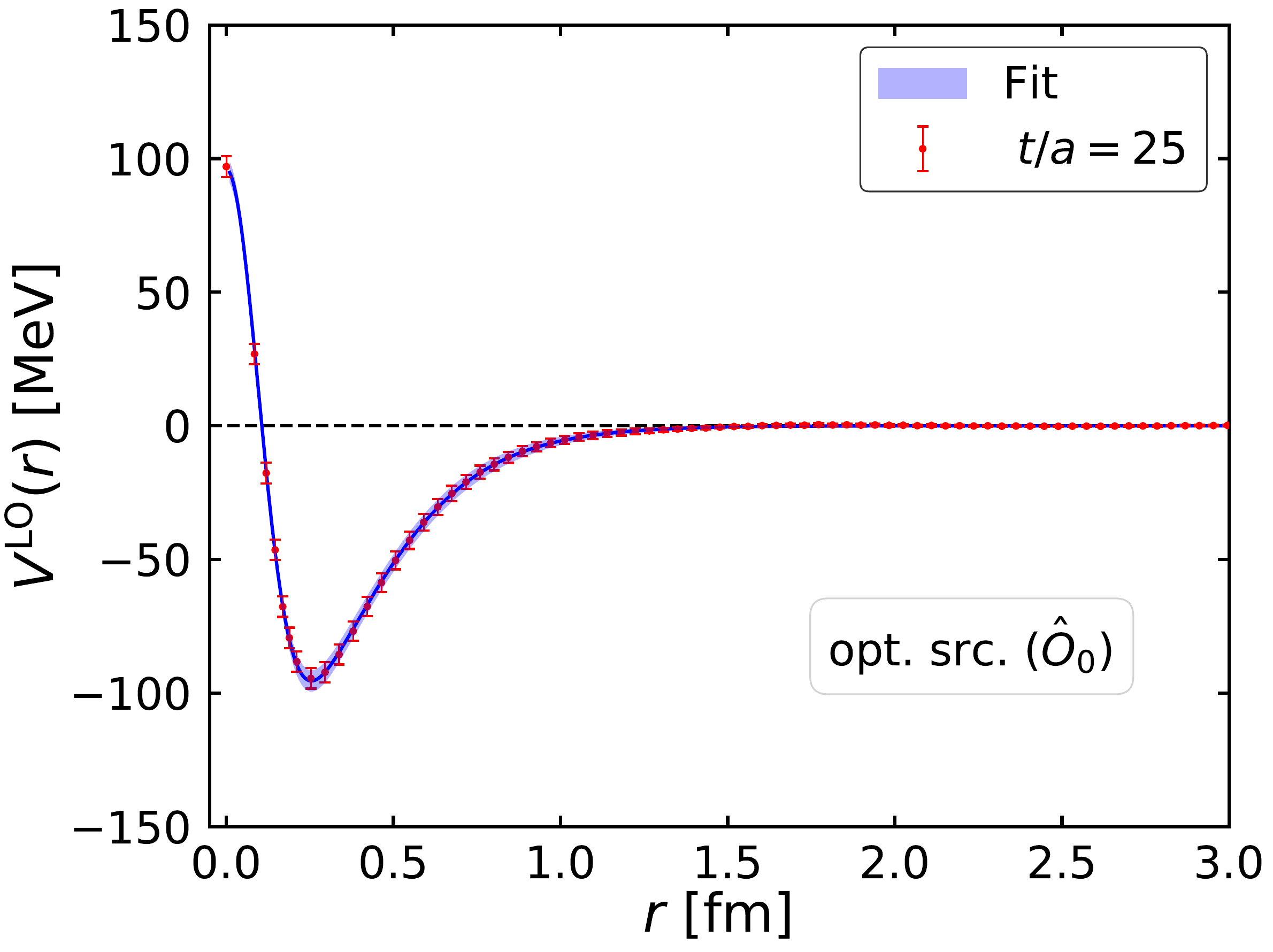}
  \caption{A comparison of the LO potentials calculated using each source at $t/a=25$ (left).
  The three-range Gaussian fit to the potential from the optimized source for the ground state $O_0$ (right).
  } \label{Fig-VLO-all_srcs}
\end{figure*}

\begin{figure}[htbp]
  \centering
  \includegraphics[width=8.7cm]{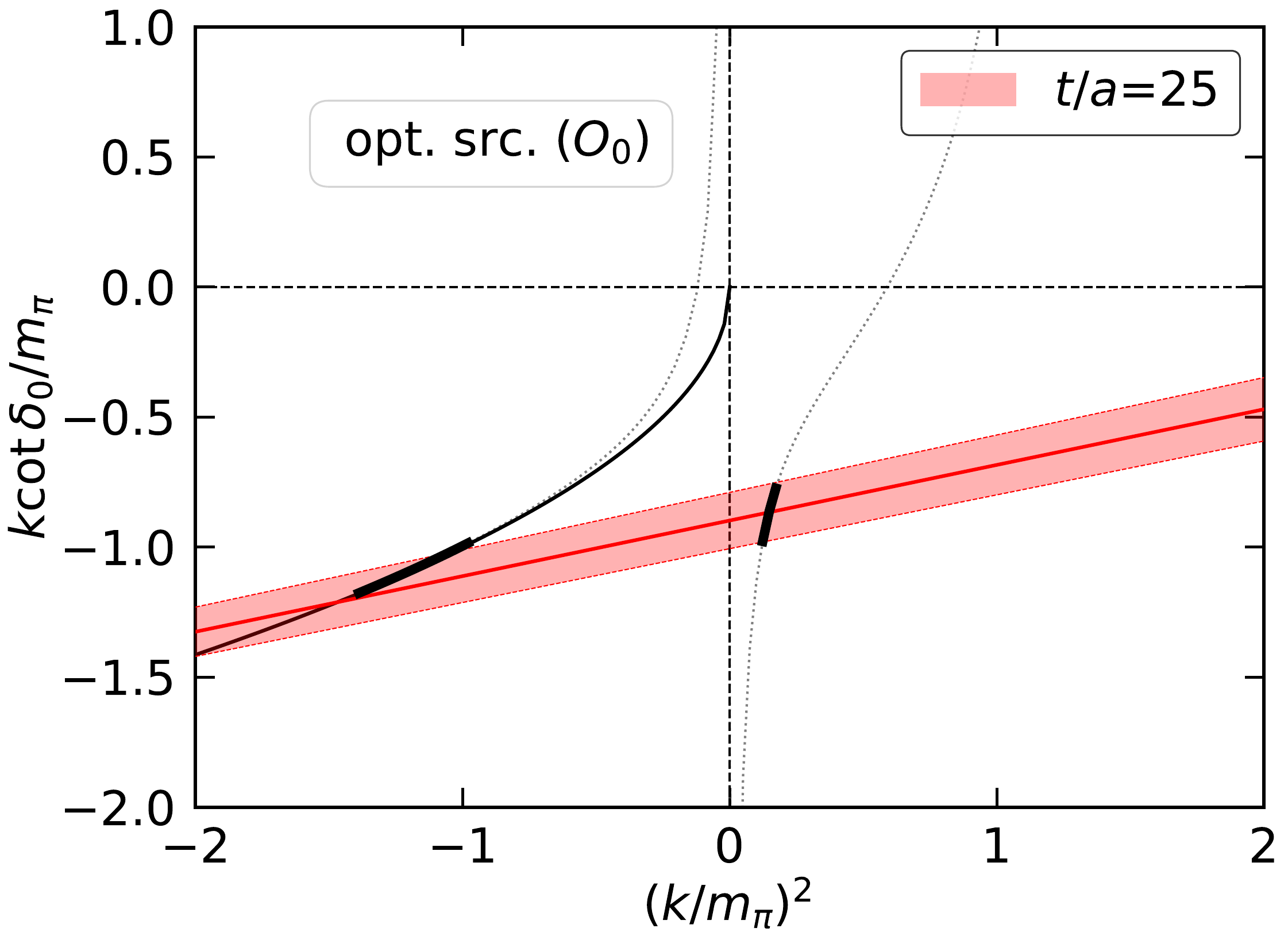}
  \caption{ 
  The scattering phase shifts calculated using the LO potential from $R_0(\vec r,t)$ at $t/a=25$.
  Two thick black line segments represent the scattering phase shifts converted from finite volume spectral in Fig.~\ref{Fig-Eeff}.
  The black solid line $-\sqrt{-(k/m_\pi)^2}$ is the bound state condition.
  } \label{Fig-phaseshift-opt0}
\end{figure}

Shown in Fig.~\ref{Fig-VLO-all_srcs} is a comparison of the local potentials extracted from $R_0(\vec r, t)$, $R_{1}(\vec r, t)$ and those calculated using the wall and compact sources at Euclidean time $t/a=25$.
These potentials show almost identical behaviors except only a slight deviation at short distances for the compact source from others. 
To determine scattering phase shifts, we fit the potentials by three-range Gaussian functions (the right panel in Fig. \ref{Fig-VLO-all_srcs} for the one from $O_0$),
\begin{align}
    V_\text{fit}(r)=\sum_{i=1,2,3} a_i e^{-(r/b_i)^2},
\end{align}
and then solve the Schr\"odinger equation in infinite volume.
The resulting scattering phase shifts calculated from the $O_0$ source are displayed in Fig.~\ref{Fig-phaseshift-opt0}.
We confirm that potentials from each source at different $t$ give consistent results, see Appendix~\ref{Sec-phaseshift-multi}.
For comparison, Fig.~\ref{Fig-phaseshift-opt0} also includes two phase shifts (thick black line segments) converted from the finite-volume spectra in Fig.~\ref{Fig-Eeff} using the finite-volume formula. 
As theoretically expected, we observe an excellent agreement between the phase shifts obtained from the potentials and those derived from the finite-volume spectra. 
These results demonstrate that the LO potentials provide an accurate description to low-energy scattering processes.

%==========================
%==========================

%==========================
%==========================

\subsubsection{Results from N$^2$LO potentials}

To see the effect of high-order terms in the derivative expansion to observables, we derive the N$^2$LO potentials according to Sec.~\ref{Sec-Dual-Func-A} by using two hadronic correlation functions $R_0(\vec r, t)$ and $R_{\text{wall}}(\vec r, t)$. 
This is because the former is dominated by the ground state, while the latter includes scattering states moderately above the two-particle threshold, which are absent in $R_1(\vec r, t)$. Consequently, the resulting N$^2$LO potentials are expected to accurately describe scattering both near and moderately above threshold.

\begin{figure*}[htbp]
  \centering
  \includegraphics[width=8.7cm]{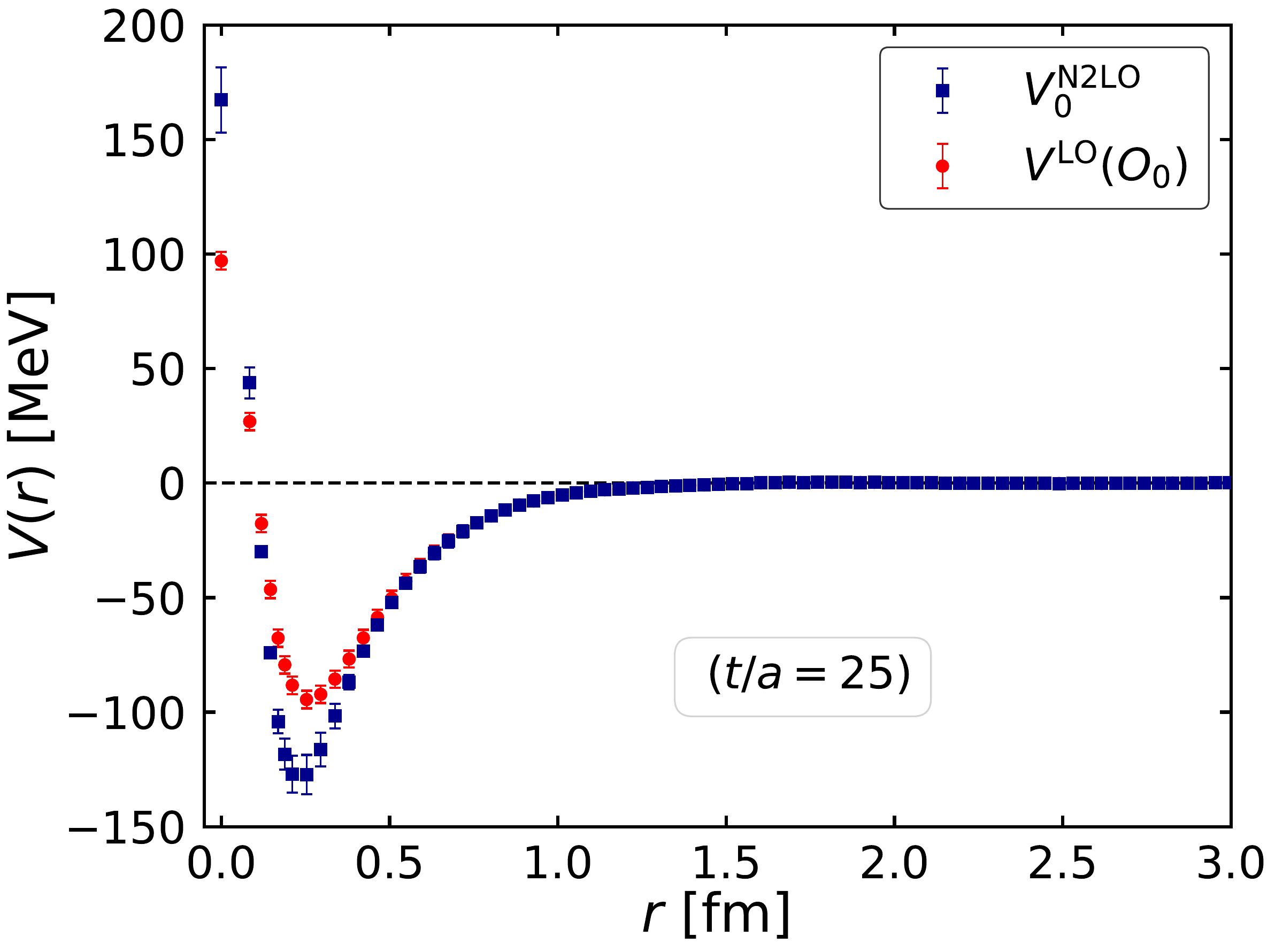}
  \includegraphics[width=8.7cm]{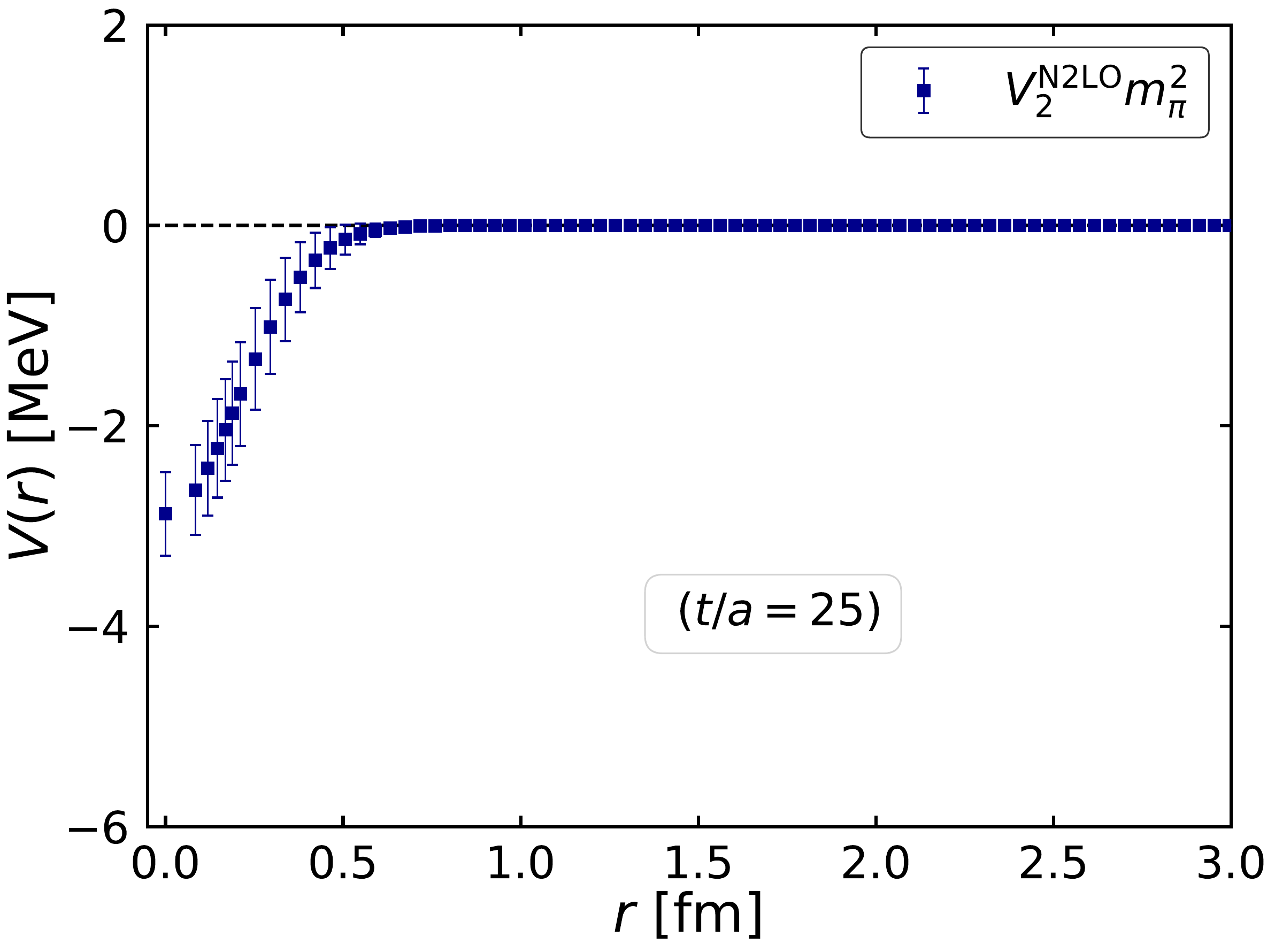}
  \caption{ The N$^2$LO potentials obtained using $R_0(\vec r, t)$ and $R_\text{wall}(\vec r, t)$ at $t/a=25$.
  $V_0^{\text{N}^2\text{LO}}$ is shown together with $V^{\text{LO}}$ computed from $R_0(\vec r, t)$ for comparison (left).
  $V_2^{\text{N}^2\text{LO}}$ is multiplied by $m^2_\pi$ (right).
  } \label{Fig-VN2LO-opt0-wall}
\end{figure*}

\begin{figure}[htbp]
  \centering
  \includegraphics[width=8.7cm]{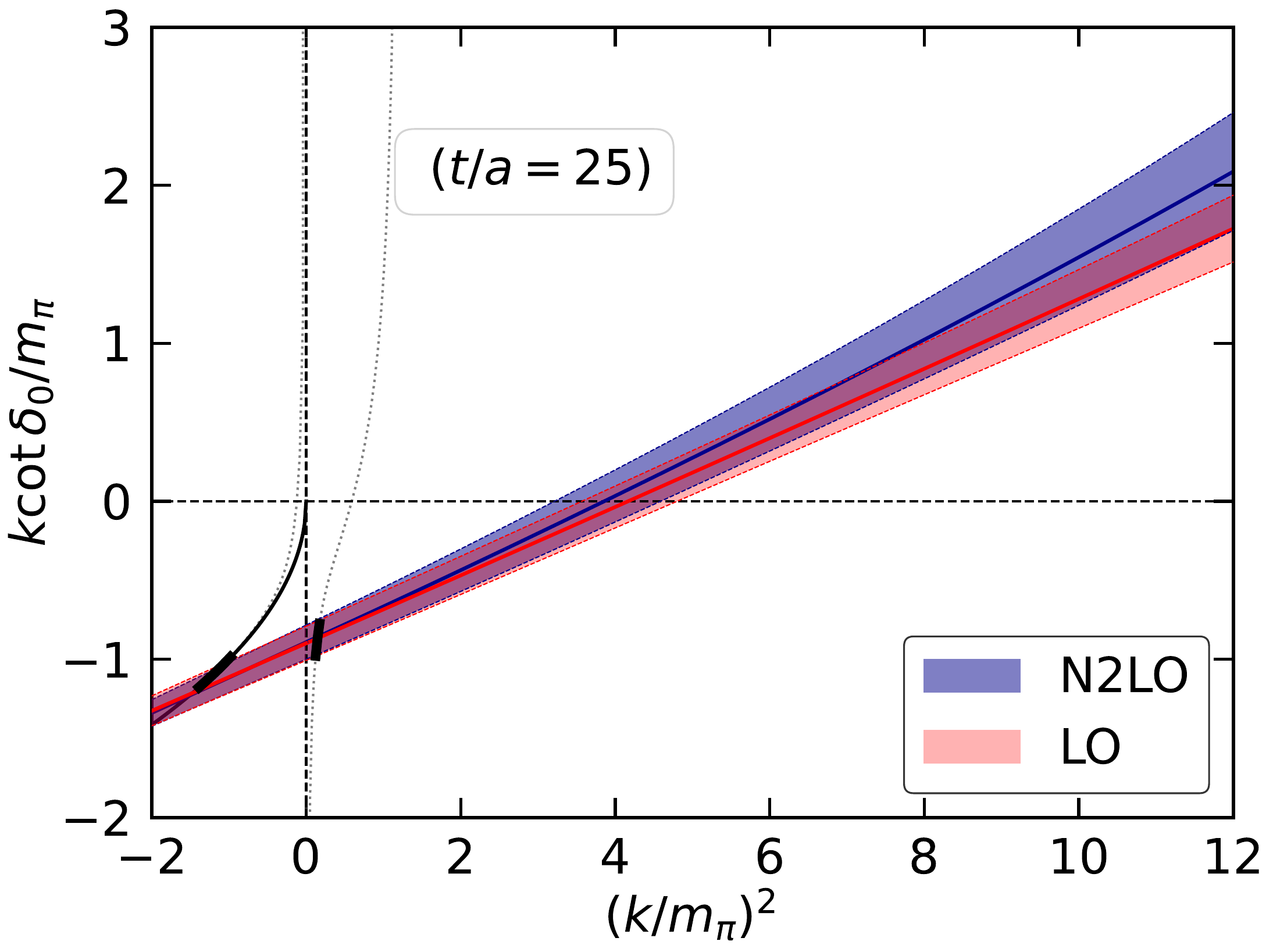}
  \caption{ 
  The scattering phase shifts calculated using the N$^2$LO potentials in Fig.~\ref{Fig-VN2LO-opt0-wall}.
  Those computed using the LO potential are also shown for comparison.
  } \label{Fig-phaseshift-N2LO}
\end{figure}

The obtained  N$^2$LO potentials
\footnote{To avoid numerical instabilities caused by the nearly vanishing $D_2$ appearing in the denominator in Eq.~(\ref{Eq-N2L0-V2}), $V^{\text{N}^2\text{LO}}_2 (\vec r)$ is determined by a fit of $V^{\text{N}^2\text{LO}}_2 (\vec r)D_2$ to $V^\text{LO}_J(\vec r)-V^\text{LO}_{J'}(\vec r)$ with the fit function being Gaussian type.}
$V^{\text{N}^2\text{LO}}_0 (\vec r)$ and $V^{\text{N}^2\text{LO}}_2 (\vec r)$
are shown in Fig.~\ref{Fig-VN2LO-opt0-wall} alongside  the LO potential $V^\text{LO}(\vec r)$ derived from $R_0(\vec r, t)$ for comparison .
The potentials $V^{\text{N}^2\text{LO}}_0 (\vec r)$ and $V^\text{LO}(\vec r)$ are identical at long distances,
with visible differences only emerging at short distances ($r < 0.4$ fm). 
The $V^{\text{N}^2\text{LO}}_2 (\vec r)$ potential is highly localized and exhibits a small magnitude.
Using these potentials, we compute the scattering phase shifts, as shown in Fig.~\ref{Fig-phaseshift-N2LO}, and the scattering phase shifts obtained with $V^\text{LO}(\vec r)$ are also included for comparison. 
We observe that the N$^2$LO corrections have an almost negligible effect on low-energy scattering, with only minor differences appearing at higher energies.

%==========================
%==========================

%==========================
%==========================

%=================================================

\section{Summary and Discussion}\label{Sec-summary}

In this paper,
we have presented a systematic approach for constructing optimized two-hadron operators in lattice QCD
by incorporating inter-hadron spatial wavefunctions.
We argued that 
realistic spatial wavefunctions
%the distinct spatial profiles of two-hadron states (represented in coordinate space using given local hadron operators)
allow for highly efficient state preparation/identification in lattice QCD calculations.
To implement the optimized operators at the source, 
we have proposed a novel quark smearing technique,
which assigns an inter-hadron wavefunction factor together with an independent $Z_3$ noise vector into quark.
By doing so, the inter-hadron spatial wavefunction can be properly included without using computational expensive all-to-all quark propagators.

As an explicit demonstration, we applied the method to $\Omega_{ccc}\Omega_{ccc}$ system in $^1S_0$ channel.
We first determined the inter-hadron spatial wavefunctions using the HAL QCD potential  
derived from the wall source.
With these wavefunctions, we constructed the optimized two-hadron operators for the ground state and the first excited state 
at the source.
Correlation functions calculated using the optimized two-hadron operators show stable spatial profiles against a long range of Euclidean time,
leading to clear identification of spectra with an energy gap as narrow as $\sim 5$ MeV, although both spectra are around $2m_{\Omega_{ccc}}\simeq9700$ MeV.
These operators, when applied at the sink of the correlation functions from the wall/compact source,
are also found to be able to identify respective states out of a tower of states. 
As a comparison, we performed a variational study using the plane-wave operators,
and found that the wavefunction-optimized operators outperform combinations of limited plane-wave operators.

Our method for operator optimization is sufficiently flexible to incorporate wavefunctions obtained from some model or EFT calculations instead of ones from the HAL QCD potential.  
These approximated wavefunctions, however, may not be accurate enough to generate/identify certain states, but they in general 
form better basis functions than simple plane wavefunctions for the variational method.
In addition, the novel quark smearing technique enables an independent treatment of single-hadron and two-hadron systems. 
It can be adapted to various localized smearing schemes (such as distillation) for single hadrons, and can accommodate different inter-hadron wavefunctions, including those with higher partial waves, for a wide range of two-hadron systems.
%Systems involving quark-annihilation diagrams may require additional considerations; nevertheless, the theoretical framework for wavefunction-based operators still exists.

Operator optimization plays a pivotal role in lattice QCD calculations. The method presented in this paper provides a new paradigm for operator optimization by incorporating spatial information, and it can be applied to both spectral and matrix-element calculations. 
In particular, it enables systematic studies of light nucleon–nucleon systems at the physical point with reasonable computational cost. Such efforts are currently underway.

%==============================================================================

%=======================================================================================vspa
%\clearpage

\begin{acknowledgments}
We thank members of the HAL QCD Collaboration for stimulating discussions. This work was partially supported by RIKEN Incentive Research Project (``Unveiling pion-exchange interactions between hadrons from first-principles lattice QCD"), RIKEN TRIP initiative, the JSPS (Grant Nos. JP22H00129, JP22H04917, JP23H05439, and JP25K17384), and the Japan Science and Technology Agency (JST) as part of Adopting Sustainable Partnerships for Innovative Research Ecosystem (ASPIRE Grant No. JPMJAP2318). The lattice QCD calculations were carried out by using Fugaku supercomputers at RIKEN through HPCI System Research Project (hp240157 and hp240213),  ``Program for Promoting Researches on the Supercomputer Fugaku'' (Simulation for basic science: from fundamental laws of particles to creation of nuclei) and (Simulation for basic science: approaching the new quantum era)
(Grants No. JPMXP1020200105, JPMXP1020230411).
\end{acknowledgments}

%=======================================================================================vspa
\appendix

%==========================
\begin{figure}[htbp]
  \centering
  \includegraphics[width=8.7cm]{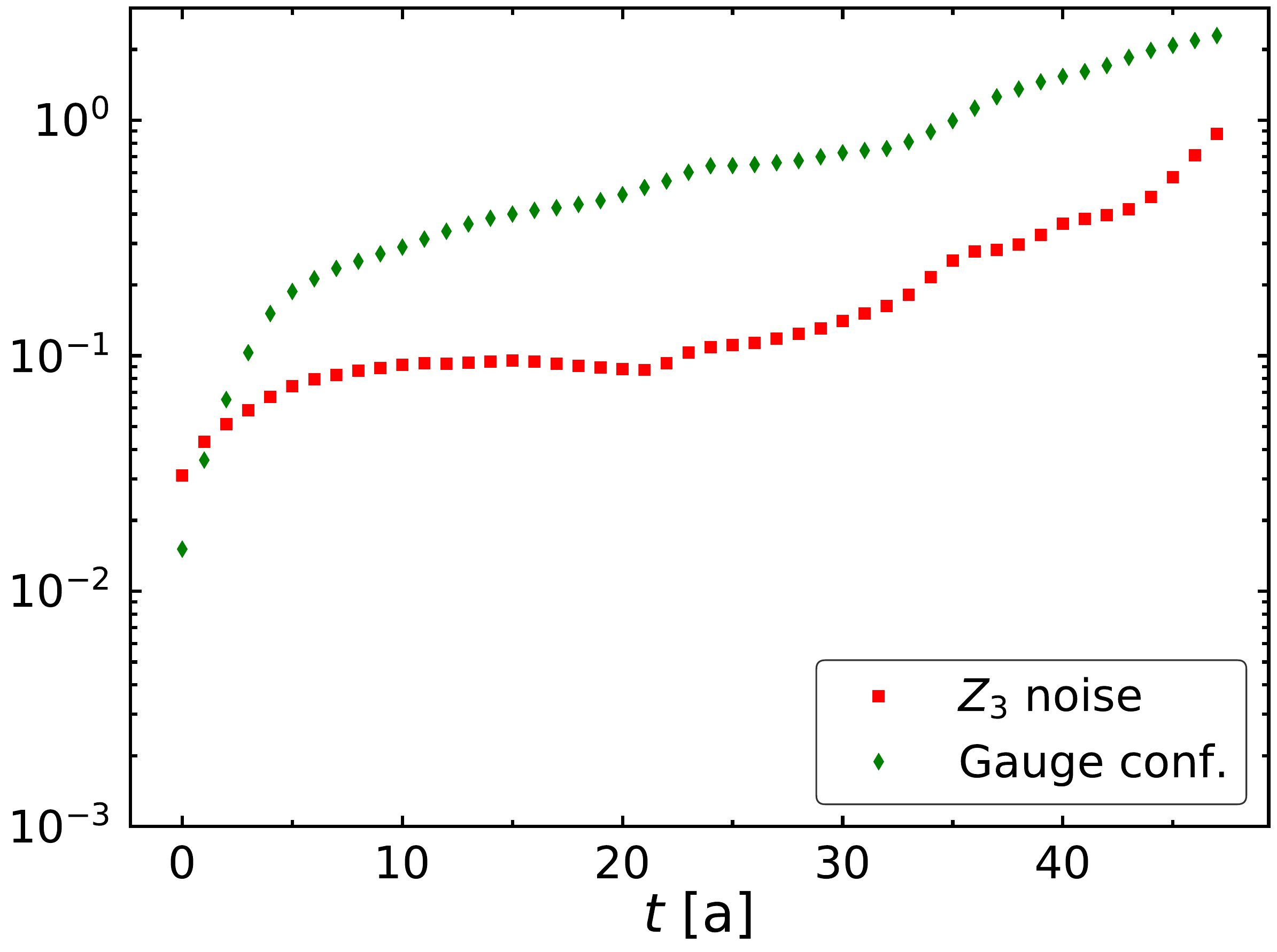}
  \caption{
A comparison on the relative error (defined as the ratio between standard deviation to the mean value $\frac{\sigma(x)}{E(x)}$) of the  $\Omega_{ccc}$ two-point function from $Z_3$ noise and that from gauge configurations.
The sub lattice is chosen with $l=8~[a]$ in Eq.~(\ref{Eq-Sub-Lattice}).
} \label{Fig-variance-2pt}
\end{figure}
%==========================

\section{Variance estimation}\label{Sec-err}
Here we perform a variance estimation, which is crucial for designing optimal calculation strategies in practice.
As shown in Eq.~(\ref{Eq-Tot-Stat-Err}), the total statistical fluctuations receive contributions from  both gauge configurations as well as the crossing terms associated with $Z_3$ noise.
When implementing the novel quark smearing technique proposed in Sec.~\ref{Sec-Smear-A}, we need to ensure that the introduction of $Z_3$
noise does not significantly increase the total fluctuations of the quantities of interest.

To quantify the variance from both sources, we employ two complementary approaches. For estimating the variance from the $Z_3$ noise, we perform multiple measurements under the same gauge configuration while randomly varying $Z_3$ noise.
For estimating the gauge configuration variance, we conduct measurements across different gauge configurations without introducing $Z_3$ noise.
This decomposition allows us to systematically evaluate and control both sources of statistical uncertainty in our calculations.

Shown in Fig.~\ref{Fig-variance-2pt} 
is a comparison on the relative error (defined as the ratio between the standard deviation to the mean value $\frac{\sigma(x)}{E(x)}$) of the  $\Omega_{ccc}$ two-point function from the $Z_3$ noise and that from gauge configurations.
The sub lattice is chosen with $l=8~[a]$ in Eq.~(\ref{Eq-Sub-Lattice}).
We observe that fluctuations from the $Z_3$ noise are smaller than that from gauge configurations by a factor of $3\sim10$ for $t/a\geq10$.
A parallel analysis of the $R(\vec r, t)$ correlation function shows similar results, with
the $Z_3$ noise-induced fluctuations remaining comparably smaller relative to gauge variations.

\section{The calculation of correlation functions}\label{Sec-wick-cont}
The calculation of the correlation function in Eq.~(\ref{Eq-Rr}) boils down to 
performing the Wick contraction and color-spinor indices summation with two $\Omega_{ccc}$ baryons located at distinct spatial positions at the source.
To this end, we consider the following expression,
\begin{align}
    & \langle  \Omega_{i\alpha}(\vec {p})\Omega_{j\beta}(-\vec {p}) \bar{Q}_{\xi'_1}(y)\bar{Q}_{\xi'_2}(y)\bar{Q}_{\xi'_3}(y) \bar{Q}_{\xi'_4}(w)\bar{Q}_{\xi'_5}(w)\bar{Q}_{\xi'_6}(w) \rangle_F \nonumber \\
    &~~\times \mathcal{F}_{i'\alpha' j' \beta'}(\xi'_1\cdots\xi'_6).  \label{Eq_wick_full}
\end{align}
Here, $ \Omega_{i\alpha}(\vec {p})$ is the $\Omega_{ccc}$ sink operator projected to momentum $\vec p$, which is used to perform $\sum\limits_{\vec x\in\Lambda}\Omega(\vec x +\vec r_\text{snk}) \Omega(\vec x) $ summation at the sink in a convolution framework.
$\xi_i=(c_i,\alpha_i)$ is an abbreviation of the color index $c_i=0,1,2$ and the spinor index $\alpha=0,1$ (only consider two components for simplicity), $\xi_i = c_i +3\alpha_i=0,\cdots,5$.
A summation is implied for repeated indices, namely, $\xi'_1,\cdots, \xi'_6$.
Two hadrons are located at $y$ and $w$ at source.
$\langle\cdots\rangle_F$ means the Wick contraction, namely permutations among quarks with the same flavor. Here, we have 6 $Q$ quarks ($Q$ denotes for heavy quark, which is the charm quark in our current context), meaning the number of permutation is $6!=720$.
The coefficient tensor $\mathcal{F}$ is defined as,
\begin{align}
    \mathcal{F}_{i'\alpha' j' \beta'}[\xi'_1\cdots\xi'_6] = & \epsilon_{c'_1c'_2c'_3} [C\gamma^*_{i'}]_{\alpha'_1\alpha'_2}  \delta_{\alpha'_3\alpha'} 
    \nonumber\\ 
    &\times \epsilon_{c'_4c'_5c'_6} [C\gamma^*_{j'}]_{\alpha'_4\alpha'_5}  \delta_{\alpha'_6\beta'}. \label{Eq_Coeff}
\end{align}
For given $i'\alpha' j' \beta'$, the 6-rank tensor $\mathcal{F}$  has $6^6=46656$ entities,
 out of which only $(6\times2)^2=144$ entities are non-vanishing given the sparse nature of $\epsilon$ and $C\gamma_i$.

We categorize all 720 permutations in Eq.~(\ref{Eq_wick_full}) into two subsets, dubbed as  $\mathcal{D}$ (Direct) and $\mathcal{C}$ (Cross). The former contains $\mathcal{D}_1$ and $\mathcal{D}_2$ terms, and the latter consists of $\mathcal{C}_1, \cdots \mathcal{C}_{18}$ terms. Each term itself is a combination of two fully-permuted baryon blocks, and therefore has $6\times6=36$ permutations. 
For instance,
\begin{align}
\mathcal{D}_1 = & -\langle \Omega_{i\alpha}(\vec {p})\bar{Q}_{\xi'_1}(y)\bar{Q}_{\xi'_2}(y)\bar{Q}_{\xi'_3}(y) \rangle_F  \nonumber\\
& \times  \langle\Omega_{j\beta}(-\vec {p}) \bar{Q}_{\xi'_4}(w)\bar{Q}_{\xi'_5}(w)\bar{Q}_{\xi'_6}(w) \rangle_F  \times\mathcal{F} \\
\mathcal{D}_2 = & +\langle \Omega_{i\alpha}(\vec {p}) \bar{Q}_{\xi'_4}(w)\bar{Q}_{\xi'_5}(w)\bar{Q}_{\xi'_6}(w) \rangle_F  \nonumber\\
& \times  \langle\Omega_{j\beta}(-\vec {p}) \bar{Q}_{\xi'_1}(y)\bar{Q}_{\xi'_2}(y)\bar{Q}_{\xi'_3}(y) \rangle_F \times\mathcal{F}, \\
\mathcal{C}_1 = & +\langle \Omega_{i\alpha}(\vec {p})\bar{Q}_{\xi'_1}(y)\bar{Q}_{\xi'_2}(y)\bar{Q}_{\xi'_4}(w) \rangle_F \nonumber\\ 
& \times  \langle\Omega_{j\beta}(-\vec {p}) \bar{Q}_{\xi'_3}(y)\bar{Q}_{\xi'_5}(w)\bar{Q}_{\xi'_6}(w) \rangle_F \times\mathcal{F}, \\
& \qquad \cdots, \nonumber\\
 \mathcal{C}_{9} = & +\langle \Omega_{i\alpha}(\vec {p})\bar{Q}_{\xi'_2}(y)\bar{Q}_{\xi'_3}(y)\bar{Q}_{\xi'_6}(w) \rangle_F \nonumber \\
& \times  \langle\Omega_{j\beta}(-\vec {p}) \bar{Q}_{\xi'_1}(y)\bar{Q}_{\xi'_4}(w)\bar{Q}_{\xi'_5}(w) \rangle_F  \times\mathcal{F},\\
\mathcal{C}_{10} = &-\langle \Omega_{i\alpha}(\vec {p})\bar{Q}_{\xi'_1}(y)\bar{Q}_{\xi'_4}(w)\bar{Q}_{\xi'_5}(w) \rangle_F \nonumber\\ 
& \times  \langle\Omega_{j\beta}(-\vec {p}) \bar{Q}_{\xi'_2}(y)\bar{Q}_{\xi'_3}(y)\bar{Q}_{\xi'_6}(w) \rangle_F \times\mathcal{F}, \\
& \qquad \cdots, \nonumber \\
\mathcal{C}_{18} = & -\langle \Omega_{i\alpha}(\vec {p})\bar{Q}_{\xi'_3}(y)\bar{Q}_{\xi'_5}(w)\bar{Q}_{\xi'_6}(w) \rangle_F \nonumber \\
& \times  \langle\Omega_{j\beta}(-\vec {p}) \bar{Q}_{\xi'_1}(y)\bar{Q}_{\xi'_2}(y)\bar{Q}_{\xi'_4}(w) \rangle_F\times\mathcal{F}. 
\end{align}
For notational simplicity, the indices of $\mathcal{F}$ defined in Eq.~(\ref{Eq_Coeff}) and those of $\mathcal{D}$ and $\mathcal C$ are omitted.
To calculate these terms, we need to prepare the following 4 fully-permuted hadron blocks, 
\begin{align}
   b_1=  & \langle \Omega_{\xi}(\vec {p})\bar{Q}_{\xi'_i}(y)\bar{Q}_{\xi'_j}(y)\bar{Q}_{\xi'_k}(y) \rangle_F , \label{Eq-hadron-block-1}\\
   b_2=  & \langle \Omega_{\xi}(\vec {p})\bar{Q}_{\xi'_i}(w)\bar{Q}_{\xi'_j}(w)\bar{Q}_{\xi'_k}(w) \rangle_F , \\
   b_3=  & \langle \Omega_{\xi}(\vec {p})\bar{Q}_{\xi'_i}(y)\bar{Q}_{\xi'_j}(y)\bar{Q}_{\xi'_k}(w) \rangle_F , \\
   b_4=  & \langle \Omega_{\xi}(\vec {p})\bar{Q}_{\xi'_i}(y)\bar{Q}_{\xi'_j}(w)\bar{Q}_{\xi'_k}(w) \rangle_F , \label{Eq-hadron-block-4}
\end{align}
for all $\vec p\in\tilde{\Lambda}$, $\xi=0, \cdots, 5$. Each hadron block has three open color-spinor indices $\xi'_{i,j,k}$ at the source, which can properly be assigned as $\xi'_{1,\cdots6}$ according to each of 20 terms.

Finally, we perform the color-spinor indices summation with the coefficient tensor $\mathcal{F}_{i'\alpha'j'\beta'}[\xi'_1,\cdots,\xi'_6]$ defined in Eq.~(\ref{Eq_Coeff}).
\begin{itemize}
    \item $\mathcal{D}_1$ term
    \begin{align}
         b_1[\xi'_1,\xi'_2,\xi'_3]\times b_2 [\xi'_4,\xi'_5,\xi'_6] \times \mathcal{F}_{D_1}[\xi'_1,\cdots,\xi'_6], \nonumber
    \end{align}
    with $\mathcal{F}_{D_1} = - \mathcal{F}_{i'\alpha'j'\beta'}[\xi'_1,\cdots,\xi'_6]$.
    
    \item $\mathcal{D}_2$ term
    \begin{align}
         b_2[\xi'_1,\xi'_2,\xi'_3]\times b_1 [\xi'_4,\xi'_5,\xi'_6] \times  \mathcal{F}_{D_2}[\xi'_1,\cdots,\xi'_6], \nonumber
    \end{align}
    with $\mathcal{F}_{D_2} = +\mathcal{F}_{i'\alpha'j'\beta'}[\xi'_4,\xi'_5,\xi'_6,\xi'_1,\xi'_2,\xi'_3]$. 
    
    \item $\mathcal{C}_{1-9}$ terms
    \begin{align}
         b_3[\xi'_1,\xi'_2,\xi'_3]\times b_4 [\xi'_4,\xi'_5,\xi'_6] \times  \mathcal{F}_{C}[\xi'_1,\cdots,\xi'_6], \nonumber
    \end{align}
    with $\mathcal{F}_{C} = \mathcal{F}_{C_1} + \cdots,+\mathcal{F}_{C_9} $. Here the coefficient tensor is defined as, 
    \begin{align}
        \mathcal{F}_{C_1} & = + \mathcal{F}_{i'\alpha'j'\beta'}[\xi'_1,\xi'_2,\xi'_4,\xi'_3, \xi'_5,\xi'_6],  \\
        \mathcal{F}_{C_2} & = - \mathcal{F}_{i'\alpha'j'\beta'}[\xi'_1,\xi'_2,\xi'_4,\xi'_5, \xi'_3,\xi'_6], \nonumber \\
        &\cdots. \nonumber
    \end{align}

    \item $\mathcal{C}_{10-18}$ terms
    \begin{align}
          b_4[\xi'_1,\xi'_2,\xi'_3]\times b_3 [\xi'_4,\xi'_5,\xi'_6] \times  \mathcal{F}_{C'}[\xi'_1,\cdots,\xi'_6], \nonumber
    \end{align}
    with $\mathcal{F}_{C'} = \mathcal{F}_{C_{10}} + \cdots,+\mathcal{F}_{C_{18}} $. Again, the coefficient tensor is defined as, 
    \begin{align}
        \mathcal{F}_{C_{10}} & = - \mathcal{F}_{i'\alpha'j'\beta'}[\xi'_1,\xi'_4,\xi'_5,\xi'_2, \xi'_3,\xi'_6],  \\
        \mathcal{F}_{C_{11}} & = + \mathcal{F}_{i'\alpha'j'\beta'}[\xi'_1,\xi'_4,\xi'_5,\xi'_2, \xi'_6,\xi'_3], \nonumber \\
        &\cdots. \nonumber
    \end{align}
\end{itemize}

After the Wick contraction and color-spin indices summation, each term has indices $i,\alpha,j, \beta, \vec p$ at the sink. For instance, $\mathcal D_1$ and $\mathcal D_2$ have the following contributions to the final correlation function,
\begin{align}
 D_{1,2}(i,\alpha,j,\beta,\vec r_\text{snk}) & = \sum_{\vec p} \mathcal D_{1,2}(i,\alpha,j,\beta,\vec p) \times e^{i\vec p \cdot\vec r_\text{snk}}.
\end{align}
Note that $\mathcal D_1(i,\alpha,j,\beta,\vec p) = -\mathcal D_2(j,\beta,i,\alpha,-\vec p)$, so that we have
\begin{align}
     D_1(i,\alpha,j,\beta,\vec r_\text{snk})  =  -D_2(j,\beta,i,\alpha,- \vec r_\text{snk}).
\end{align}
The spin-projection to $s=0$ is performed as $\sum\limits_{i,\alpha,j,\beta} D_1(i,\alpha,j,\beta,\vec r_\text{snk}) f_{s=0}(i,\alpha,j,\beta)$,
with the coefficient $f_{s=0}(i,\alpha,j,\beta) =- f_{s=0}(j,\beta, i,\alpha)$. Therefore, we have
\begin{align}
 D^{s=0}_1(\vec r_\text{snk}) = D^{s=0}_2(-\vec r_\text{snk}).
\end{align}
After $A^+_1$ projection, $D_1$ and $D_2$ will be exactly same. 
The above arguments also apply to $\mathcal C_{1}$ and  $\mathcal C_{18}$, $\mathcal C_{2}$ and  $\mathcal C_{17}$, etc. Eventually, we find that contributions from $\mathcal C_{1-9}$ equal to those from  $\mathcal C_{10-18}$.
Therefore, it is sufficient to calculate $\mathcal{D}_1$ and $\mathcal C_{1-9}$ in practice.

In summary, the calculation requires constructing four hadron blocks defined in Eqs.~(\ref{Eq-hadron-block-1})-(\ref{Eq-hadron-block-4}) and performing contraction/summation associated with $\mathcal{D}_1$ and $\mathcal C_{1-9}$.
The computational cost is approximately twice that of the conventional calculation with identical source positions ($y=w$), where $\mathcal{D}_1$ and $\mathcal C_{1-9}$ can be calculated in a unified framework.

\section{Sink-operator dependence of the HAL QCD potential}\label{Sec-sink-opr}

The correlation functions $R(\vec r,t)$ in Figs.~\ref{Fig_R_smr_p01} and ~\ref{Fig_R_smr_p01_pt} calculated using smeared sink and point sink allow us to investigate the sink-operator dependence of the HAL QCD potential.

\begin{figure*}[htbp]
  \centering
  \includegraphics[width=8.7cm]{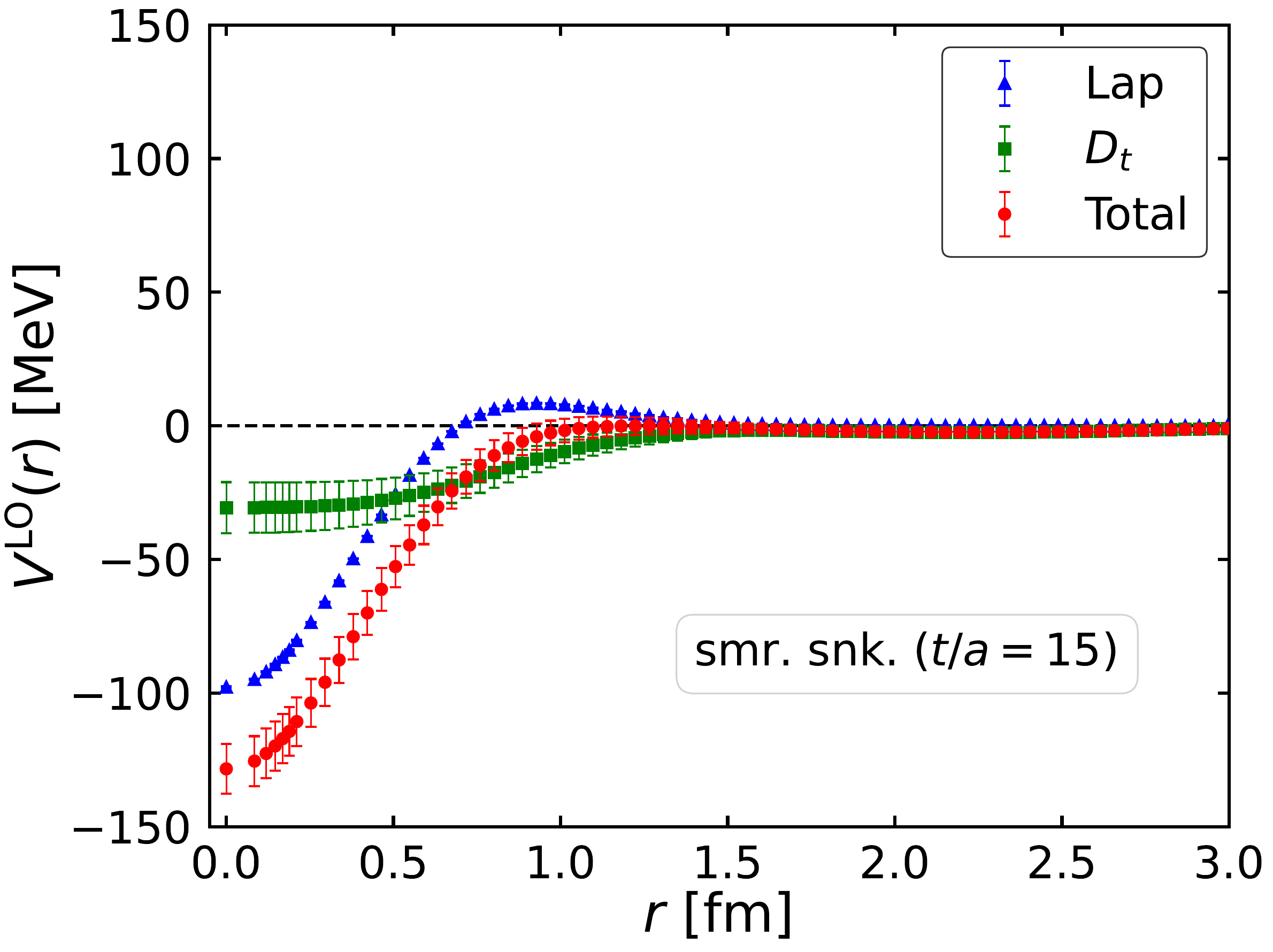}
  \includegraphics[width=8.7cm]{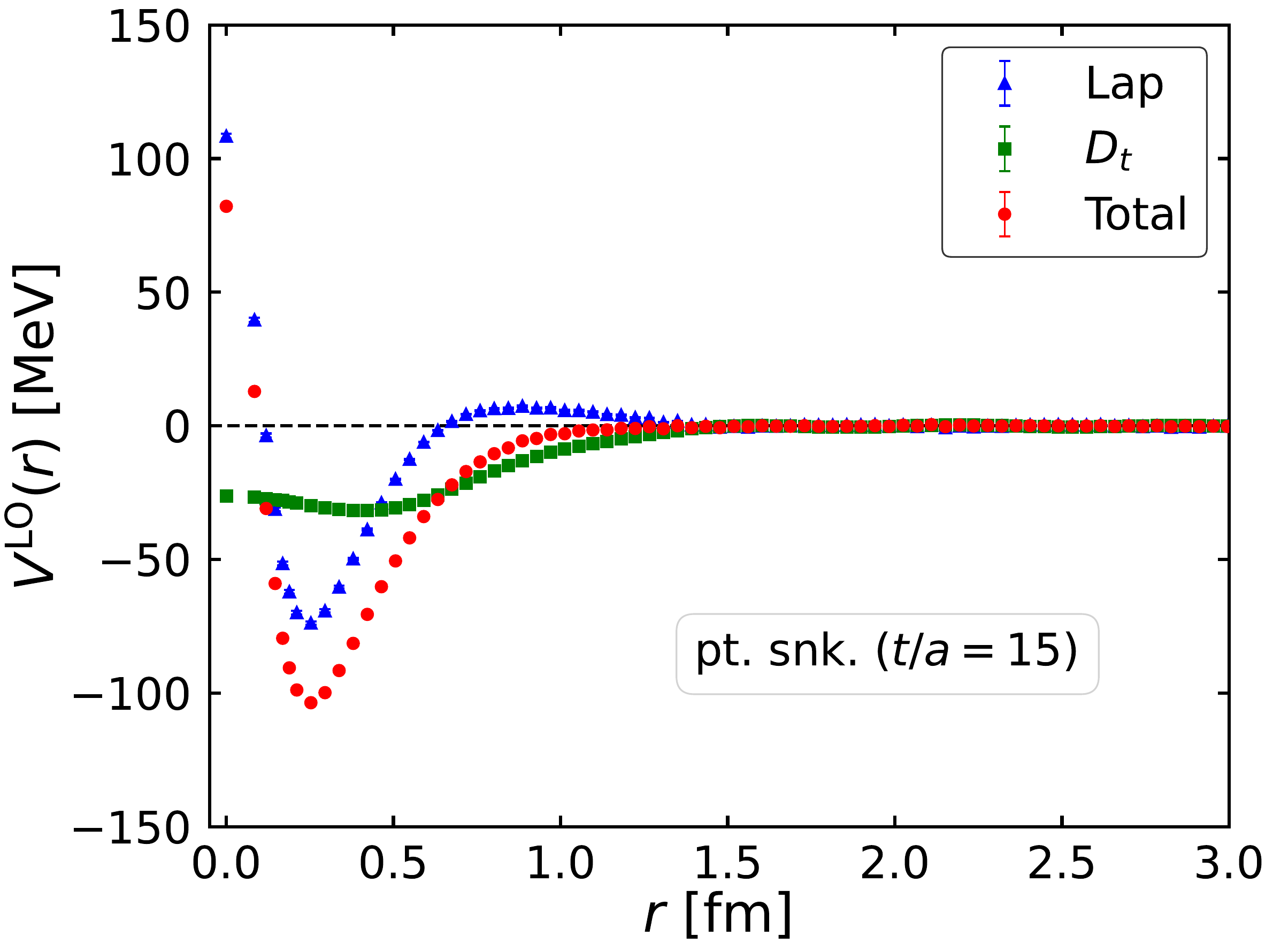}
  \caption{ 
  The LO potentials calculated from the zero-momentum source together with the smeared-sink (left) and point-sink (right), respectively.} \label{Fig-V-snk-opr}
\end{figure*}

\begin{figure*}[htbp]
  \centering
  \includegraphics[width=8.7cm]{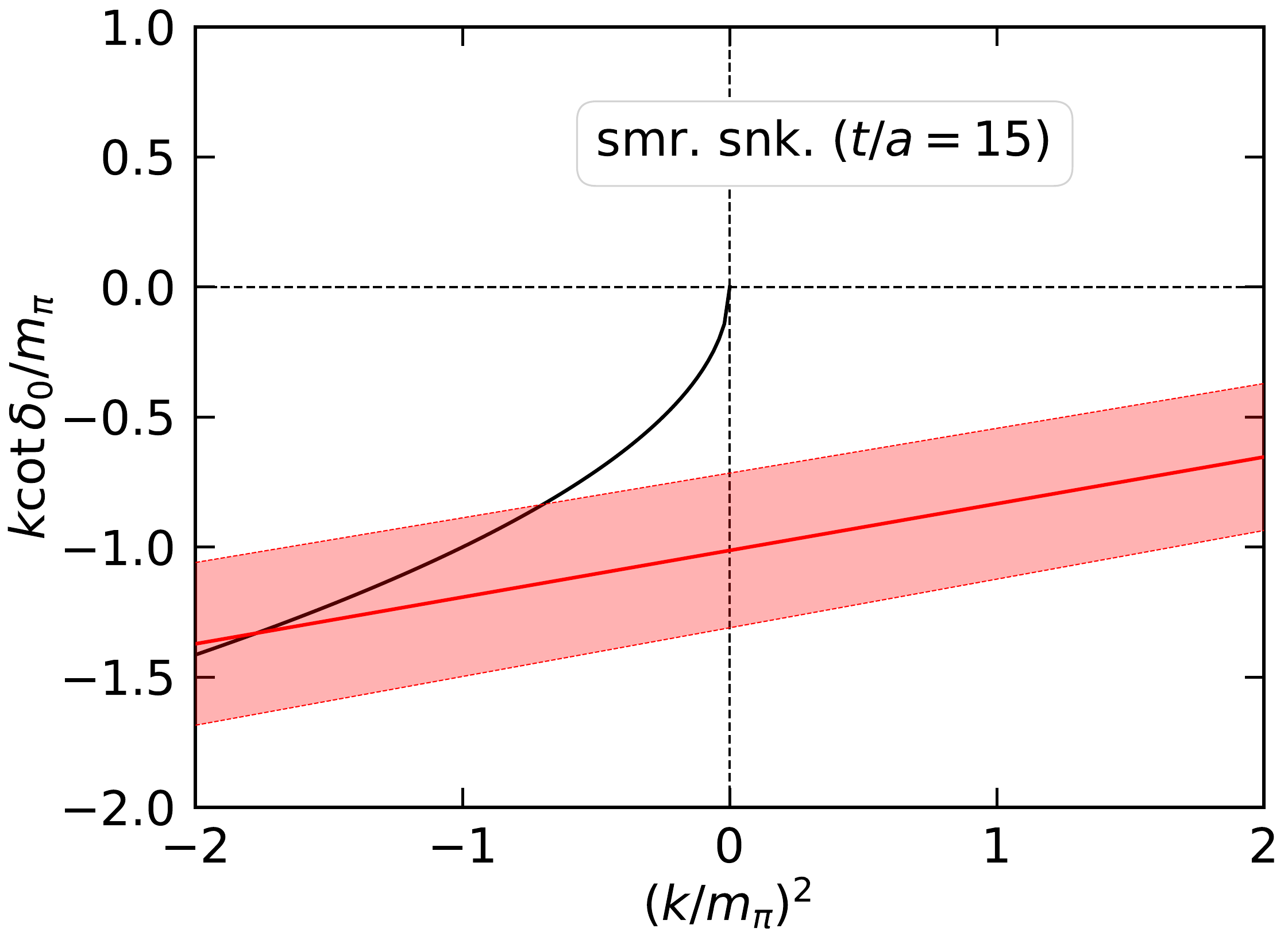}
  \includegraphics[width=8.7cm]{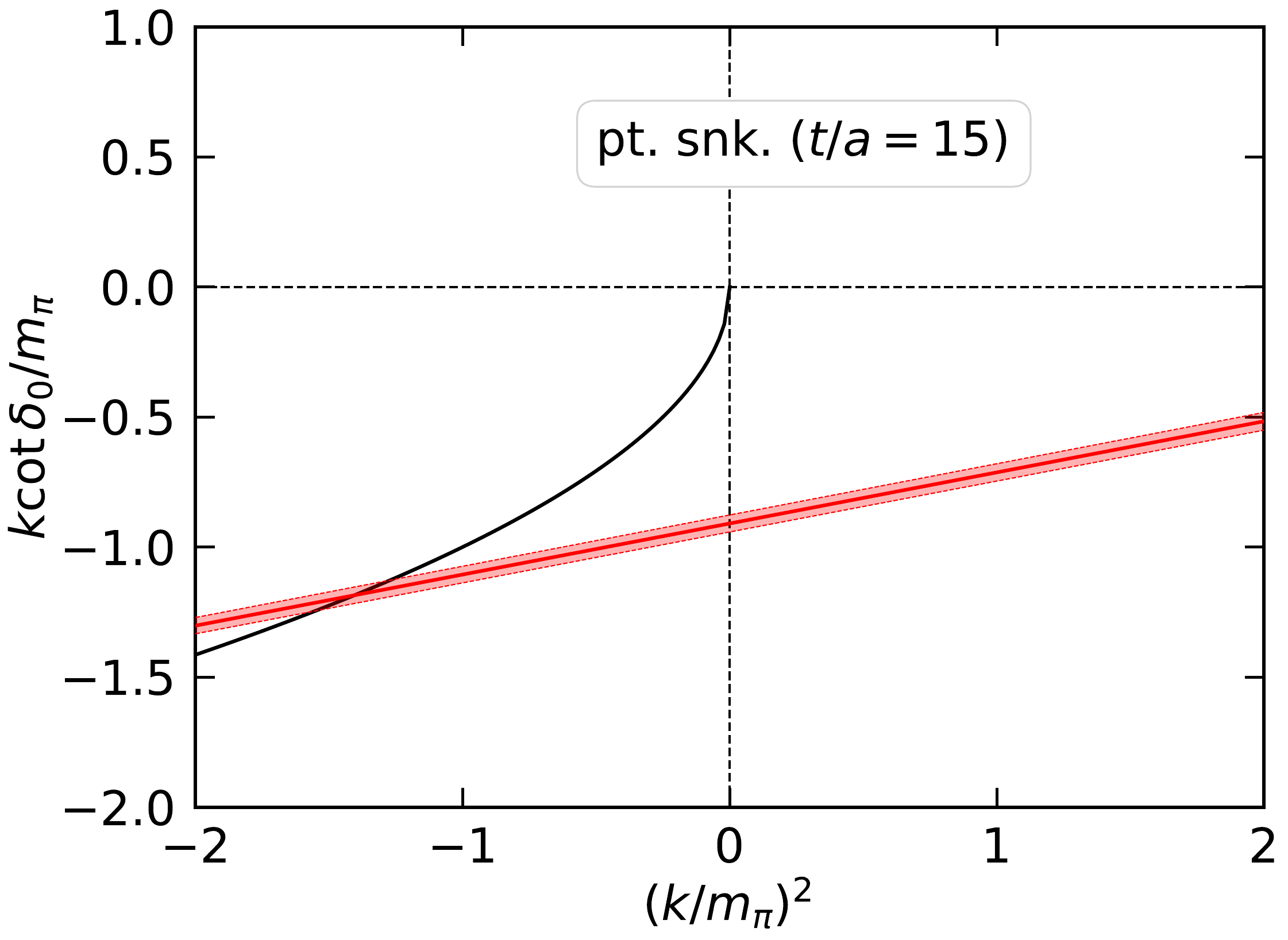}
  \caption{
  The scattering phase shifts calculated using potentials in Fig.~\ref{Fig-V-snk-opr}.} \label{Fig-delta-snk-opr}
\end{figure*}

Fig.~\ref{Fig-V-snk-opr} presents the LO potentials obtained from the zero-momentum source with the smeared sink (left) and the point sink (right), respectively.
Both Lap term and $D_t$ term in the two potentials are very similar except in the region $r\lesssim0.3$ fm,
where the total potential from the smeared sink exhibits attraction 
while that from the point sink shows repulsion.
The sink-operator dependence of the potential observed $r\lesssim0.3$ fm is natural, since the size of $\Omega_{ccc}$, $\sqrt{\langle r^2\rangle}\simeq 0.28$ fm \cite{Can2015}, is comparable with this distance. Thus, the two baryons strongly overlap in this region. 
As is well known, although the potential provides important information for understanding the underlying dynamics, it is not itself a physical observable. 
Therefore, despite the differences at short distances, the two potentials should yield consistent physical observables.
As shown in Fig.~\ref{Fig-delta-snk-opr}, we confirm that the scattering phase shifts extracted from the two potentials are indeed consistent with each other, although the statistical uncertainty is larger in the smeared-sink case than in the point-sink case.

%==========================
%==========================

%==========================
%==========================

\section{Scattering phase shifts}\label{Sec-phaseshift-multi}
\begin{figure*}[htbp]
  \centering
  \includegraphics[width=8.7cm]{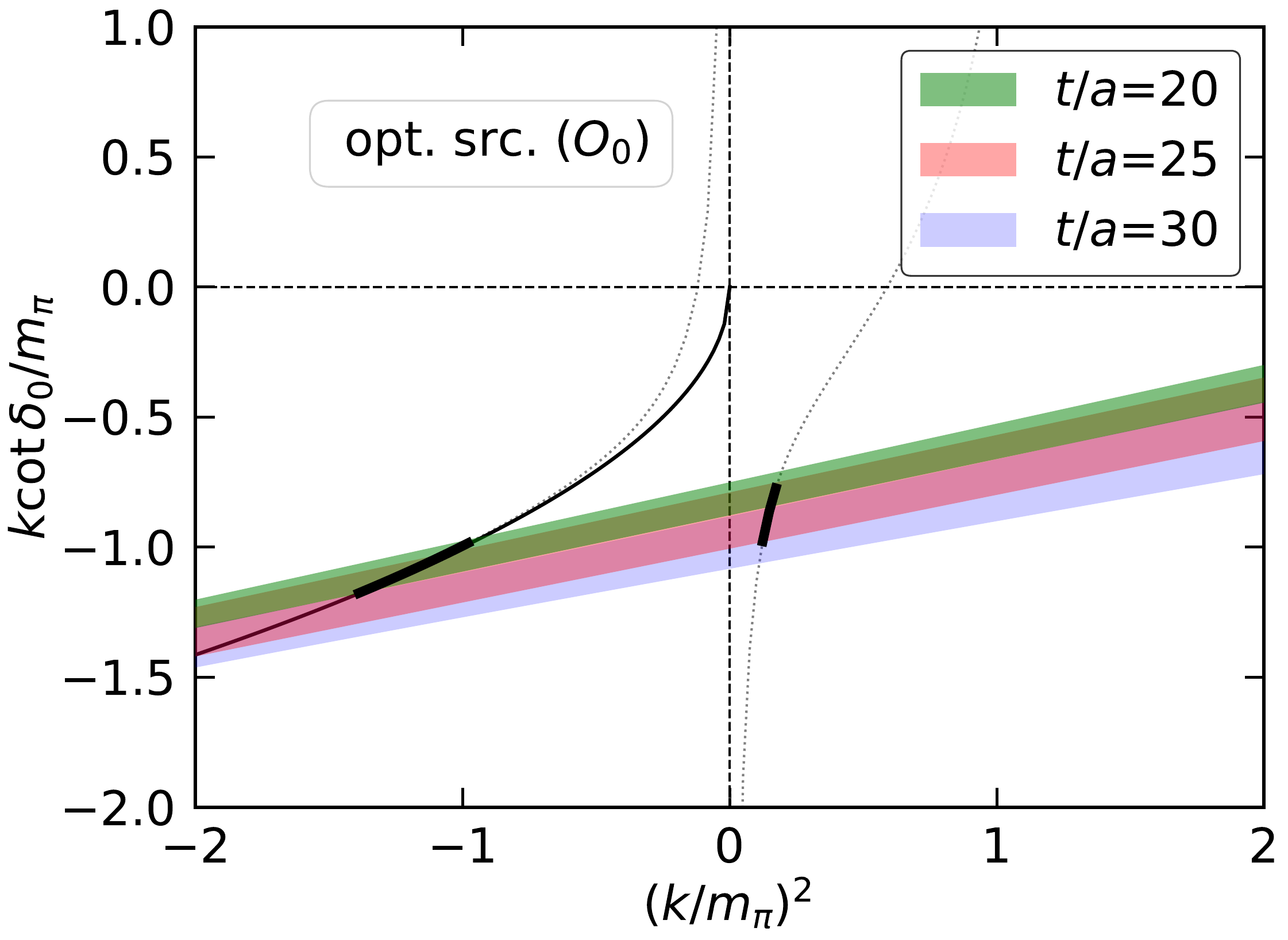}
  \includegraphics[width=8.7cm]{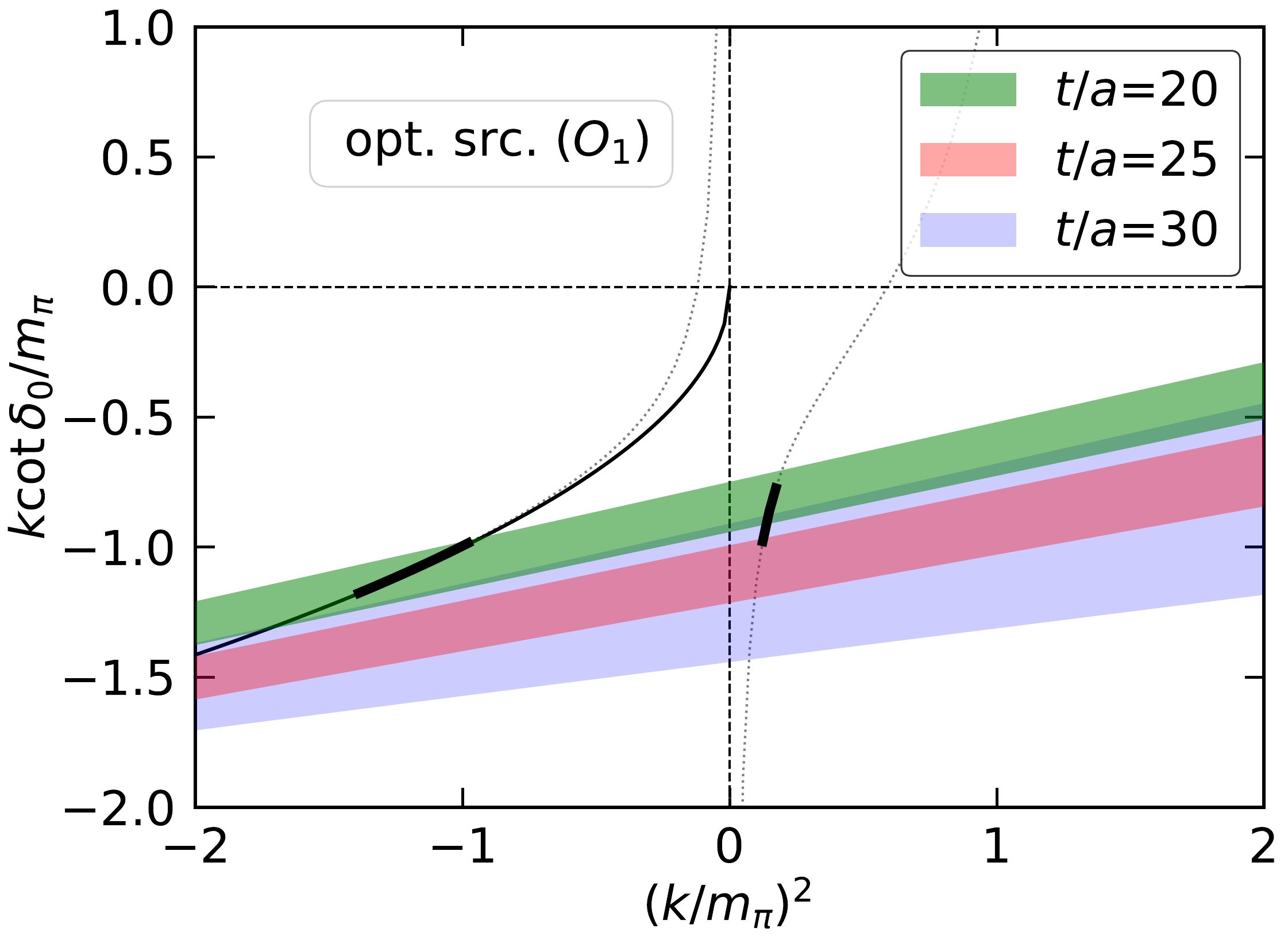}
  \includegraphics[width=8.7cm]{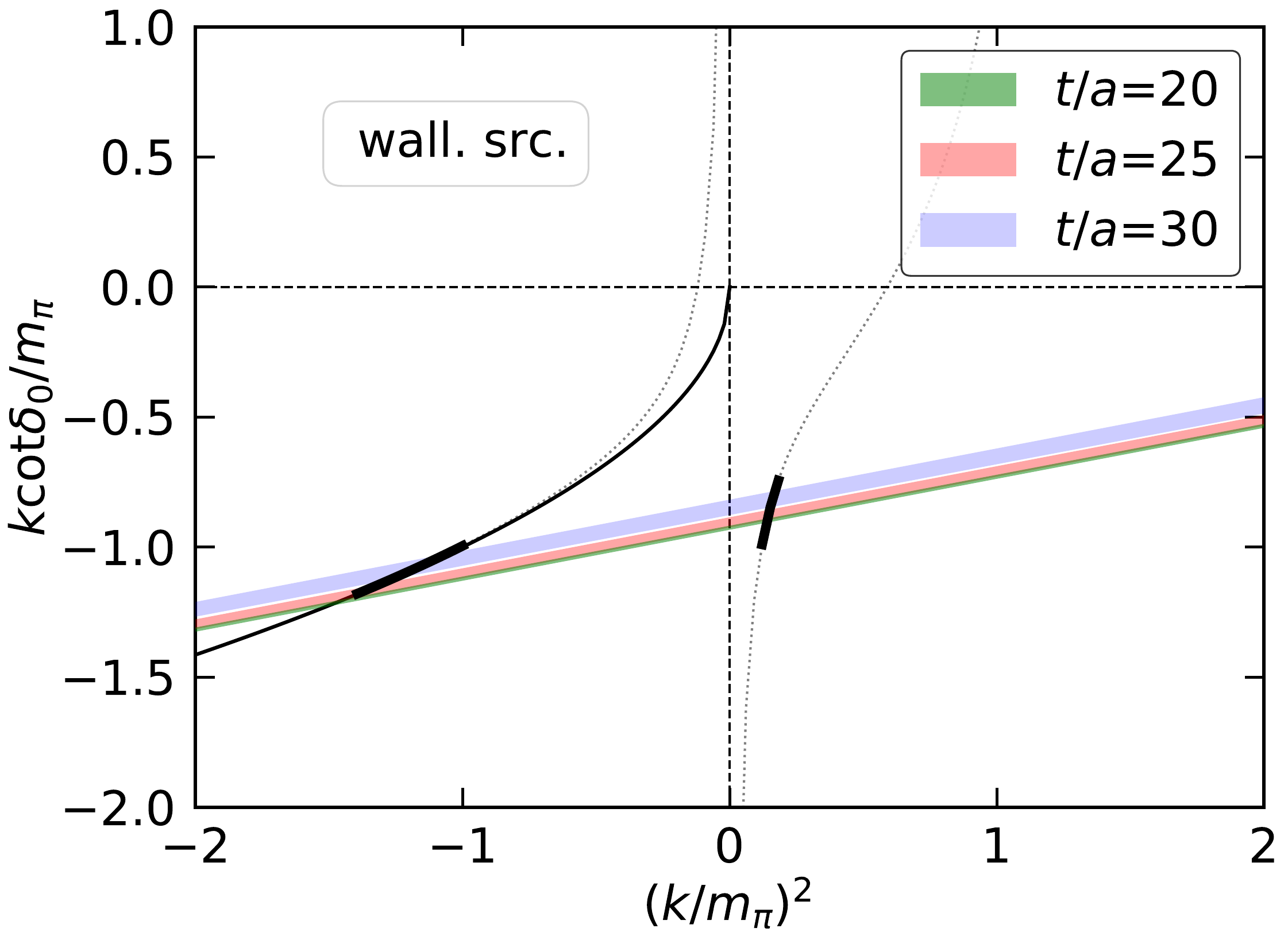}
  \includegraphics[width=8.7cm]{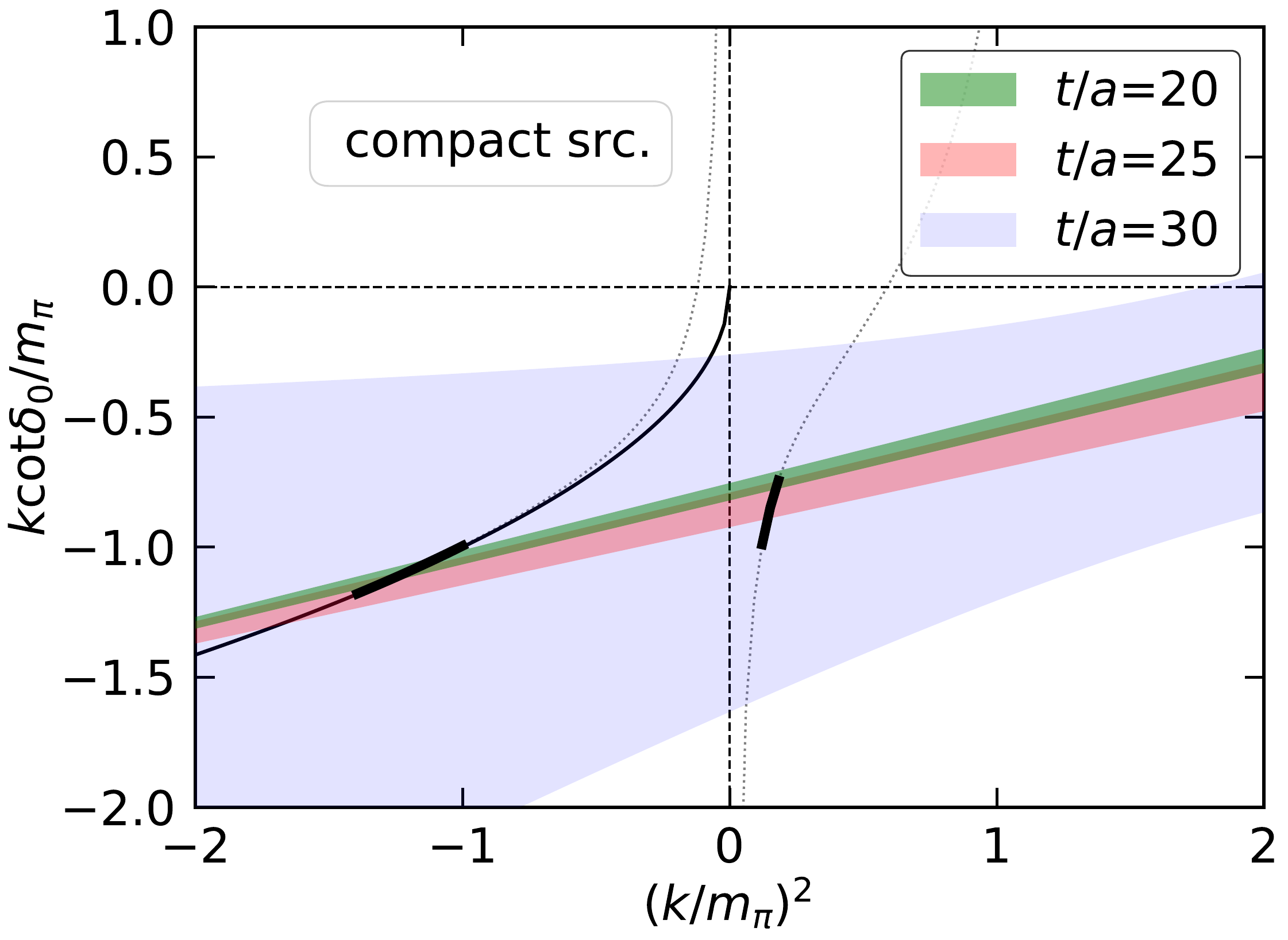}
  \caption{ 
  The scattering phase shifts calculated using the LO potentials for each source at multiple Euclidean time $t/a=20$, $25$, and $30$.
  Two thick black line segments represent the scattering phase shifts converted from finite volume spectral in Fig.~\ref{Fig-Eeff}.
  The black solid line $-\sqrt{-(k/m_\pi)^2}$ is the bound state condition.
  } \label{Fig-phaseshift-all-srcs}
\end{figure*}

 In Fig.~\ref{Fig-phaseshift-all-srcs}, we show the scattering phase shifts calculated using potentials from each source at multiple Euclidean time.
Results from the wall source show least statistical fluctuation, while those from the compact source exhibit large uncertainties in particular at $t/a=30$.
We find all results are consistent with each other within statistical fluctuations.

%==========================
%==========================

\

%\bibliography{Reference}
%=======================================================================================

%apsrev4-2.bst 2019-01-14 (MD) hand-edited version of apsrev4-1.bst
%Control: key (0)
%Control: author (8) initials jnrlst
%Control: editor formatted (1) identically to author
%Control: production of article title (0) allowed
%Control: page (0) single
%Control: year (1) truncated
%Control: production of eprint (0) enabled
%

\end{document}